\documentclass[useAMS,usenatbib,fleqn]{mn2e}
\usepackage{amsmath}
\usepackage{amssymb}
\usepackage{graphicx}
\usepackage{caption}
\usepackage{hyperref}
\usepackage{subcaption}
\usepackage{graphicx,epstopdf}
\usepackage{times}
\usepackage{balance} 
\usepackage{verbatim}

\title[A Sparse GP Framework for Photometric Redshift]{A Sparse Gaussian Process Framework for Photometric Redshift Estimation}
\author[Almosallam et al.]
{\parbox{\textwidth}{Ibrahim A. Almosallam,$^{1,2}$\thanks{E-mail: ialmosallam@kacst.edu.sa} Sam N. Lindsay,$^{3}$ Matt J. Jarvis$^{3,4}$ and Stephen J. Roberts$^{2}$
}
\vspace{0.4cm}\\
\parbox{\textwidth}{
$^1$King Abdulaziz City for Science and Technology, Riyadh, Saudi Arabia\\
$^2$Information Engineering, Parks Road, Oxford, OX1 3PJ, UK\\
$^3$Oxford Astrophysics, Department of Physics, Keble Road, Oxford, OX1 3RH, UK\\
$^4$Department of Physics, University of the Western Cape, Bellville 7535, South Africa\\
}}

\begin{document}

\date{\today}

\pagerange{\pageref{firstpage}--\pageref{lastpage}} \pubyear{2015}

\maketitle

\label{firstpage}

\begin{abstract}
Accurate photometric redshifts are a lynchpin for many future experiments to pin down the cosmological model and for studies of galaxy evolution. In this study, a novel sparse regression framework for photometric redshift estimation is presented. Synthetic dataset simulating the {\em Euclid} survey and real data from SDSS DR12 are used to train and test the proposed models. We show that approaches which include careful data preparation and model design offer a significant improvement in comparison with several competing machine learning algorithms. Standard implementations of most regression algorithms use the minimization of the sum of squared errors as the objective function. For redshift inference, this induces a bias in the posterior mean of the output distribution, which can be problematic. In this paper we directly minimize the target metric $\Delta z = (z_\textrm{s} - z_\textrm{p})/(1+z_\textrm{s})$ and address the bias problem via a distribution-based weighting scheme, incorporated as part of the optimization objective. The results are compared with other machine learning algorithms in the field such as Artificial Neural Networks (ANN), Gaussian Processes (GPs) and sparse GPs. The proposed framework reaches a mean absolute $\Delta z = 0.0026(1+z_\textrm{s})$, over the redshift range of $0 \le z_\textrm{s} \le 2$ on the simulated data, and $\Delta z = 0.0178(1+z_\textrm{s})$ over the entire redshift range on the SDSS DR12 survey, outperforming the standard {\sc ANNz} used in the literature. We also investigate how the relative size of the training sample affects the photometric redshift accuracy. We find that a training sample of \textgreater 30 per cent of total sample size, provides little additional constraint on the photometric redshifts, and note that our GP formalism strongly outperforms {\sc ANNz} in the sparse data regime for the simulated data set.

\end{abstract}


\section{Introduction}
The radial component of the position of a distant object is inferred from its cosmological redshift, induced by the expansion of the Universe; the light observed from a distant galaxy appears to us at longer wavelengths than in the rest frame of that galaxy. The most accurate determination of the exact redshift, $z$, comes from directly observing the spectrum of an extragalactic source and measuring a consistent multiplicative shift, relative to the rest frame, of various emission (or absorption) features. The rest-frame wavelengths of these emission lines are known to a high degree of accuracy which can be conferred onto the measured spectroscopic redshifts, $z_\textrm{s}$. However, the demand on telescope time to obtain spectra for every source in deep, wide surveys is prohibitively high, and only relatively small area spectroscopic campaigns can reach faint magnitudes \citep[e.g.][]{Lilly2009,LeFevre2013,LeFevre2015}, or at the other extreme, relatively bright magnitudes over larger areas \citep[e.g.][]{2dfgrs,GAMA,SDSS3}.
This forces us towards the use of photometric observations to infer the redshift by other means. Rather than individual spectra, the emission from a distant galaxy is observed in several broad filters, facilitating the characterization of the spectral energy distribution (SED) of fainter sources, at the expense of fine spectral resolution.

Photometric redshift methods largely fall into two categories, based on either SED template fitting or machine learning. Template fitting software such as {\sc Hyperz}; \citep[][]{Hyperz}, {\sc ZEBRA}; \citep{ZEBRA}, {\sc EAZY}; \citep[][]{EAZY} and {\sc Le Phare} \citep[][]{Ilbert2006} rely on a library of SED templates for a variety of different galaxy types, which (given the transmission curves for the photometric filters being used) can be redshifted to fit the photometry. This method can be refined in various ways, often with the use of simulated SEDs rather than only those observed at low redshift, composite SEDs, and through calibration using any available spectroscopic redshifts. Machine learning methods such as artificial neural networks \citep[e.g. {\sc ANNz};][]{Firth2003,Collister04}, nearest-neighbour (NN) \citep{Ball2008}, genetic algorithms \citep[e.g.][]{Hogan2015}, self-organized maps \citep[][]{Geach2012} and random forest \citep[][]{kind2013}, to name but a few, rely on a significant fraction of sources in a photometric catalogue having spectroscopic redshifts. These `true' redshifts are used to train the algorithm. In addition to providing a point estimate, machine learning methods can provide the degree of uncertainty in their prediction \citep{kind2013,bonnett2015,rau2015}.

Both methods have their strengths and weaknesses, with the best performance often depending on the available data and the intended science goals. As such, future surveys may well depend on contributions from both in tandem, but there has been extensive work on comparing the current state of the art in public software using a variety of techniques \citep{hildebrandt10,abdalla11,sanchez14,bonnett2015}. Artificial neural networks motivate the most commonly used machine learning software \citep{Firth2003,vanzella2004photometric,brescia2014catalogue}, however Gaussian Processes \citep[e.g.][]{Way2009} have not yet become well established in this area, despite comparison by \citet{bonfield10} suggesting that they may outperform the popular {\sc ANNz} code, using the rms error as a metric.

In this paper, we introduce a novel sparse kernel regression model that greatly reduces the number of basis (kernel) functions required to model the data considered in this paper. This is achieved by allowing each kernel to have its own hyper-parameters, governing its shape. This is in contrast to the standard kernel-based models in which a set of global hyper-parameters are optimized (such as is typical in Gaussian Process (GP) methods). The complexity cost of such a kernel-based regression model is $O\left(n^{3}\right)$, where $n$ is the number of basis functions. This cubic time complexity arise from the cost of inverting an $n$ by $n$ covariance matrix. In a standard Gaussian Process model \citep{rasmussen2006gaussian}, seen as a kernel regression algorithm, we may regard the basis functions, as located at the $n$ points in the training sample. This renders such an approach unusable for many large training data applications where scalability is a major concern. Much of the work done to make GPs more scalable is either to (a) make the inverse computation faster or (b) use a smaller representative set from the training sample to reduce the rank and ease the computation of the covariance matrix. Examples of (a) include methods such as structuring the covariance matrix such that it is much easier to invert, using Toeplitz \citep{zhang2005time} or Kronecker decomposition \citep{tsiligkaridis2013}, or inverse approximation as an optimization problem \citep{gibbs97}. To reduce the number of representative points (b), an $m \ll n$ subset of the training sample can be selected which maximizes the accuracy or the numerical stability of the inversion \citep{foster2009}. Alternatively, one may search for ``inducing'' points not necessarily present in the training sample, and not necessarily even lying within the data range, to use as the basis set such that the probability of the data being generated from the model is maximized \citep{snelson2005}. Approaches such as Relevance Vector Machines \citep[RVM;][]{tipping2001} and Support Vector Machines \citep[SVM;][]{smola1997} are basis-function models. However, unlike sparse GPs, they do not learn the basis functions' locations but rather apply shrinkage to a set of kernels in the form of weight-decay on the linear weights that couple the kernels, located at training data points, to the regression. 

In this paper, we propose a non-stationary sparse Gaussian model to target photometric redshift estimation. The key difference between the proposed approach and other basis function models, is that our model does not use shrinkage (automatic relevance determination) external to the kernel, but instead has a length-scale parameter in each kernel. This allows for parts of the input-output regression mapping to have different characteristic length-scales. We can see this as allowing for shrinkage and reducing the need for more basis functions, as well as allowing for non-stationary mappings. A regular GP, sparse GP or RVM does not do this, and we demonstrate that this is advantageous to photometric redshift estimation. Furthermore, the model is presented within a framework with components that address other challenges in photometric redshift estimation such as incorporating a weighting scheme as an integral part of the process to remove, or introduce, any systematic bias, and a prior mean function to enhance the extrapolation performance of the model. The results are demonstrated on photometric redshift estimation for a simulated {\em Euclid}-like survey \citep{laureijs2011} and on observational data from the 12th Data Release of the Sloan Digital Sky Survey (SDSS) \citep{SDSS3}\footnote{\url{www.sdss.org}} . In particular, we use the weighting scheme to remove any distribution bias and introduce a linear bias to directly target the mission's requirement. 

The paper is organised as follows, a brief introduction to Gaussian Processes for regression is presented in Section \ref{sec-gaussian-process} followed by an introduction to sparse GPs in Section \ref{sec-sparse-gaussian-processes}. The proposed approach is described in Section \ref{sec-proposed-approach} followed by an application to photometric redshift estimation in Section \ref{sec-application}, where the details of the mock dataset are described. The experiments and results are discussed in Section \ref{sec-experiments} on the simulated survey, and in Section \ref{sec-experiments-sdss} we demonstrate the performance of the proposed model and compare it to ANNz on the SDSS 12th Data Release. Finally, we summarize and conclude in Section \ref{sec-summary}.

\section{Gaussian Processes}
\label{sec-gaussian-process}
In many modelling problems, we have little prior knowledge of the explicit functional form of the function that maps our observables onto the variable of interest. Imposing, albeit sensible, parametric models, such as polynomials, makes a tacit bias. For this reason, much of modern function modelling is performed using \emph{non-parametric} techniques. For regression, the most widely used approach is that of \emph{Gaussian Processes} \citep{rasmussen2006gaussian}.
A Gaussian Process is a supervised non-linear regression algorithm that makes few explicit \emph{parametric} assumptions about the nature of the function fit. For this reason, Gaussian Processes are seen as lying within the class of Bayesian non-parametric models. The underlying assumption in a GP is that, given a set of input $\mathbf{X}=\left\{\mathbf{x}_{i}\right\}_{i=1}^{n}\in \mathbb{R}^{n\times d}$ and a set of target outputs $\mathbf{y}=\left\{y_{i}\right\}_{i=1}^{n}\in \mathbb{R}^{n}$, where $n$ is the number of samples in the dataset and $d$ is the dimensionality of the input, the observed target $y_{i}$ is generated by a function $\mathbf{f}$ of the input $\mathbf{x}_{i}$ plus additive noise $\epsilon_{i}$:
\begin{equation}
y_{i} = \mathbf{f}\left(\mathbf{x}_{i}\right)+\epsilon_{i}.
\end{equation}
The noise $\epsilon$ is taken to be normally distributed with a mean of zero and variance $\sigma_{n}^{2}$, or $\epsilon\sim\mathcal{N} \left(0,\sigma_{n}^{2}\right)$. To simplify the notation, it is assumed that $\mathbf{y}\sim\mathcal{N} \left(0,1\right)$ (this can readily be achieved without loss of generality, via a linear whitening process) and univariate, although the derivation can be readily extended to multivariable problems. The conditional probability of the observed variable given the function is hence distributed as follows:
\begin{equation}
p\left(\mathbf{y}|\mathbf{f}\right)\sim\mathcal{N} \left(\mathbf{f},\sigma_{n}^{2}\right).
\end{equation}
A GP then proceeds by applying a \emph{Bayesian} treatment to the problem to infer a probability distribution over the space of possible functions $\mathbf{f}$ given the data:
\begin{equation}
p\left(\mathbf{f}|\mathbf{y},\mathbf{X}\right) = \frac{p\left(\mathbf{y}|\mathbf{f}\right)p\left(\mathbf{f}|\mathbf{X}\right)}{p\left(\mathbf{y}|\mathbf{X}\right)}.
\end{equation}

This requires us to define a prior, $p\left(\mathbf{f}|\mathbf{X}\right)$, over the function space. The function is normally distributed with a mean of zero, to match the mean of the normalized variable $\mathbf{y}$, with a covariance \emph{function} $\mathbf{K}$, i.e. $p\left(\mathbf{f}|\mathbf{X}\right)\sim\mathcal{N} \left(0,\mathbf{K}\right)$. The covariance function captures prior knowledge about the relationships between the observables. Most widely used covariance functions assume that there is local similarity in the data, such that nearby inputs are mapped to similar outputs. The covariance $\mathbf{K}$ can therefore be modelled as a function of the input $\mathbf{X}$, $\mathbf{K}=k\left(\mathbf{X},\mathbf{X}\right)$, where each element $\mathbf{K}_{ij}=k\left(\mathbf{x}_{i},\mathbf{x}_{j}\right)$ and $k$ is the covariance function. For $\mathbf{K}$ to be a valid covariance it has to be symmetric and positive semi-definite matrix; arbitrary functions for $k$ cannot guarantee these constraints. A class of functions that guarantees these structural constraints are referred to as \emph{Mercer kernels} \citep{mercer1909}. A commonly used kernel function, which is the focus of this work, is the squared exponential kernel, defined as follows:
\begin{equation}
k\left(\mathbf{x}_{i},\mathbf{x}_{j}\right) = \sigma^{2}\exp\left(-\frac{1}{2\lambda^{2}}\left\|\mathbf{x}_{i}-\mathbf{x}_{j}\right\|^{2}\right),
\label{eq-squared-exponential}
\end{equation}
where $\sigma$ and $\lambda$ are referred to as the height (output, or variance) and characteristic length (input) scale respectively, which correspond to tunable \emph{hyper-parameters} of the model. The similarity between two input vectors, under the squared exponential kernel, is a non-linear function of the Euclidean distance between them. We note that this choice of kernel function guarantees continuity and smoothness in the function and all its derivatives. For a more extensive discussion of covariances, the reader is referred to \citep{rasmussen2006gaussian} or \citep{roberts2012rs}. With the likelihood $p\left(\mathbf{y}|\mathbf{f}\right)$ and prior $p\left(\mathbf{f}|\mathbf{X}\right)$, the marginal likelihood $p\left(\mathbf{y}|\mathbf{X}\right)$ can be computed as follows:
\begin{equation}
p\left(\mathbf{y}|\mathbf{X}\right) = \int p\left(\mathbf{y}|\mathbf{f}\right)p\left(\mathbf{f}|\mathbf{X}\right) \mathrm{d}\mathbf{f},
\end{equation}
By multiplying the likelihood and the prior and completing the square over $\mathbf{f}$, we can express the integration as a normal distribution independent of $\mathbf{f}$ multiplied by a another normal distribution over $\mathbf{f}$. The distribution independent of $\mathbf{f}$ can then be taken out of the integral, and the integration of the second normal distribution with respect to $\mathbf{f}$ will be equal to one. The resulting distribution of the marginal likelihood is distributed as follows \citep{rasmussen2006gaussian}:
\begin{equation}
p\left(\mathbf{y}|\mathbf{X}\right) \sim\mathcal{N} \left(0,\mathbf{K}+\sigma_{n}^{2}\mathbf{I}\right).
\label{eq-marginal-likelihood}
\end{equation}
The marginal likelihood of the full data set can hence be computed as follows:
\begin{equation}
p\left(\mathbf{y}|\mathbf{X}\right) = \prod_{i=1}^{n}\frac{1}{\sqrt{2\pi\left |\mathbf{K}+\sigma_{n}^{2}\mathbf{I}\right |}}\exp\left({-\frac{1}{2}y_{i}^{T}\left(\mathbf{K}+\sigma_{n}^{2}\mathbf{I}\right)^{-1}y_{i}}\right).
\label{eq-marginal-likelihood-complete}
\end{equation}

The aim of a GP, is to maximize the probability of observing the target $\mathbf{y}$ given the input $\mathbf{X}$, Eq. \eqref{eq-marginal-likelihood-complete}. Note that the only free parameters to optimize in the marginal likelihood are the parameters of the kernel and the noise variance, collectively referred to as the \emph{hyper-parameters} of the model. It is more convenient however to maximize the log of the marginal likelihood, Eq. \eqref{eq-log-marginal-likelihood}, since the log function is a monotonically increasing function, maximizing the log of a function is equivalent to maximizing the original function. The log likelihood is given as:
\begin{align}
\label{eq-log-marginal-likelihood}
\log\text{ }p(\mathbf{y}|\mathbf{X}) &= -\frac{1}{2}\mathbf{y}^{T}\left(\mathbf{K}+\mathbf{I}\sigma_{n}^{2} \right)^{-1}\mathbf{y} \nonumber \\
&\qquad -\frac{1}{2} \log\left | \mathbf{K}+\mathbf{I}\sigma_{n}^{2}\right|-\frac{n}{2}\log(2\pi).
\end{align}

We search for the optimal set of hyper-parameters using a gradient-based optimization, hence we require the derivatives of the log marginal likelihood with respect to each hyper-parameter. In this paper, the L-BFGS algorithm was used to optimize the objective which uses a Quasi-Newton method to compute the search direction in each step by approximating the inverse of the Hessian matrix from the history of gradients in previous steps \citep{jorge1980,schmidt2005}. It is worth mentioning that non-parametric models require the optimization of few hyper-parameters that do not grow with the size of the data and are less prone to overfitting. The distinction between parameters and hyper-parameters of a model is that the former directly influence the input-output mapping, for example the linear coupling weights in a basis function model, whereas the latter affect properties of distributions in the probabilistic model, for example the widths of kernels. Although this distinction is somewhat semantic, we keep to this nomenclature as it is standard in the statistical machine learning literature.

Once the hyper-parameters have been inferred, the conditional distribution of future predictions $\mathbf{f}_{*}$ for test cases $\mathbf{X}_{*}$ given the training sample can be inferred from the joint distribution of $\mathbf{f}_{*}$ and the observed targets $\mathbf{y}$. If we assume that the joint distribution is a multivariate Gaussian, then the joint probability is distributed as follows:
\begin{equation}
p\left ( \mathbf{y},\mathbf{f}_{*}|\mathbf{X},\mathbf{X}_{*}\right) \sim \mathcal{N} \left ( \mathbf{0}, \begin{bmatrix}\mathbf{K_{xx}}+\sigma_{n}^{2}\mathbf{I} & \mathbf{K_{x*}}\\\mathbf{K_{*x}} & \mathbf{K_{**}} \end{bmatrix}\right ),
\end{equation}
where we introduce the shorthand notations $\mathbf{K_{xx}}=k\left(\mathbf{X},\mathbf{X}\right)$, $\mathbf{K_{x*}}=k\left(\mathbf{X},\mathbf{X}_{*}\right)$, $\mathbf{K_{*x}}=k\left(\mathbf{X}_{*},\mathbf{X}\right)$ and $\mathbf{K_{**}}=k\left(\mathbf{X}_{*},\mathbf{X}_{*}\right)$. The conditional distribution $p\left(\mathbf{f}_{*}|\mathbf{y},\mathbf{X},\mathbf{X}_{*}\right)$ is therefore distributed as follows:
\begin{equation}
\begin{array}{rcl}
p\left(\mathbf{f}_{*}|\mathbf{y},\mathbf{X},\mathbf{X}_{*}\right)		&	\sim		&	\mathcal{N} \left ( \bmu, \mathbf\Sigma \right ),\\
\bmu			&	=		&	\mathbf{K_{*x}}\left(\mathbf{K_{xx}}+\sigma_{n}^{2}\mathbf{I}\right)^{-1}\mathbf{y},\\
\mathbf\Sigma		&	=		&	\mathbf{K_{**}}-\mathbf{K_{*x}}\left(\mathbf{K_{xx}}+\sigma_{n}^{2}\mathbf{I}\right)^{-1}\mathbf{K_{x*}}.
\end{array}
\label{eq-posterior-of-f}
\end{equation}
If we assume a non-zero prior mean $\bmu_{\mathbf{f}}$ over the function, $p(\mathbf{f})\sim\mathcal{N}\left(\bmu_{\mathbf{f}},\mathbf{K}\right)$, and an un-normalized $\mathbf{y}$ with mean $\bmu_{\mathbf{y}}$, the mean of the posterior distribution will be equal to:
\begin{equation}
\bmu	 = \bmu_{\mathbf{f}}+\mathbf{K_{*x}}\left(\mathbf{K_{xx}}+\sigma_{n}^{2}\mathbf{I}\right)^{-1}\left(\mathbf{y}-\bmu_{\mathbf{y}}\right),
\label{eq-conditional-distribution}
\end{equation}

The main drawback of GPs is the $O\left(n^{3}\right)$ computational cost required to invert the $n\times n$ matrix $\mathbf{K_{xx}}+\sigma_{n}^{2}\mathbf{I}$. The \emph{sparse Gaussian Process} allows us to reduce this computational cost and is detailed in the following section.

\section{Sparse Gaussian Processes}
\label{sec-sparse-gaussian-processes}
Gaussian processes are often described as non-parametric regression models due to the lack of an explicit parametric form. Indeed GP regression can also be viewed as a functional mapping $\mathbf{X}\in \mathbb{R}^{n\times d}:\rightarrow \mathbf{K}\in \mathbb{R}^{n\times n}$ parameterized by the data and the kernel function, followed by linear regression via optimization of the following objective:
\begin{equation}
\label{eq-linear-regression-objective}
\begin{array}{lcl}
\underset{\mathbf{w}}{\text{min}} &\frac{1}{2}\left ( \mathbf{K}\mathbf{w}-\mathbf{y} \right )^{T}\left( \mathbf{K}\mathbf{w}-\mathbf{y} \right )+\frac{1}{2}\sigma_{n}^{2}\mathbf{w}^{T}\mathbf{w},
\end{array}
\end{equation}
where $\mathbf{w}$ are the set of coefficients for the linear regression model that maps the transformed features $\mathbf{K}$ to the desired output $\mathbf{y}$. The feature transformation $\mathbf{K}$ evaluate how ``similar'' a datum is to every point in the training sample, where the similarity measure is defined by the kernel function. If two points have a high kernel response via Eq. \eqref{eq-squared-exponential}, this will result in very correlated features, adding extra computational cost for very little or no added information. Selecting a subset of the training sample that maximizes the preserved information is a research question addressed in \citet{foster2009}, whereas in \citet{snelson2005} the basis functions are treated as a search problem rather than a selection problem and their locations are treated as hyper-parameters which are optimized. These approaches result in a transformation $\mathbf{X}\in \mathbb{R}^{n\times d}:\rightarrow \mathbf{K}\in \mathbb{R}^{n\times m}$, in which $m\ll n$ is the number of basis functions used. The transformation matrix $\mathbf{K}$ will therefore be a rectangular $n$ by $m$ matrix and the solution for $\mathbf{w}$ in Eq. \eqref{eq-linear-regression-objective} is calculated via standard linear algebra as:
\begin{equation}
\label{eq-linear-regression-objective-rectangular}
\mathbf{w} = \left(\mathbf{K}^{T}\mathbf{K}+\mathbf{I}\sigma_{n}^{2} \right)^{-1}\mathbf{K}^{T}\mathbf{y}.
\end{equation}

Even though these models improve upon the computational cost of a standard GP, very little is done to compensate for the reduction in modelling power caused by the ``loss'' of basis functions. The selection method is always bounded by the full GP's accuracy, on the \emph{training} sample, since the basis set is a subset of the full GP basis set. On the other hand, the sparse GP's ability to place the basis set freely across the input space does go some way to compensate for this reduction, as the kernels can be optimized to describe the distribution of the data. In other words, instead of training a GP model with all data points as basis functions, or restricting it to a subset of the training sample which require some cost to select them, a set of inducing points is used in which their locations are treated as hyper-parameters of the model to be optimized. In both a full and a low rank approximation GP, a global set of hyper-parameters is used for all basis functions, therefore limiting the algorithm's local modelling capability. Moreover, the objective in Eq. \eqref{eq-linear-regression-objective} minimizes the sum of squared errors, therefore for any non-uniformly distributed output, the optimization routine will bias the model towards the mean of the output distribution and will seek to fit preferentially the region of space where there are more data. Hence, the model might allow for very poor predictions for few points in poorly represented regions, e.g. the high redshift range, in order to produce good predictions for well represented regions. Therefore, the error distribution as a function of redshift is not uniform unless the training sample is well balanced, producing a model that is sensitive to how the target output is distributed. 

In the next section, a method is proposed which addresses the above issues by parametrizing each basis function with bespoke hyper-parameters which account for variable density and/or patterns across the input space. This is particularly pertinent to determining photometric redshifts, where complete spectroscopic information may be restricted or biased to certain redshifts or galaxy types, depending on the target selection for spectroscopy of the training sample.
This allows the algorithm to learn more complex models with fewer basis functions. In addition, a weighting mechanism to remove any distribution bias from the model is directly incorporated into the objective.

\section{Proposed Approach}
\label{sec-proposed-approach}

In this paper, we extend the sparse GP approach by modelling each basis (kernel) function with its own set of hyper-parameters. The kernel function in Eq. \eqref{eq-squared-exponential} is hence redefined as follows:
\begin{equation}
\label{eq-squared-exponential-extension}
k(\mathbf{x}_{i},\mathbf{p}_{j}) = \exp{\left(-\frac{1}{2\blambda_{j}^{2}}\left\| \mathbf{x}_{i}-\mathbf{p}_{j}\right\|^{2}\right)},
\end{equation}
where $\mathbf{P}=\{\mathbf{p}_{j}\}_{j=1}^{m} \in \mathbb{R}^{m\times d}$ are the set of basis coordinates and $\blambda_{j}$ is the corresponding length scale for basis $j$. The multivariate input is denoted as $\mathbf{X}=\{\mathbf{x}_{i}\}_{i=1}^{n} \in \mathbb{R}^{n\times d}$. Throughout the rest of the paper, $\mathbf{X}_{i,*}$ denotes the $i$-th row of matrix $\mathbf{X}$, or $\mathbf{x}_{i}$ for short, whereas $\mathbf{X}_{*,j}$ denotes the $j$-th column and $\mathbf{X}_{i,j}$ refers to the element at row $i$ and column $j$ in matrix $\mathbf{X}$, and similarly for other matrices. Note that the hyper-parameter $\sigma$ has been dropped, as it interferes with the regularization objective. This can be seen from the final prediction equation $\hat y_{i}=\sum_{j=1}^{m}\mathbf{w}_{j}\bsigma_{j}^{2}\exp{\left(-\left\| \mathbf{x}_{i}-\mathbf{p}_{j}\right\|^{2}/2\blambda_{j}^{2}\right)}$, the weights are always multiplied by their associated $\sigma$. Therefore, the optimization process will always compensate for decreasing $\mathbf{w}_{j}^{2}$ by increasing $\bsigma_{j}^{2}$. Dropping the height variance ensures that the kernel functions do not grow beyond control and delegates learning the linear coefficients and regularization to the weights in $\mathbf{w}$. The derivatives with respect to each length scale and position are provided in equations Eq. \eqref{eq-dfdl} and Eq. \eqref{eq-dfdp} respectively:
\begin{subequations}
\begin{align}
\mathbf{E}\phantom{_{i,j}} &= \left(\left(\mathbf{K}\mathbf{w}-\mathbf{y}\right)\mathbf{w}^{T}\right)\circ\mathbf{K},\\
\mathbf\Delta_{j\phantom{,i}} &= \mathbf{X}-\vec{\mathbf{1}}_{n}\mathbf{p}_{j},\\
\mathbf{D}_{i,j} &= \left \| \mathbf{x}_{i}-\mathbf{p}_{j}\right\|^{2},\\
\label{eq-dfdl}
\frac{\partial f(\mathbf{X},\mathbf{y},\mathbf{w})}{\partial \blambda_{j}} &=\mathbf{E}_{*,j}^{T}\mathbf{D}_{*,j}\blambda_{j}^{-3},\\
\label{eq-dfdp}
\frac{\partial f(\mathbf{X},\mathbf{y},\mathbf{w})}{\partial \mathbf{p}_{j}} &=\mathbf{E}_{*,j}^{T}\mathbf\Delta_{j}\blambda_{j}^{-2}.
\end{align}
\end{subequations}
The symbol $\circ$ denotes the Hadamard product, i.e. element-wise matrix multiplication and $\vec{\mathbf{1}}_{n}$ denotes a column vector of length $n$ with all elements set to 1. Finding the set of hyper-parameters that optimizes the solution, is in effect finding the set of radial basis functions defined by their positions $\mathbf{p}$ and radii $\blambda$ that jointly describe the patterns across the input space. By parametrizing them differently, the model is more capable to accommodate different regions of the space more specifically. A global variance model assumes that the relationship between the input and output is global or equal across the input space, whereas a variable variance model, or non-stationary GP, makes no assumptions and learns the variable variances for each basis function which reduces the need for more basis functions to model the data. The kernel in Eq. \eqref{eq-squared-exponential-extension} can be further extended to, not only model each basis function with its own radius $\blambda_{j}$, but also model each one with its own covariance $\mathbf{C}_{j} \in \mathbb{R}^{d\times d}$. This enables the basis function to have any arbitrary shaped ellipses giving it more flexibility. The kernel in Eq. \eqref{eq-squared-exponential-extension} can be extended as follows:
\begin{equation}
\label{eq-squared-exponential-covariance-extension}
k(\mathbf{x}_{i},\mathbf{p}_{j}) = \exp{\left(-\frac{1}{2}\left(\mathbf{x}_{i}-\mathbf{p}_{j}\right)\mathbf{C}_{j}^{-1}\left(\mathbf{x}_{i}-\mathbf{p}_{j}\right)^{T}\right)}.
\end{equation}

Furthermore, to make the optimization process faster and simpler, we define the additional variables:
\begin{subequations}
\begin{align}
\label{eq-Cinv}
\mathbf{C}_{j}^{-1} &= \mathbf\Lambda_{j}\mathbf\Lambda_{j}^{T},\\
\label{eq-V_j}
\mathbf{V}_{j} &= \mathbf\Delta_{j}\mathbf\Lambda_{j},
\end{align}
\end{subequations}
where $\mathbf\Lambda_{j} \in \mathbb{R}^{d\times d}$ is a local affine transformation matrix for basis function $j$ and $\mathbf{V}_{j}$ is the application of the local transformation to the data. Optimizing with respect to $\mathbf\Lambda_{j}$ directly ensures that the covariance matrix is positive definite. This makes it faster from a computational perspective as the kernel functions for all the points with respect to a particular basis can be computed more efficiently as follows:
\begin{equation}
\label{eq-squared-exponential-covariance-extension-simplified}
k(\mathbf{X},\mathbf{p}_{j}) = \exp{\left(-\frac{1}{2}\left(\mathbf{V}_{j}\circ \mathbf{V}_{j}\right)\vec{\mathbf{1}}_{d}\right)}.
\end{equation}
The exponent in Eq. \eqref{eq-squared-exponential-covariance-extension-simplified} basically computes the sum of squares in each row of $\mathbf{V}_{j}$. This allows for a more efficient computation of the kernel functions for all the points in a single matrix operation. The derivatives with respect to each $\mathbf\Lambda_{j}$ and $\mathbf{p}_{j}$ are shown in Eq. \eqref{eq-dfdL} and Eq. \eqref{eq-dfdP} respectively.

\begin{subequations}
\begin{align}
\label{eq-dfdL}
\frac{\partial f(\mathbf{X},\mathbf{y},\mathbf{w})}{\partial \mathbf\Lambda_{j}} &= -\left( \mathbf\Delta_{j}^{T}\circ \left(\vec{\mathbf{1}}_{d}\mathbf{E}_{*,j}^{T}\right) \right)\mathbf{V}_{j},\\
\label{eq-dfdP}
\frac{\partial f(\mathbf{X},\mathbf{y},\mathbf{w})}{\partial \mathbf{p}_{j}} &= \mathbf{E}_{*,j}^{T}\mathbf{V}_{j}\mathbf\Lambda_{j}^{T}.
\end{align}
\end{subequations}

Setting up the problem in this manner allows the setting of matrix $\mathbf\Lambda_{j}$ to be of any size $d$ by $q$, where $q<d$ which can be considered as a low rank approximation to $\mathbf{C}_{j}^{-1}$ without affecting the gradient calculations. In addition, the inverse of the covariance can be set to $\mathbf{C}_{j}^{-1}=\mathbf\Lambda_{j}\mathbf\Lambda_{j}^{T}+diag(\blambda_{j})^{-2}$ in the low rank approximation case to ensure that the final covariance can model a diagonal covariance. This is referred to as \emph{factor analysis distance} \citep[p. 107]{rasmussen2006gaussian} but previously used to model a global covariance as opposed to variable covariances as is the case here.

\subsection{Prior Mean Functions}

In the absence of observations, all Bayesian models, Gaussian processes included, rely on their priors to provide function estimation. For the case of Gaussian processes this requires us to consider the prior over the function, especially the prior mean. For example, the first term in the mean prediction in Eq. \eqref{eq-conditional-distribution}, $\bmu_{\mathbf{f}}$, is our prior mean in which we learn the deviation from using a GP. Similarly, we may consider a mean \emph{function} that is itself a simple linear regression from the independent to dependent variable. The parameters of this function are then inferred and the GP infers non-linear deviations. In the absence of data, e.g. in extrapolative regions, the GP will fall back to the linear regression prediction \citep{roberts2012rs}. We can incorporate this directly into the optimization objective instead of having it as a separate preprocessing step by redefining $\mathbf{K}$ as a concatenation of the linear and non-linear features, or setting $\mathbf{\hat K}=[\mathbf{K}|\mathbf{X}|\vec{\mathbf{1}}_{n}]$ and $\mathbf{\hat w} = \left [\mathbf{w}|\mathbf{w}_{L}|b \right]$, where $\mathbf{w}_{L}$ is the linear regression's coefficients and $b$ is the bias. The prediction can then be formulated as follows:

\begin{equation}
\label{eq-joint-concatenation}
\mathbf{\hat K}\mathbf{\hat w} = \mathbf{K}\mathbf{w}+\mathbf{X}\mathbf{w}_{L}+b.
\end{equation}

Furthermore, the regularization matrix, $\mathbf{I}\sigma_{n}^{2}$, in Eq. \eqref{eq-linear-regression-objective-rectangular} can be modified so that it penalises the learning of high coefficients for the non-linear terms, $\mathbf{w}$, but small or no cost for learning high linear terms, $\mathbf{w}_{L}$ and $b$, by setting the corresponding elements in the diagonal of $\mathbf{I}$ to 0 instead of $\sigma_{n}^{2}$, or the last $d+1$ elements. Therefore, as $\sigma_{n}^{2}$ goes to infinity, the model will approach a simple linear regression model instead of fallen back to zero.

\subsection{Cost-Sensitive Learning}

Thus far in the discussion, we make the tacit assumption that the objective of the inference process is to minimize the sum of squared errors between the model and target function values. Although this is a suitable objective for many applications, it is intrinsically biased by uneven distributions of training data, sacrificing accuracy in less represented regions of the space. Ideally we would like to train a model with a balanced data distribution to avoid such bias. This however, is a luxury that we often do not have. For example, the lack of strong emission lines that are detectable with visible-wavelength spectrographs in the ``redshift-desert'' at $1.2 < z <1.8$ means that this redshift range is often under-represented in spectroscopic samples. A common technique is to either over-sample or under-sample the data to achieve balance \citep{weiss2007}. In under-sampling, samples are removed from highly represented regions to achieve balance, over-sampling on the other hand duplicates under represented samples. Both approaches come with a cost; in the former good data are wasted and in the latter more computation is introduced due to the data size increase. In this paper, we perform cost-sensitive learning, which increases the intrinsic error function in under-represented regions. In regression tasks, such as we consider here, the output can be either discretized and treated as classes for the purpose of cost assignment, or a specific bias is used such as $1/\left(1+z_{\rm s}\right)$. To mimic a balanced data set in our setup, the galaxies were grouped by their spectroscopic redshift using non-overlapping bins of width 0.1. The weights are then assigned as follows for balanced training:
\begin{equation}
\label{eq-balanced-weights}
w_{i} = \frac{\mbox{max}\left(\left \{f_{1},\hdots,f_{B}\right\}\right)}{\left\{f_{b}:i\in S_{b}\right\}},
\end{equation}
where $w_{i}$ is the error cost for sample $i$, $f_{b}$ is the frequency of samples in bin number number $i$, $B$ is the number of bins and $S_{b}$ is the set of samples in set number $b$. Eq. \eqref{eq-balanced-weights} assigns a weight to each training point which is the maximum bin frequency over the frequency of the bin in which the source belongs. This ensures that the error cost of source $i$ is inversely proportional to its spectroscopic redshift frequency in the training sample. The normalized weights are assigned as follows:

\begin{equation}
\label{eq-normalized-weights}
w_{i} = \left(\frac{1}{1+z_{\rm s}^{\left(i\right)}}\right)^{2}.
\end{equation}

After the weights have been assigned, they can be incorporated directly into the objective as follows:
\begin{equation}
\label{eq-weighted-linear-regression-objective}
\begin{array}{lcl}
\underset{\mathbf{w}}{\text{min}} &\frac{1}{2}\left ( \mathbf{K}\mathbf{w}-\mathbf{y} \right )^{T} \mathbf{W}\left( \mathbf{K}\mathbf{w}-\mathbf{y} \right )+\frac{1}{2}\sigma_{n}^{2}\mathbf{w}^{T}\mathbf{w}.
\end{array}
\end{equation}

The difference between the objectives in Eq. \eqref{eq-linear-regression-objective} and Eq. \eqref{eq-weighted-linear-regression-objective} is the introduction of the diagonal matrix $\mathbf{W}$, where each element $\mathbf{W}_{ii}$ is the corresponding cost $w_{i}$ for sample $i$. The first term in Eq. \eqref{eq-weighted-linear-regression-objective} is a matrix operation form for a weighted sum of squares $\sum_{i=1}^{n}w_{i}\left(\mathbf{K}_{i,*}\mathbf{w}-\mathbf{y}_{i}\right)^{2}$, where the solution can be found analytically as follows:
\begin{equation}
\label{eq-weighted-linear-regression-objective-rectangular}
\mathbf{w} = \left(\mathbf{K}^{T}\mathbf{WK}+\mathbf{I}\sigma_{n}^{2} \right)^{-1}\mathbf{K}^{T}\mathbf{W}\mathbf{y}.
\end{equation}
The only modification to the gradient calculation is to set the matrix $\mathbf{E}=\mathbf{W}\left(\left(\mathbf{K}\mathbf{w}-\mathbf{y}\right)\mathbf{w}^{T}\right)\circ\mathbf{K}$. In standard sum of squared errors, $ \mathbf{W}= \mathbf{I}$ or the identity matrix. It is worth emphasising that this component of the framework does not attempt to weight the training sample in order to match the distribution of the test sample, or matching the spectroscopic distribution to the photometric distribution as proposed in \citet{Lima2008}, \citet{Cunha2009} and applied to photometric redshift in \citet{sanchez14}, but rather gives the user of the framework the ability to control the cost per sample to serve different science goals depending on the application. In this paper, the weighting scheme was used for two different purposes, the first was to virtually balance the data to mimic training on a uniform distribution, and the second was to directly target the weighted error of $1/\left(1+z_{\rm s}\right)$.

\section{Application to Photometric Redshift Estimation}
\label{sec-application}

In this section, we specifically target the photometric bands and depths planned for {\em Euclid}. {\em Euclid} aims to provide imaging data in a broad $RIZ$ band and the more standard near-infrared $Y$, $J$ and $H$ bands, while ground-based ancillary data are expected in the optical $g$, $r$, $i$ and $z$ bands \citep{laureijs2011}.

\subsection{Simulated Dataset}
\label{sec-dataset}

\begin{figure}
        \centering
        \begin{subfigure}[b]{1\columnwidth}
                 \includegraphics[width=\textwidth]{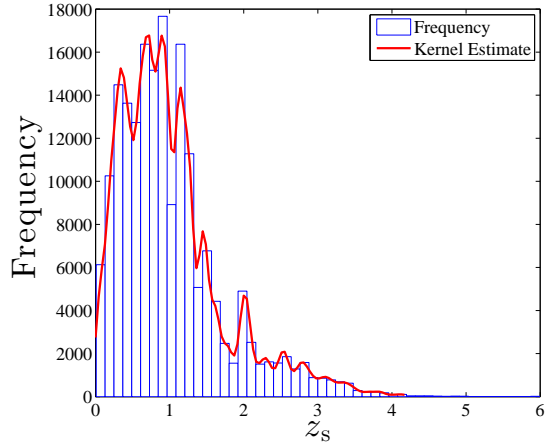}
                 \caption{Full}
        \end{subfigure}
        ~
        \begin{subfigure}[b]{1\columnwidth}
                 \includegraphics[width=\textwidth]{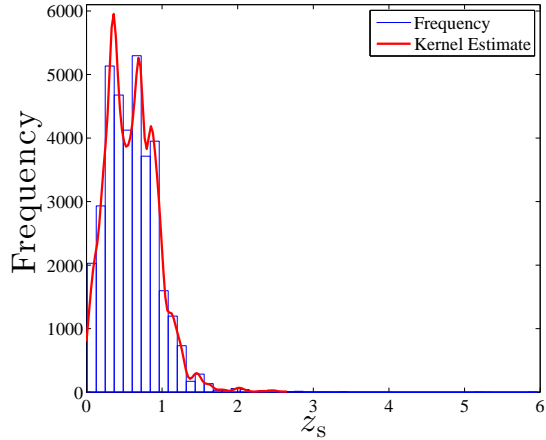}
                 \caption{$RIZ<$23}
        \end{subfigure}
        ~
        \begin{subfigure}[b]{1\columnwidth}
                 \includegraphics[width=\textwidth]{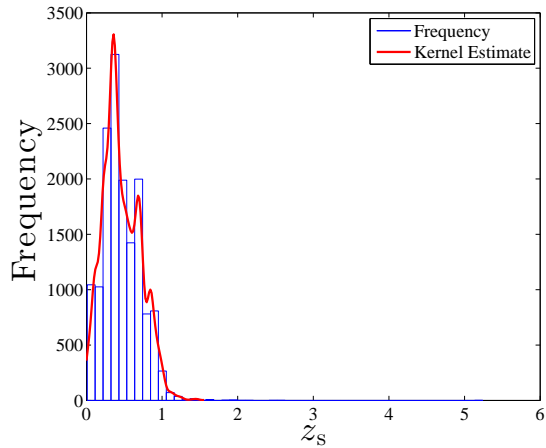}
                 \caption{$RIZ<$22}
        \end{subfigure}
        
       \caption{The spectroscopic redshift distribution of the (a) full dataset, (b) sources with RIZ magnitude\textless23 and (c) sources with RIZ magnitude\textless22 from the simulated data}
	 \label{fig-zspec-histogram}
\end{figure}

We use a mock dataset from \citet{jouvel09}, consisting of the $g$, $r$, $i$, $z$, $RIZ$, $Y$, $J$ and $H$ magnitudes (to 10$\sigma$ depths of 24.6, 24.2, 24.4, 23.8, 25.0 for the former, and 5$\sigma$ depth of 24.0 for each of the latter three near-infrared filters) for 185,253 simulated sources. We remove all sources with $RIZ>25$ to simulate the target magnitudes set for {\em Euclid}. In addition, we remove any sources with missing measurements in any of their bands prior to training (only 15 sources). No additional limits on any of the bands were used, however in Section \ref{subsec-prior-mean} we do explicitly impose limits on the RIZ band to test the extrapolation performance of the models. The distribution of the spectroscopic redshift is provided in Figure \ref{fig-zspec-histogram}. For all experiments on the simulated data, we ignore the uncertainties on the photometry in each band and train only on the magnitudes since in the simulated data set, unlike real datasets, the log of the associated errors are linearly correlated with the magnitudes, especially in the targeted range, therefore adding no information. However, they were fed as input to {\sc ANNz} to satisfy the input format of the code.

We preprocess the data using Principle Component Analysis \citep[PCA;][]{jolliffe1986} to de-correlate the features prior to learning, but retain all features with no dimensionality reduction. De-correlation accelerates the convergence rate of the optimization routine especially when using a logistic-type kernel machines such as Neural Networks \citep{lecun1998}. To understand this, consider a simple linear regression example where we would like to solve for $\mathbf{w}$ in $\mathbf{A}\mathbf{w}=\mathbf{b}$, the solution for this is $\mathbf{w}=\left(\mathbf{A}^{T}\mathbf{A}\right)^{-1}\mathbf{A}^{T}\mathbf{b}$. Note that if $\mathbf{A}$ is de-correlated $\mathbf{A}^{T}\mathbf{A}=\mathbf{I}$, therefore learning $\mathbf{w}_{i}$ depends only on the $i$-th column of $\mathbf{A}$ and it is independent from learning $\mathbf{w}_{j}$, where $i\ne j$. In an optimization approach, the convergence rate is a function of the condition number of the $\mathbf{A}^{T}\mathbf{A}$ matrix, which is minimized in the case of de-correlated data. This represents a quadratic error surface which helps accelerate the search. This is particularly important in the application addressed in this paper because the magnitudes measured in each filter are strongly correlated with each other. 

\section{Results on the Simulated Data}
\label{sec-experiments}

Five algorithms are considered to model the data; Artificial Neural Networks \citep[{\sc ANNz};][]{Collister04}, a GP with low rank approximation \citep[{\sc stableGP};][]{foster2009}, a sparse GP with global length scale (GP-GL), a GP with variable length scale (GP-VL) and a GP with variable covariances (GP-VC). For {\sc ANNz}, a single layer network is used, and to satisfy the input format for the code, the data were not de-correlated and the uncertainties on photometry for each band were used as part of the training input. For {\sc stableGP}, we use the SR-VP method proposed in \citet{foster2009}. In subsequent tests, the variable $m$ refers to the number of hidden units in {\sc ANNz}, the rank in {\sc stableGP}, and the number of basis functions in GP-GL, GP-VL and GP-VC. The time complexities for each algorithm are shown in Table \ref{table-time-complexity}. The data were split at random into 80 per cent for training, 10 per cent for validation and 10 per cent for testing. We note that we investigate the accuracy for various training sample sizes in Section~\ref{sec-sizetraining}. All models were trained using the entire redshift range available, but we only report the performance on the redshift range of $0 \le z_\textrm{s} \le 2$ to target the parameter space set out in \cite{laureijs2011}. We train each model for 500 iterations in each run and the validation sample was used for model selection and parameter tuning, but all the results here are reported on the test sample, which is not used in any way during the training process. Table \ref{table-metrics} shows the metrics used to report the performance of each algorithm.

\begin{table}
\caption{The time complexity of each approach.}
\begin{center}
  \begin{tabular}{| l | l |}
     	Method		&	Time Complexity					\\	\hline				\\
	{\sc ANNz} ($l$-layers)			&	$O\left(nmd+(l-1)(nm^{2})\right)$					\\
	{\sc stableGP}		&	$O\left(nm^{2}\right)$				\\
	GP-GL		&	$O\left(nmd+nm^{2}\right)$		\\	
	GP-VL		&	$O\left(nmd+nm^{2}\right)$		\\	
	GP-VC		&	$O\left(nmd^{2}+nm^{2}\right)$	\\	\hline
  \end{tabular}
\end{center}
\label{table-time-complexity}
\end{table}

 \begin{table*}
 \caption{Performance metrics used to evaluate the models. The number of samples is denoted by $n$.}
\begin{center}
  \begin{tabular}{| l | l | l |}
     	Metric				&	Equation	& Description\\	\hline
	$\delta^{\left(i\right)}$		&	$z_\textrm{s}^{\left(i\right)}-z_\textrm{p}^{\left(i\right)}	$		&Error for the $i$-th object\\

	$\delta^{\left(i\right)}_{norm}$		&	$\delta^{\left(i\right)}/\left(1+z_\textrm{s}^{\left(i\right)}\right)$	&Normalized error for the $i$-th object		\\

	$\Delta z$		&	$\sqrt{\frac{1}{n}\sum_{i=1}^{n}\left(\delta^{\left(i\right)}\right)^{2}}$	&Root mean squared error		\\

	$\Delta z_{norm}$		&	$\sqrt{\frac{1}{n}\sum_{i=1}^{n}\left(\delta_{norm}^{\left(i\right)}\right)^{2}}$	&Normalized root mean squared error		\\

	max$_{z}$	& max$\left(\left\{\left |\delta^{\left(1\right)}\right |,\hdots,\left |\delta^{\left(n\right)}\right |\right\}\right)$& Maximum error \\
	
	max$_{norm}$	& max$\left(\left\{\left |\delta_{norm}^{\left(1\right)}\right |,\hdots,\left |\delta_{norm}^{\left(n\right)}\right |\right\}\right)$& Maximum normalized error \\
	
	$\mu_{z}$		&	$\frac{1}{n}\sum_{i=1}^{n}\delta^{\left(i\right)}$	&Bias		\\

	$\mu_{norm}$		&	$\frac{1}{n}\sum_{i=1}^{n}\delta_{norm}^{\left(i\right)}$	&Normalized bias		\\

	$\sigma_{z}$		&	$\sqrt{\frac{1}{n}\sum_{i=1}^{n}\left(\delta^{\left(i\right)}-\mu_{z}\right)^{2}}$	&Standard deviation of the errors		\\

	$\sigma_{norm}$		&	$\sqrt{\frac{1}{n}\sum_{i=1}^{n}\left(\delta_{norm}^{\left(i\right)}-\mu_{norm}\right)^{2}}$	&Standard deviation of the normalized errors \\
	
	out$_{z}$		&	$\frac{1}{n}\left | \left\{i: \delta^{\left(i\right)}>2\sigma_{z}\right\}\right |$	&Fraction of errors above two standard deviations from the mean \\
	out$_{norm}$		&	$\frac{1}{n}\left | \left\{i: \delta_{norm}^{\left(i\right)}>2\sigma_{norm}\right\}\right |$	&Fraction of normalized errors above two standard deviations from the mean\\ \hline
	
  \end{tabular}
\end{center}
\label{table-metrics}
\end{table*}

\subsection{Modelling Performance}

\begin{figure*}
        \centering
        \begin{subfigure}[b]{0.3\textwidth}
                \includegraphics[width=\textwidth]{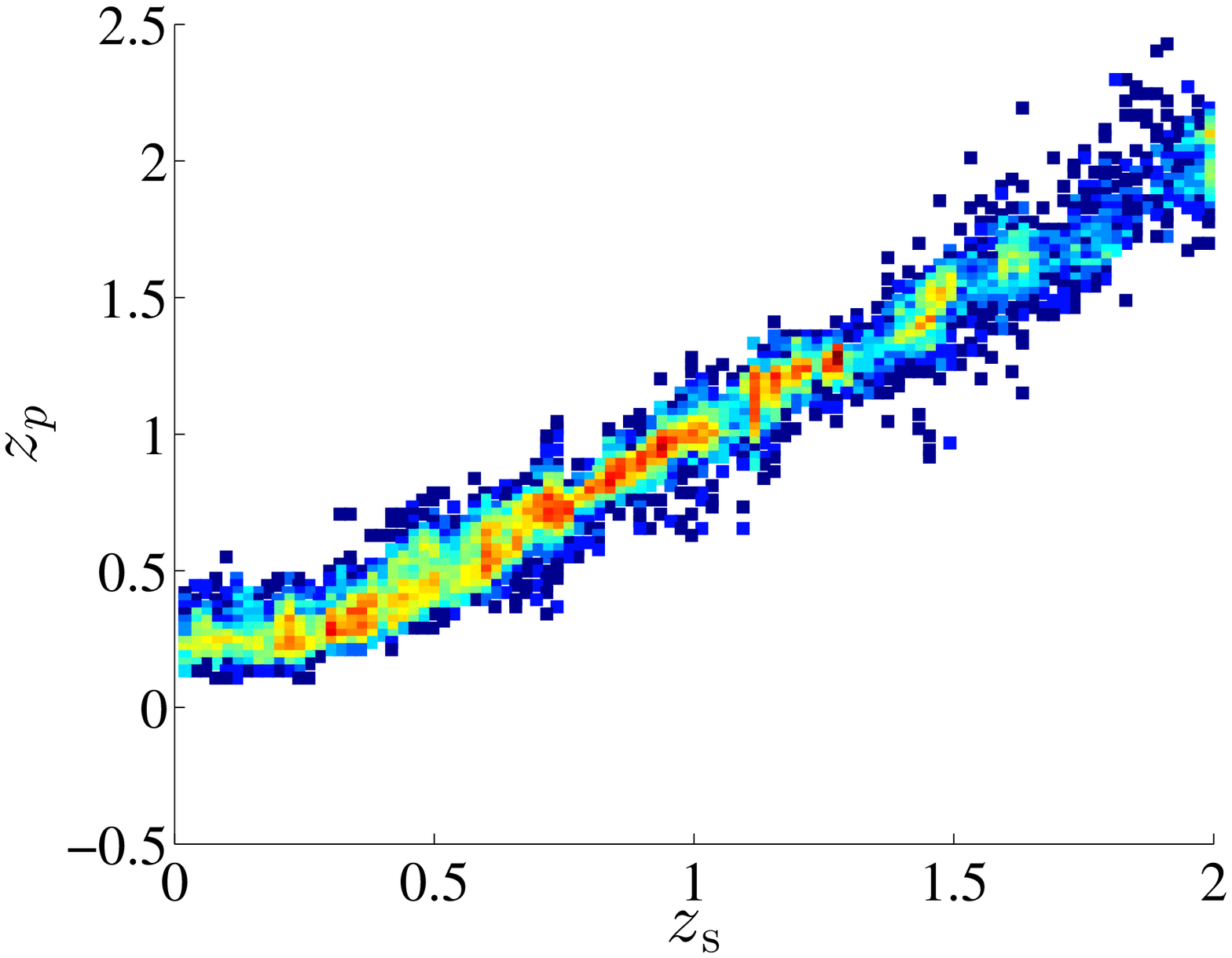}
                \caption{{\sc ANNz}}
                \label{annz-plot}
        \end{subfigure}
        ~
        \begin{subfigure}[b]{0.3\textwidth}
                \includegraphics[width=\textwidth]{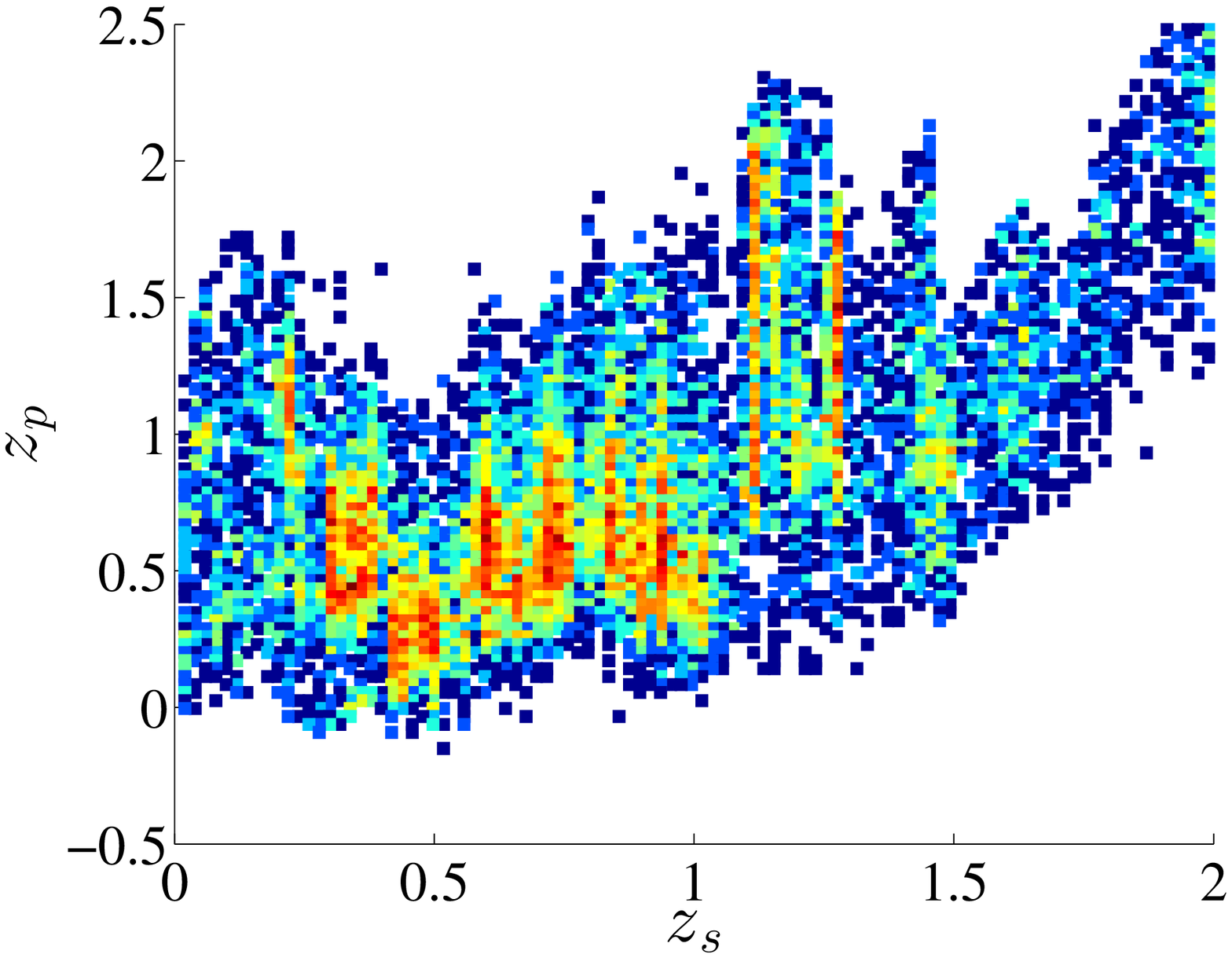}
                \caption{{\sc stableGP}}
                \label{stableGP-plot}
        \end{subfigure}
        ~
        \begin{subfigure}[b]{0.3\textwidth}
                \includegraphics[width=\textwidth]{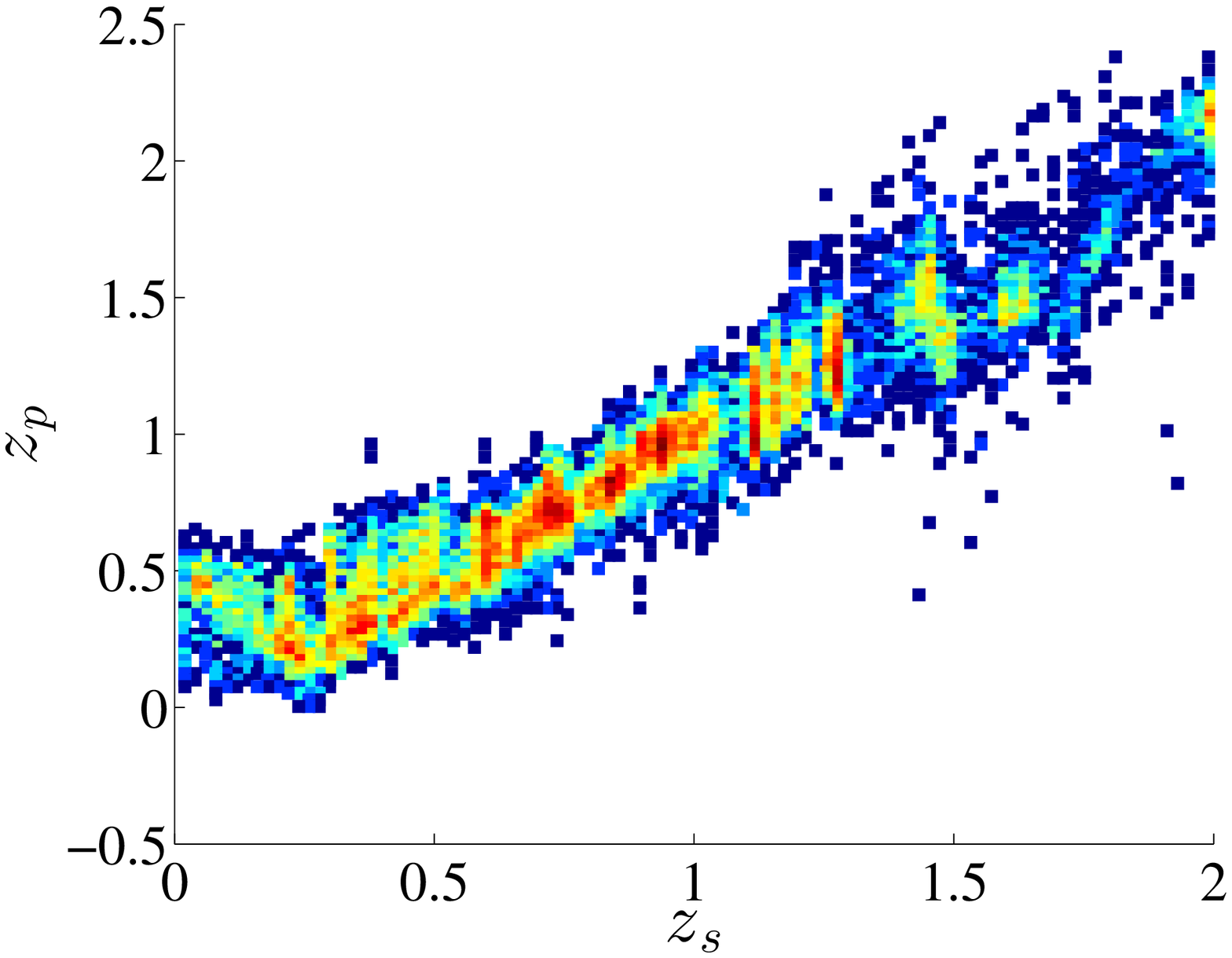}
                \caption{GP-GL}
                \label{GPGL-plot}
        \end{subfigure}
        
        \begin{subfigure}[b]{0.3\textwidth}
               \includegraphics[width=\textwidth]{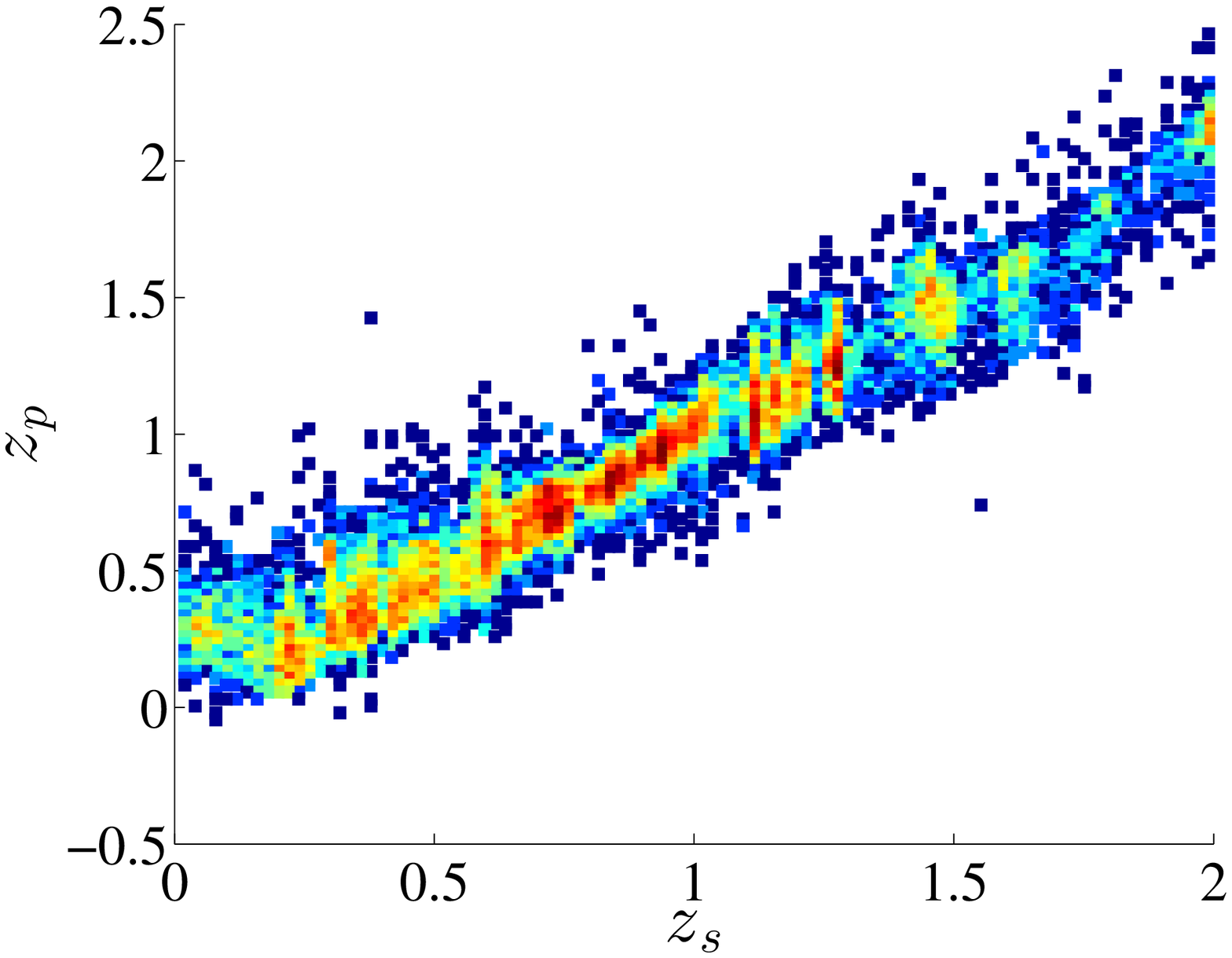}
                \caption{GP-VL}
                \label{GPVL-plot}
        \end{subfigure}
        ~
        \begin{subfigure}[b]{0.3\textwidth}
                \includegraphics[width=\textwidth]{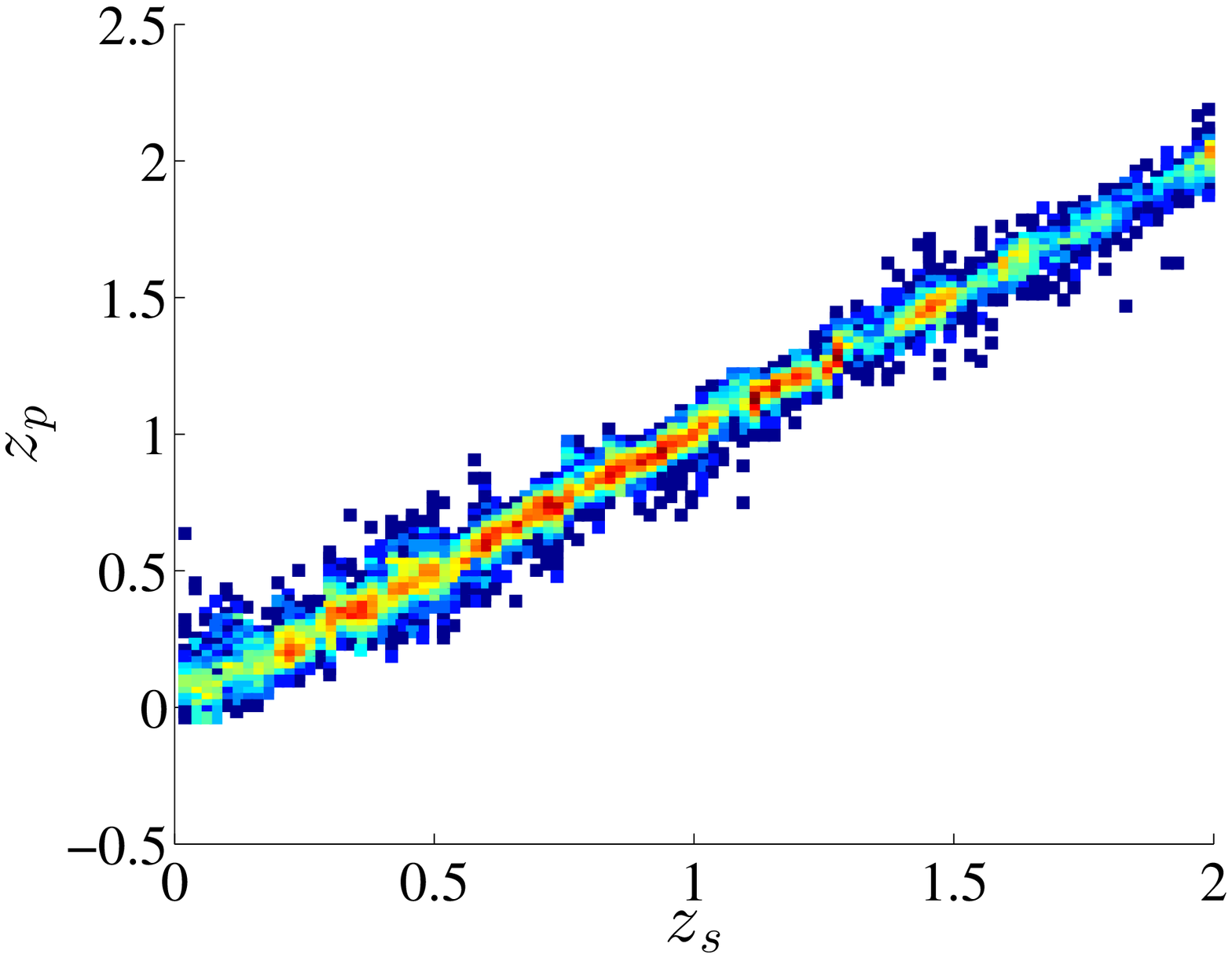}
                \caption{GP-VC}
                \label{GPVC-plot}
        \end{subfigure}
        
        \caption{Density scatter plots of the true $z_\textrm{s}$ vs the predicted $z_\textrm{p}$ for (a) {\sc ANNz}, (b) {\sc stableGP}, (c) GP-GL, (d) GP-VL and (e) GP-VC using $m=10$ basis functions trained on the simulated dataset. The plots shows the performance on the same test sample, the colours however are scaled differently according to the density range in each plot to avoid colour saturation}
        \label{fig-methods-plots}
\end{figure*}

In the first test, all models were trained using a fixed $m=10$ to cross-compare the performance of the methods using the same number of basis functions. The number of basis functions was set deliberately low to highlight the sparse-limit modelling capabilities of each algorithm, as for large values of $m$ the performance gap between the algorithms reduces making it harder to compare the strengths and weaknesses of each algorithm. The standard sum of squares objective was used, without cost-sensitive learning or a prior mean function to keep the comparison as simple as possible. The $z_\textrm{s}$ versus $z_\textrm{p}$ density scatter plots are shown in Figure \ref{fig-methods-plots} and their performance scores of each algorithm are reported in Table \ref{table-10-basis}. Figure \ref{fig-methods-plots} clearly shows the advantage of using inducing points over an active set when we compare GP-GL's performance in Figure \ref{stableGP-plot} with {\sc stableGP}'s performance in Figure \ref{GPGL-plot}. The problem formulation of the two are identical, except that GP-GL's basis set is learned as part of the optimization objective whereas {\sc stableGP}'s basis set is pre-selected in an unsupervised fashion.

 \begin{table*}
\caption{Performance measures for each algorithm trained on the simulated survey using $m=10$ basis functions. The best-performing algorithm is highlighted in bold font}
\begin{center}
\begin{tabular}{| l | c | c |  c | c |  c | c |  c | c |  c | c | }
     				&	$\Delta z$	&	$\Delta z_\textrm{norm}$	&	max$_{z}$ & max$_{norm}$		&	$\mu_{z}$&	$\mu_{norm}$	& $\sigma_{z}$ & $\sigma_{norm}$ & out$_{z}$&out$_{norm}$\\	\hline
	{\sc ANNz}	&	0.0848		&	0.0568	&	1.0126			&	\textbf{0.4199}&	-0.0025			&	-0.0048&	0.0847		&	0.0566&	0.0505		&	0.0532	\\
	{\sc stableGP}	&	0.4399		&	0.2836	&	1.5906			&	1.4441&	-0.0085			&	-0.0365&	0.4399		&	0.2812&	0.0509		&	0.0539	\\
	GP-GL		&	0.1420		&	0.0952	&	1.1183			&	0.5953&	-0.0064			&	-0.0106&	0.1418		&	0.0946&	0.0548		&	0.0530	\\
	GP-VL		&	0.1251	&	0.0833	&	1.0349			&	0.7953&	-0.0074			&	-0.0094&	0.1249		&	0.0828&	0.0549		&	0.0552	\\
	GP-VC	&	\textbf{0.0435}	&	\textbf{0.0294}	 &	\textbf{0.5488	}		&	0.5380 &	\textbf{-0.0005}			&	\textbf{-0.0007}&	\textbf{0.0435	}	&	\textbf{0.0294} &	\textbf{0.0487}		&	\textbf{0.0473}	 \\\hline
  \end{tabular}
\end{center}
\label{table-10-basis}
\end{table*}

\subsection{Prior Mean}
\label{subsec-prior-mean}

\begin{figure*}
        \centering
        \begin{subfigure}[b]{0.24\textwidth}
                \includegraphics[width=\textwidth]{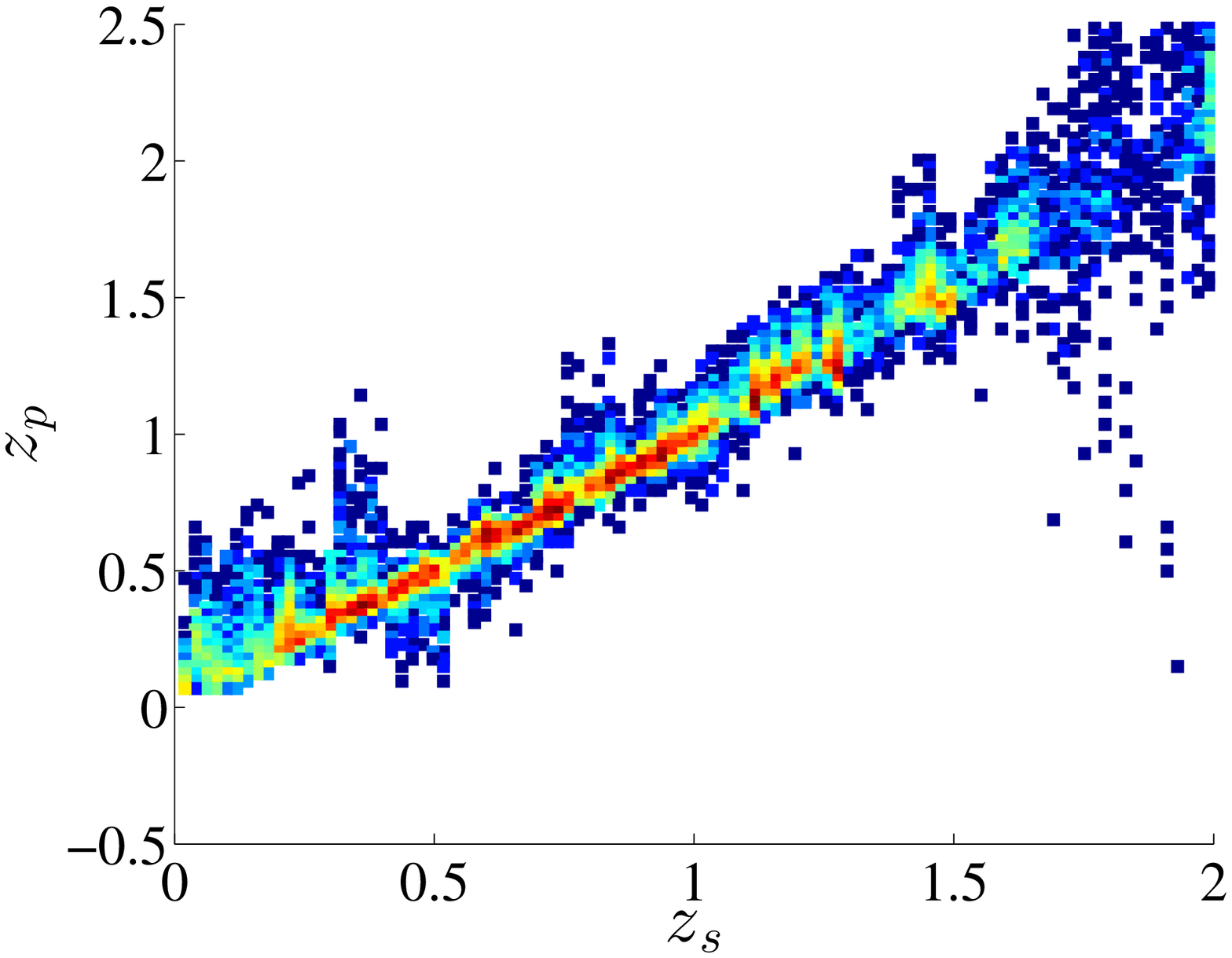}
        \end{subfigure}
        ~
        \begin{subfigure}[b]{0.24\textwidth}
                \includegraphics[width=\textwidth]{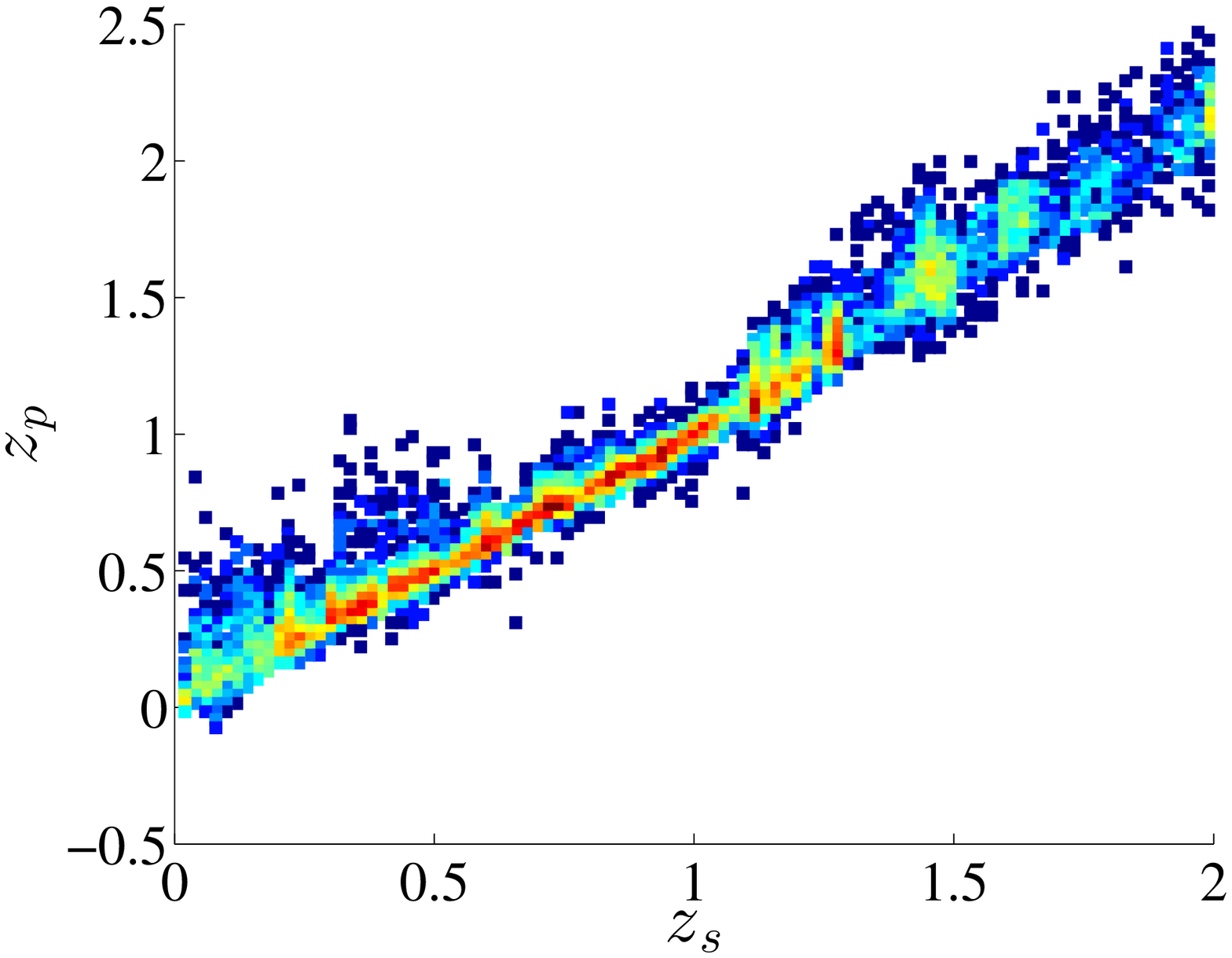}
        \end{subfigure}
        ~
        \begin{subfigure}[b]{0.24\textwidth}
                \includegraphics[width=\textwidth]{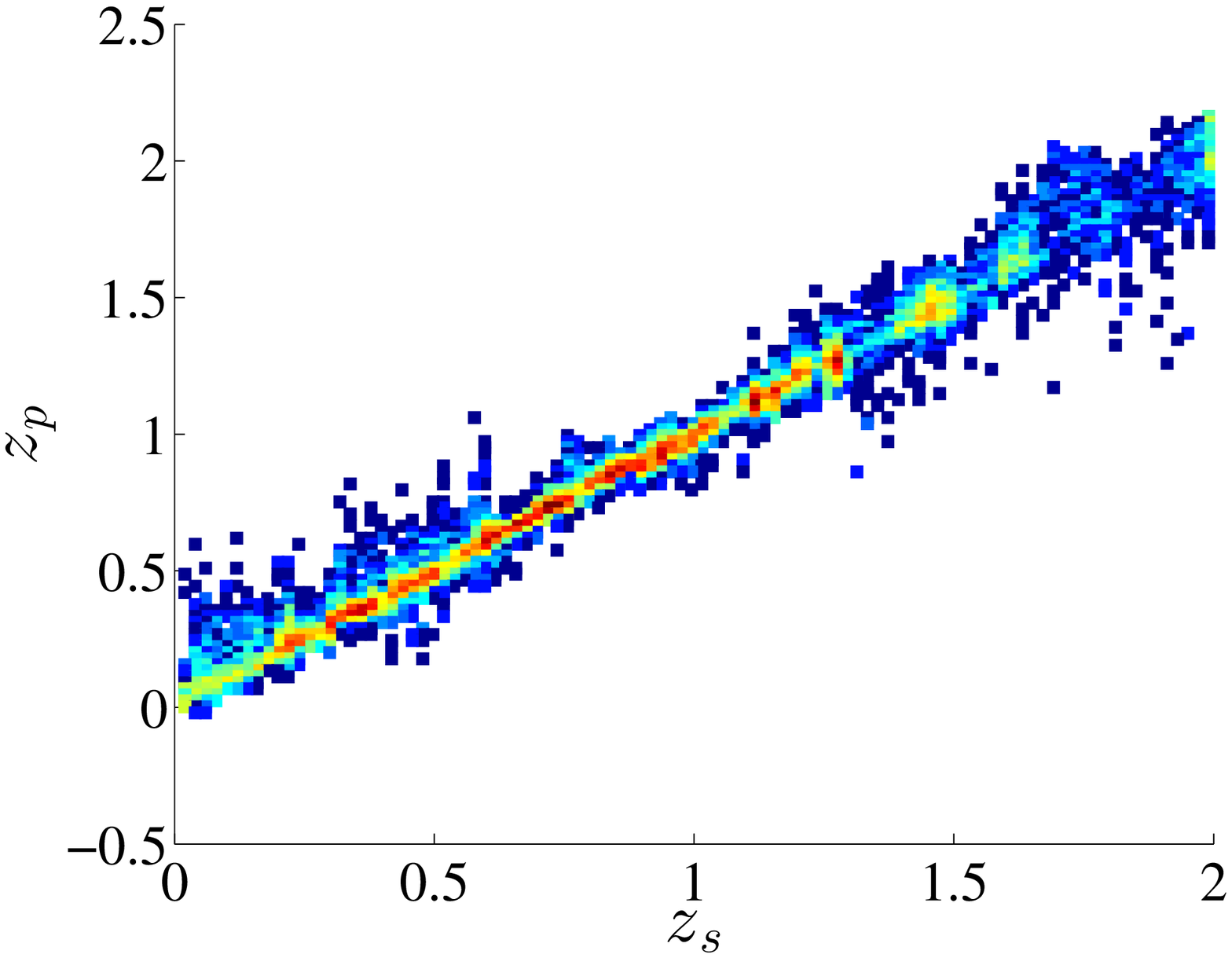}
        \end{subfigure}
         ~
        \begin{subfigure}[b]{0.24\textwidth}
                \includegraphics[width=\textwidth]{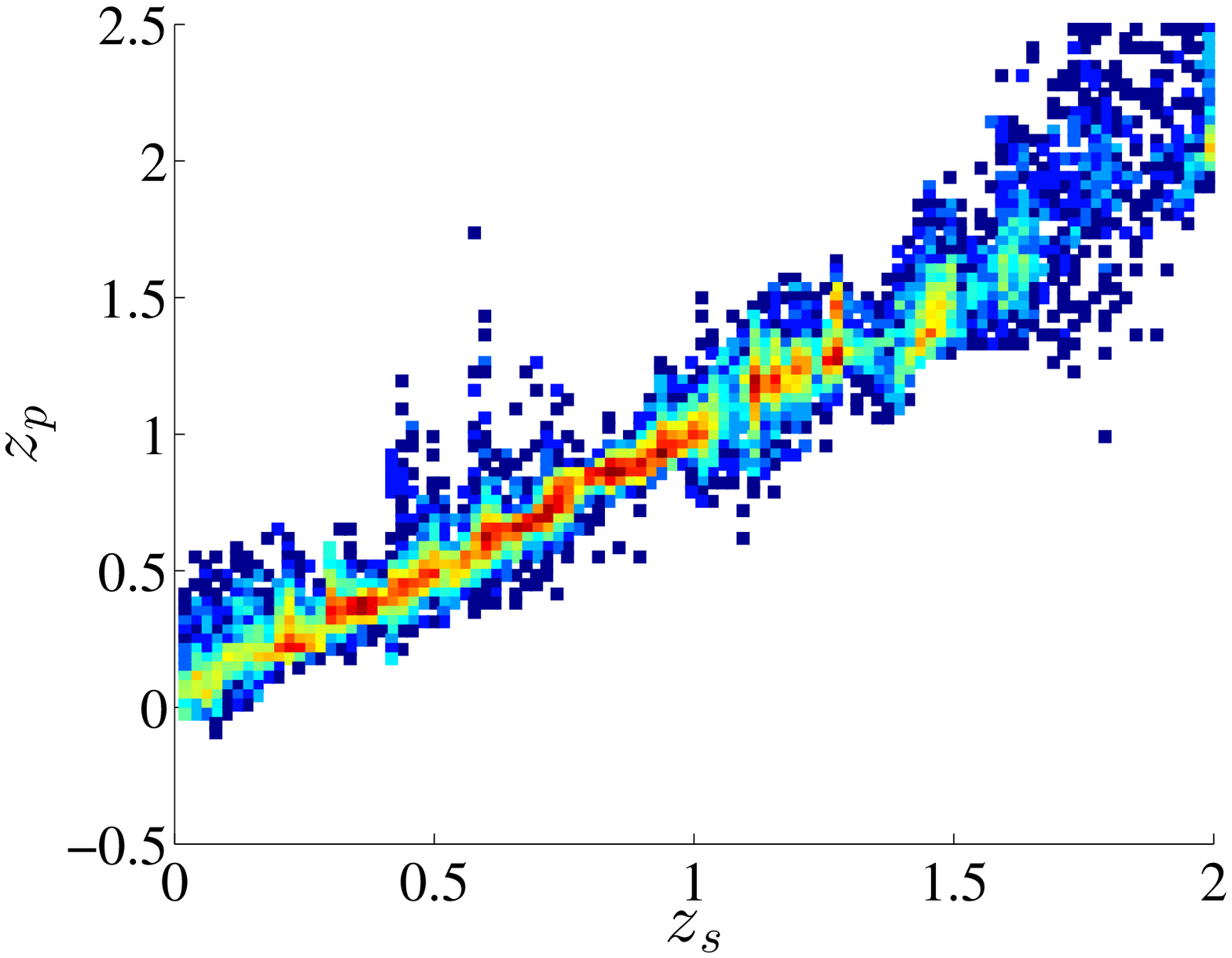}
        \end{subfigure}
        
       \begin{subfigure}[b]{0.24\textwidth}
                \includegraphics[width=\textwidth]{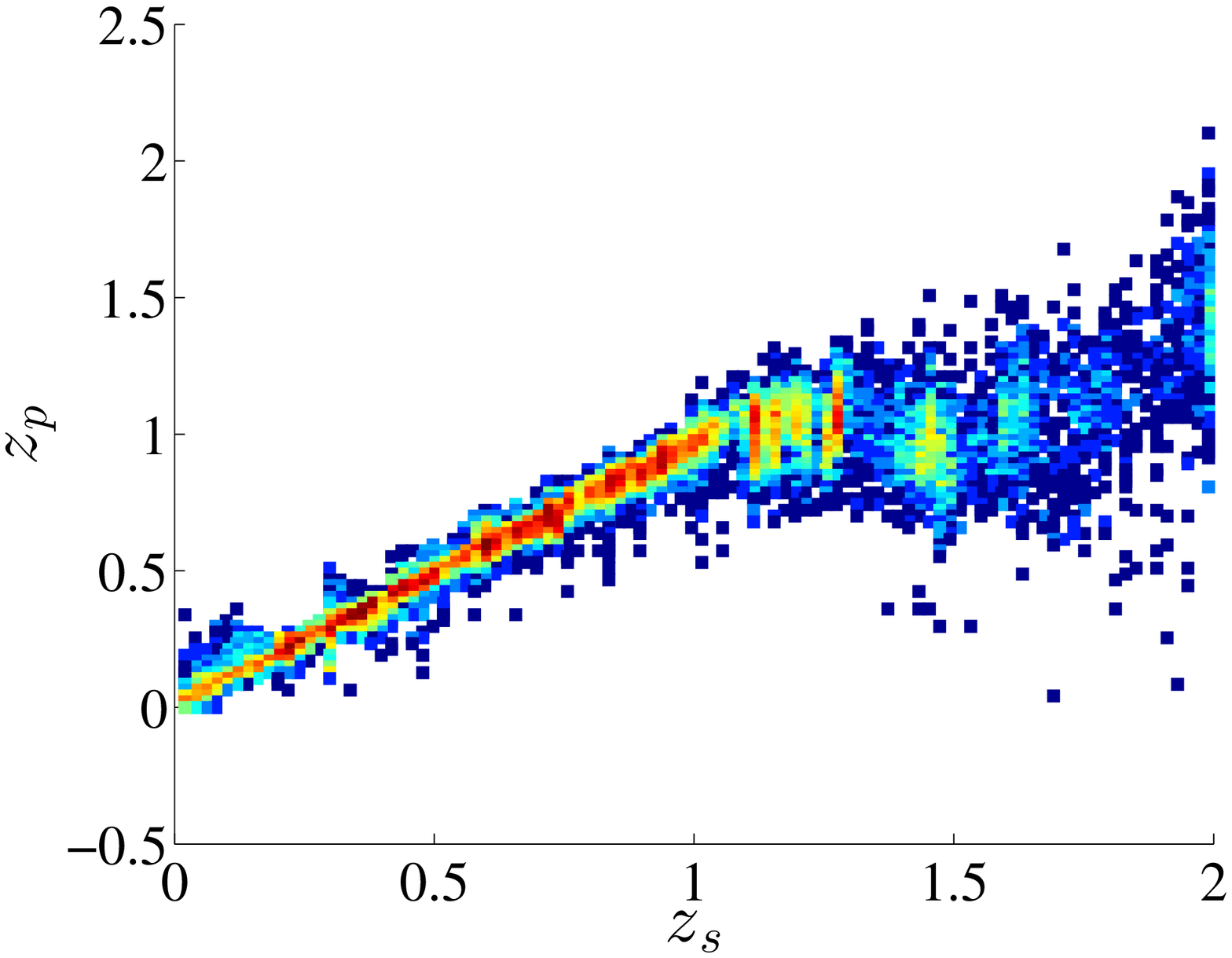}
                \caption{Zero}
        \end{subfigure}
        ~
        \begin{subfigure}[b]{0.24\textwidth}
                \includegraphics[width=\textwidth]{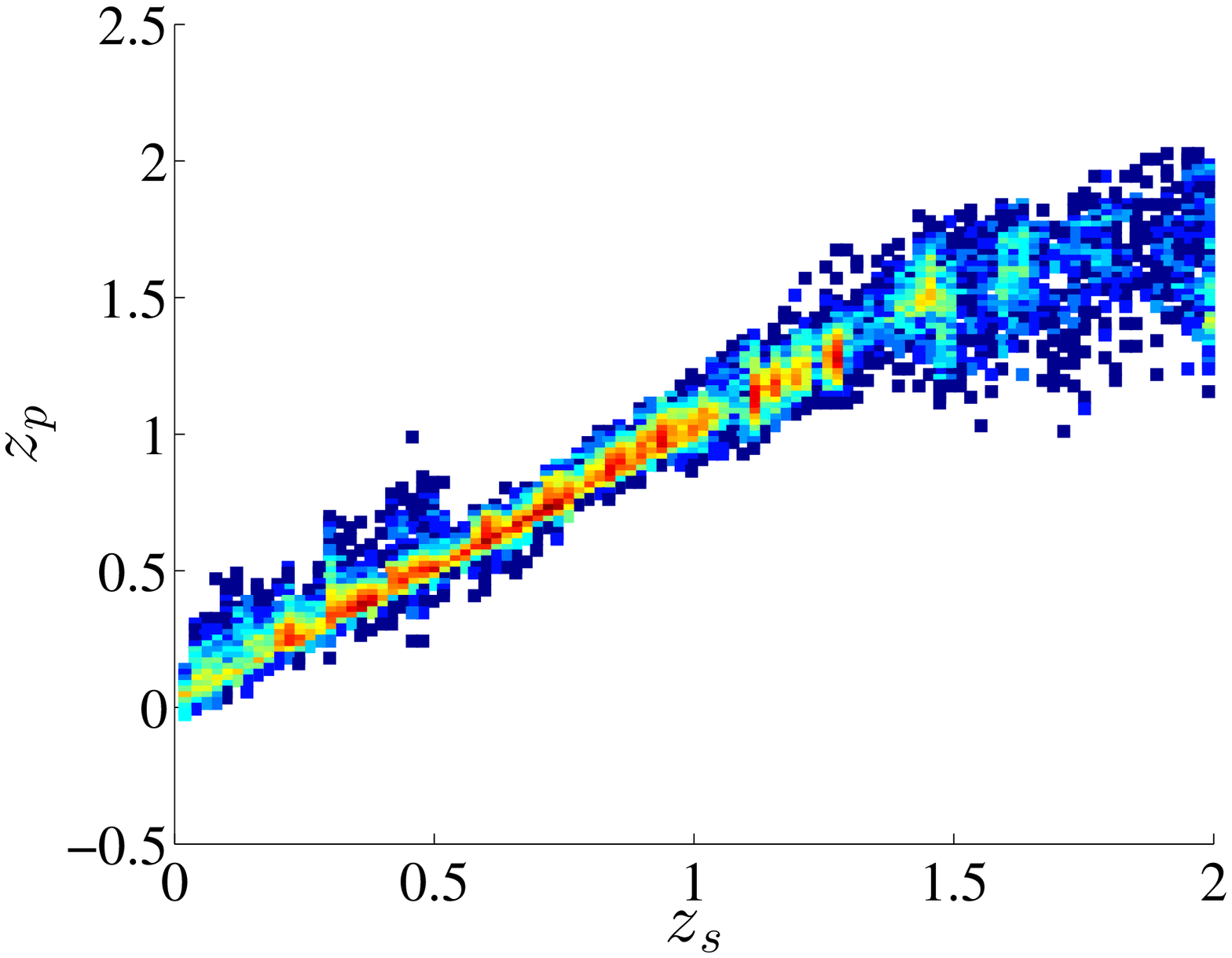}
                \caption{Linear}
        \end{subfigure}
        ~
        \begin{subfigure}[b]{0.24\textwidth}
                \includegraphics[width=\textwidth]{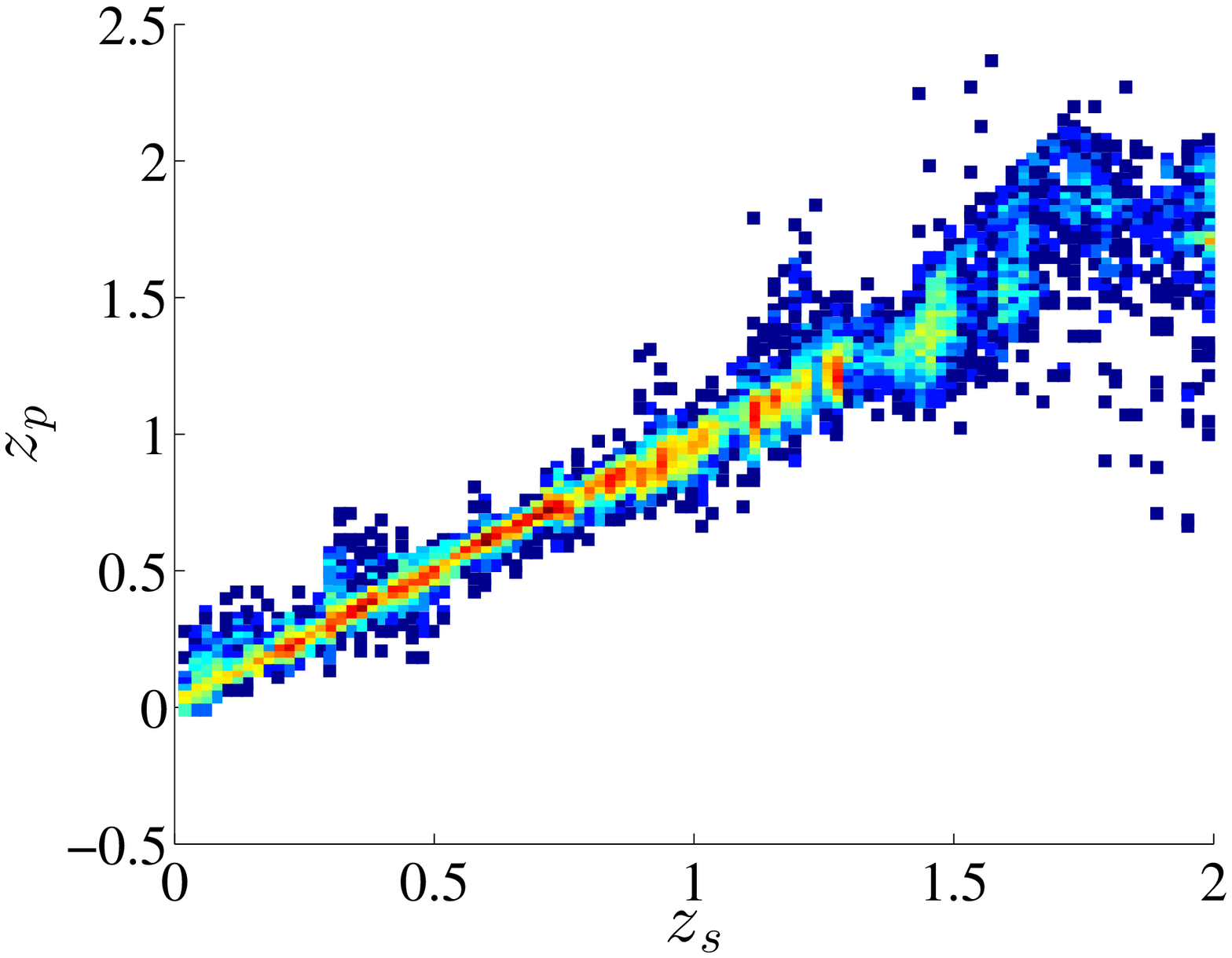}
                \caption{Joint}
        \end{subfigure}
       ~
        \begin{subfigure}[b]{0.24\textwidth}
                \includegraphics[width=\textwidth]{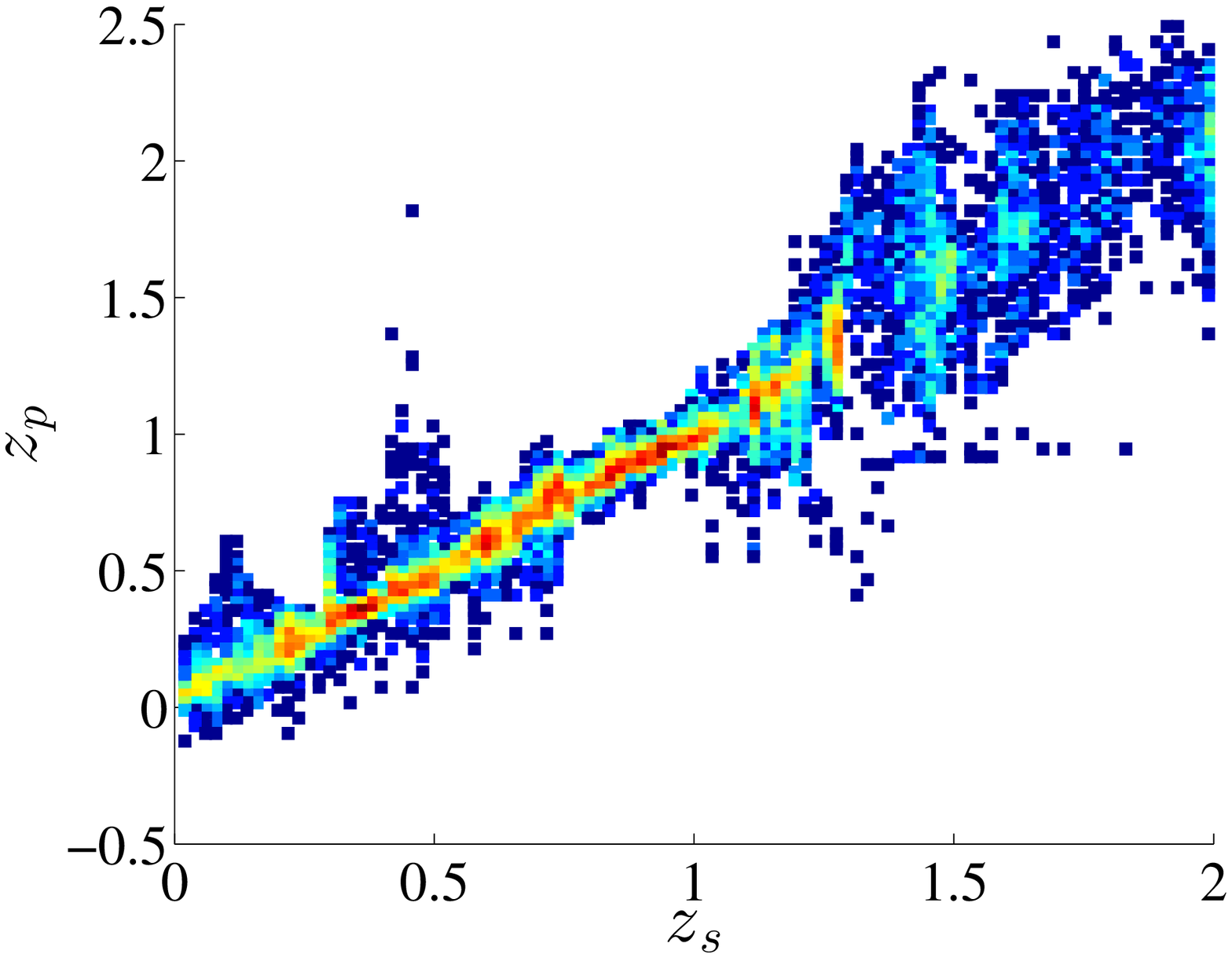}
                \caption{{\sc ANNz}}
        \end{subfigure}
        
        \caption{Density scatter plots of the true $z_\textrm{s}$ versus the predicted $z_\textrm{p}$ after training the GP-VC model on samples with $RIZ<23$ (top) and $RIZ<22$ (bottom) from the simulated data using $m=10$ basis functions with (a) a zero mean prior, (b) linear regression prior and (c) a joint prior optimization and (d) {\sc ANNz}. The plots shows the performance on the same test sample, the colours however are scaled differently according to the density range in each plot to avoid colour saturation}
        \label{fig-RIZ-splits}
\end{figure*}

 \begin{table*}
\caption{The value of $\Delta z$ for the GP-VC model when trained on the simulated survey using $m=10$ basis functions with different prior mean functions and $RIZ$ splits. The results for {\sc ANNz} are shown for comparison}
\begin{center}
  \begin{tabular}{| l | c | c | c | c | c | c | c | c | c |}
  	Trained on				& 	\multicolumn{3}{|c|}{ $RIZ<22$}				&	& 	\multicolumn{3}{c}{$RIZ<23$}  &  & Full\\ \cline{2-4} \cline{6-8} \cline{10-10} 
     	Tested on					&	$<22$			&	$\ge 22$		&	Full				&	&	$<23$	&	$\ge 23$	&	Full	& & Full\\	\hline
	{\sc ANNz}						&	0.0385			&	0.1383			&	0.1325			&	&	0.0537&	0.1458	&	0.1315 &  & 0.0848				\\
	Zero						&	0.0233			&	0.2539			&	0.2424			&	&	0.0362&	0.1261	&	0.1129 &  & 0.0435				\\
	Linear						&	0.0199			&	0.1043			&	0.0997			&	&	0.0321	&	0.1097	&	0.0983 &  & 0.0412				\\
	Joint						&	\textbf{0.0192}	&	\textbf{0.0982}	&	\textbf{0.0939}	&	&	\textbf{0.0277}	&	\textbf{0.0653}	&	\textbf{0.0593} &  & \textbf{0.0298}	\\	\hline
  \end{tabular}
\end{center}
\label{table-RIZ-splits}
\end{table*}

We also test the extrapolation performance of the GP-VC model using different prior means, namely a zero mean, a linear regression prior and a joint optimization approach that learns the linear and non-linear features simultaneously by regularizing the non-linear features more aggressively than linear features and compares them with {\sc ANNz}. The difference between the linear regression prior and the joint optimization approach, is that the former first fits a linear model to the data then subtracts the predictions from the ground truth before training a GP model, whereas the latter learns both the linear model and the non-linear deviations from it jointly. To test this more effectively, the models were trained using sources, with $RIZ<23$ (29,024 objects from the training sample) and tested on the unseen samples with $RIZ<23$, $RIZ\ge23$, and the entire test sample. A similar test was also conducted using a split of $RIZ<22$ (12,056 objects from the training sample). This also demonstrates the effectiveness of the algorithms in the scenario where the brightest sources dominate the training sample, as may be true in practice. The results are reported in Table \ref{table-RIZ-splits} and the density scatter plots are shown for comparison in Figure \ref{fig-RIZ-splits}. The results show that the ``Joint'' method consistently outperformed the other methods in extrapolation as well as in interpolation, especially when trained with a small sample size as in the $RIZ<22$ case. Moreover, upon examining the density scatter plots in Figure \ref{fig-RIZ-splits}, it has fewer systematic and catastrophic errors than the other methods with a factor of $\sim 2$ improvement over {\sc ANNz} where the training data are limited in magnitude/flux-density.

\subsection{Cost-Sensitive Learning}

We now perform a comparison between cost-sensitive learning and the normal sum of squared errors for the GP-VC model. Two different weight configurations are tested, the first is to assign an error cost to each sample as in Eq. \eqref{eq-normalized-weights} (Normalized), and the second experiment is to weight each sample according to the frequency of their true redshift to ensure balanced learning (Balanced) as in Eq. \eqref{eq-balanced-weights}, in addition to the (Normal) sum of squared errors. The algorithms were trained such that they have equal $\Delta z$ score, to examine the differences between the other metrics and the resulting error distributions. The box plots for the ``Normal'', ``Balanced'' and ``Normalized'' are shown in Figure \ref{fig-normal-balanced}. The figures show the performance on the held out test sample for the entire range to demonstrate the effect more clearly. Cost-sensitive learning is more consistent across the redshift range as opposed to the normal sum of squares, especially in the high redshift regions where there are less data. The confidence intervals are also considerably smaller for the ``Balanced'' case. The ``Normalized'' training on the other hand results in a systematic bias, as expected, towards the lower part of the redshift range. The performance comparison for the ``Normal'', ``Balanced'' and ``Normalized'' training are summarized in Table \ref{table-normal-balanced}, but the metrics are reported on the desired range of $0<z_{\rm s}<2$. Balanced training shows a better generalization performance, as it outperforms the normal sum of squares objective on the test sample and has lower maximum errors, although the differences are generally small.

\begin{figure}
        \centering
        \begin{subfigure}[b]{\columnwidth}
                \includegraphics[width=\textwidth]{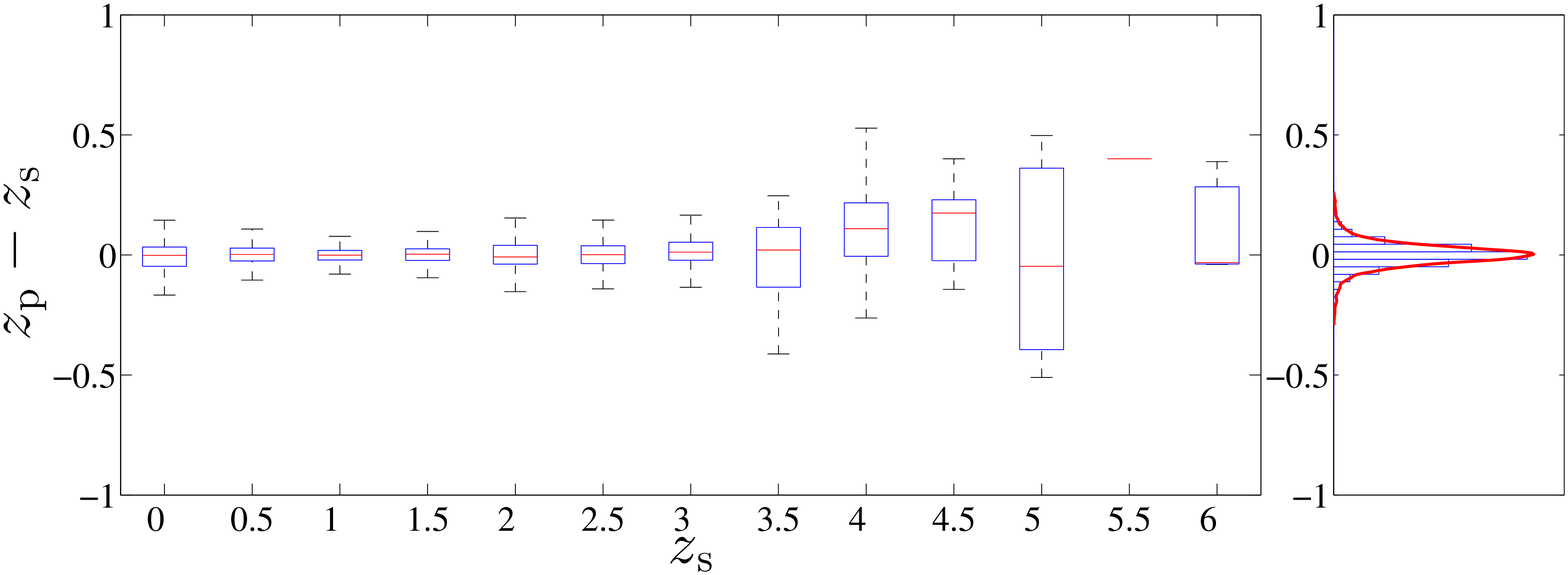}
                \caption{Normal}
                \label{fig-normal}
        \end{subfigure}	
        \begin{subfigure}[b]{\columnwidth}
                \includegraphics[width=\textwidth]{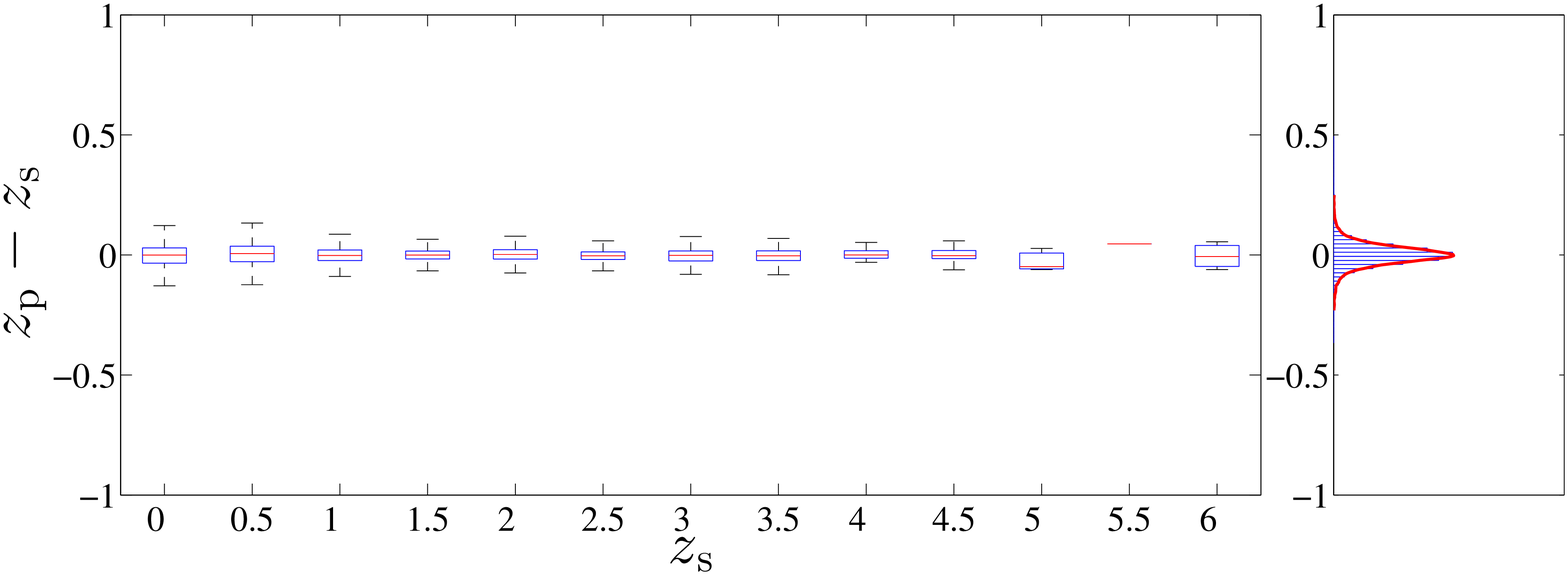}
                \caption{Balanced}
                \label{fig-balanced}
        \end{subfigure}
       \begin{subfigure}[b]{\columnwidth}
                \includegraphics[width=\textwidth]{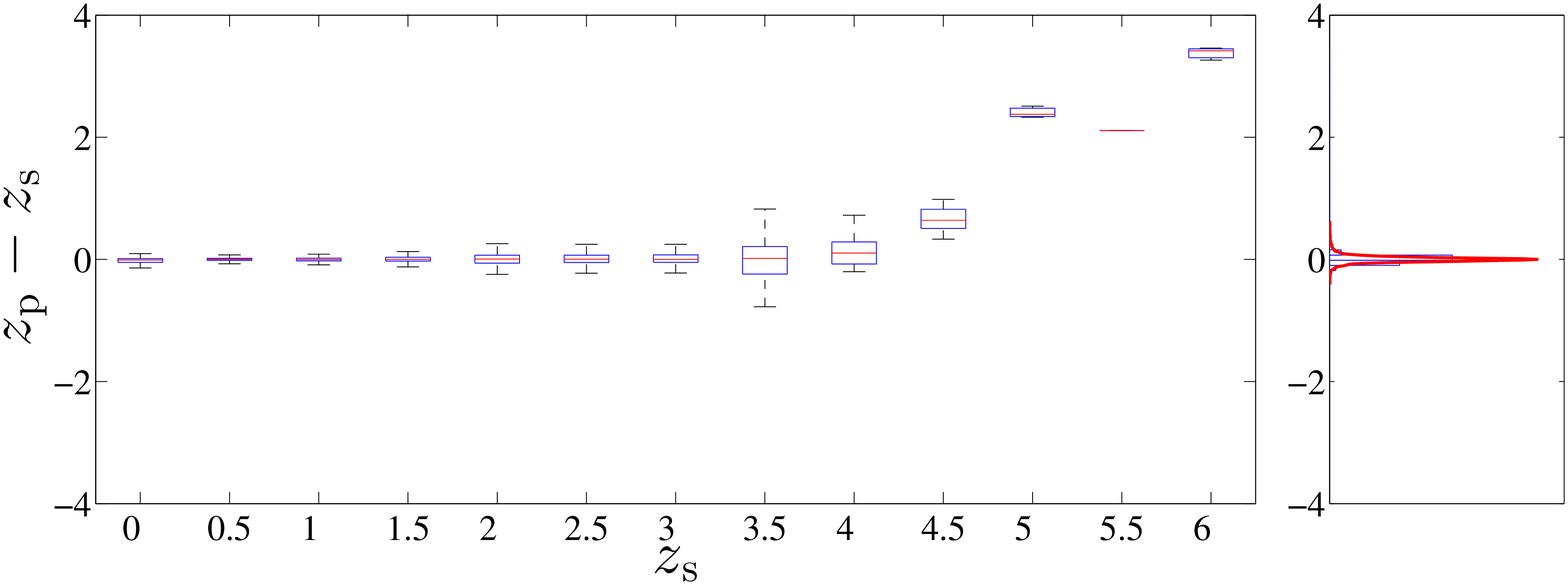}
                \caption{Normalized}
                \label{fig-normalized}
        \end{subfigure}
       \caption{Box plots of residual errors on the hold-out test sample, showing median (bar), inter-quartile range (box) and range (whiskers) for (a) the direct sum of squared errors, (b) the balanced cost-sensitive learning and (c) the normalized cost learning for the GV-VC model trained on the simulated data using $m=10$ basis functions. The right-most histograms are the empirical densities of errors. Figures (a) and (b) have similar scales, but note the scale difference in (c).}
	\label{fig-normal-balanced}
\end{figure}

 \begin{table*}
\caption{Performance measures of training the GP-VC model using $m=10$ basis functions and different weighting schemes trained on the simulated survey.}
\begin{center}
\begin{tabular}{| l | c | c |  c | c |  c | c |  c | c |  c | c | }
     				&	$\Delta z$	&	$\Delta z_\textrm{norm}$	&	max$_{z}$ & max$_{norm}$		&	$\mu_{z}$&	$\mu_{norm}$	& $\sigma_{z}$ & $\sigma_{norm}$ & out$_{z}$&out$_{norm}$\\	\hline
	Normal		&	0.0500	&	0.0337		&	0.6128		&	0.6008&	-0.0017		&	-0.0016&	0.0500		&	0.0336&	0.0507		&	0.0507\\
	Balanced		&	0.0500 	&	0.0324		&	\textbf{0.4933}	&	0.3419&	\textbf{0.0007}		&	\textbf{0.0001}&	0.0500		&	0.0324&	0.0510 	&	0.0510	\\
	Normalized	&	0.0500 	&	\textbf{0.0280}	&	0.6389		&	\textbf{0.2862}&	0.0008			&	-0.0005&	0.0500		&	\textbf{0.0280}&	\textbf{0.0458}	&	\textbf{0.0498	}\\\hline
  \end{tabular}
\end{center}
\label{table-normal-balanced}
\end{table*}

\subsection{Size of the Training Sample}
\label{sec-sizetraining}

\begin{figure}
        \centering
        \begin{subfigure}[b]{\columnwidth}
                \includegraphics[width=\textwidth]{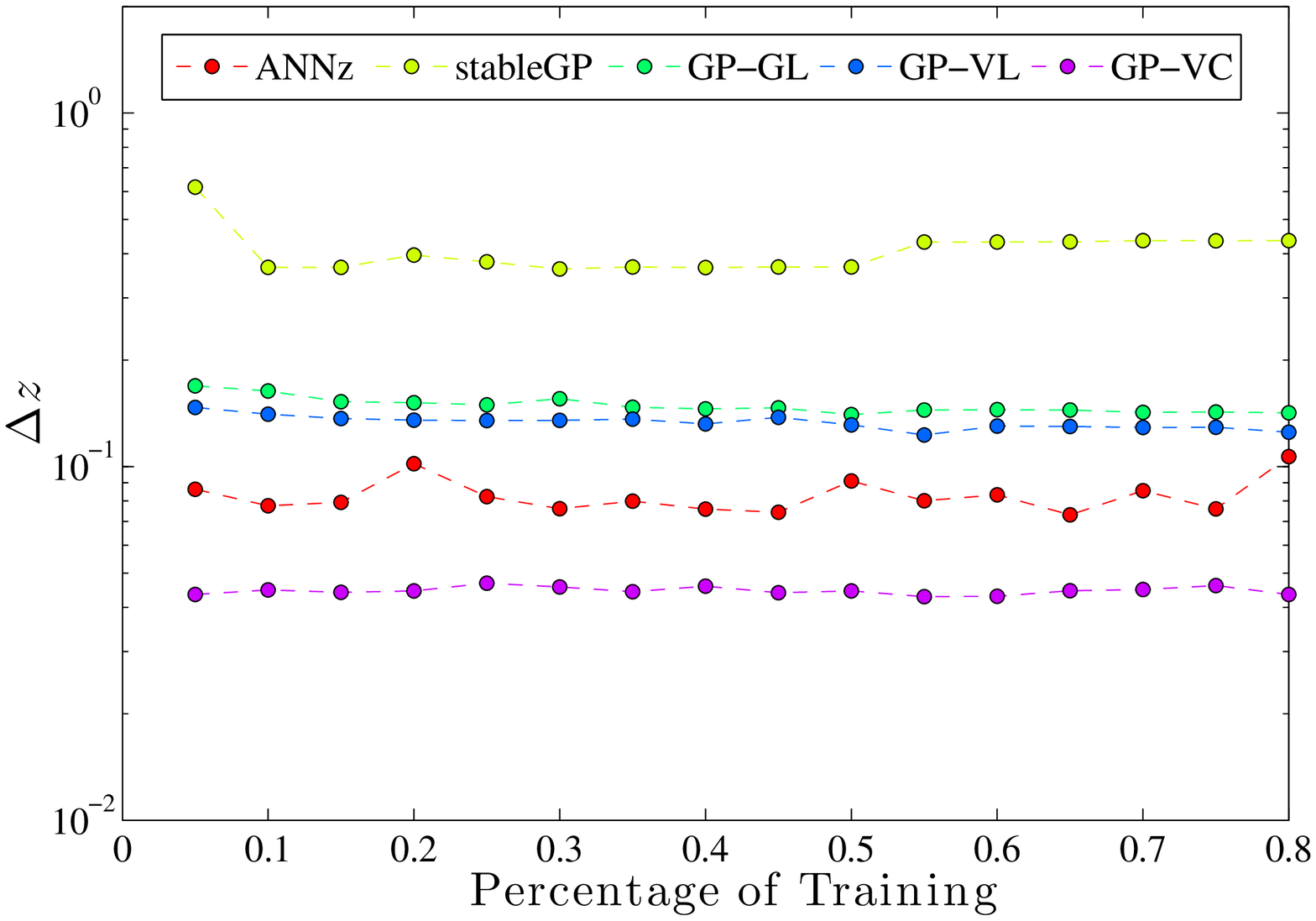}
                \caption{10 Basis Functions}
                \label{fig-training-percentage-10}
        \end{subfigure}	
        \begin{subfigure}[b]{\columnwidth}
                \includegraphics[width=\textwidth]{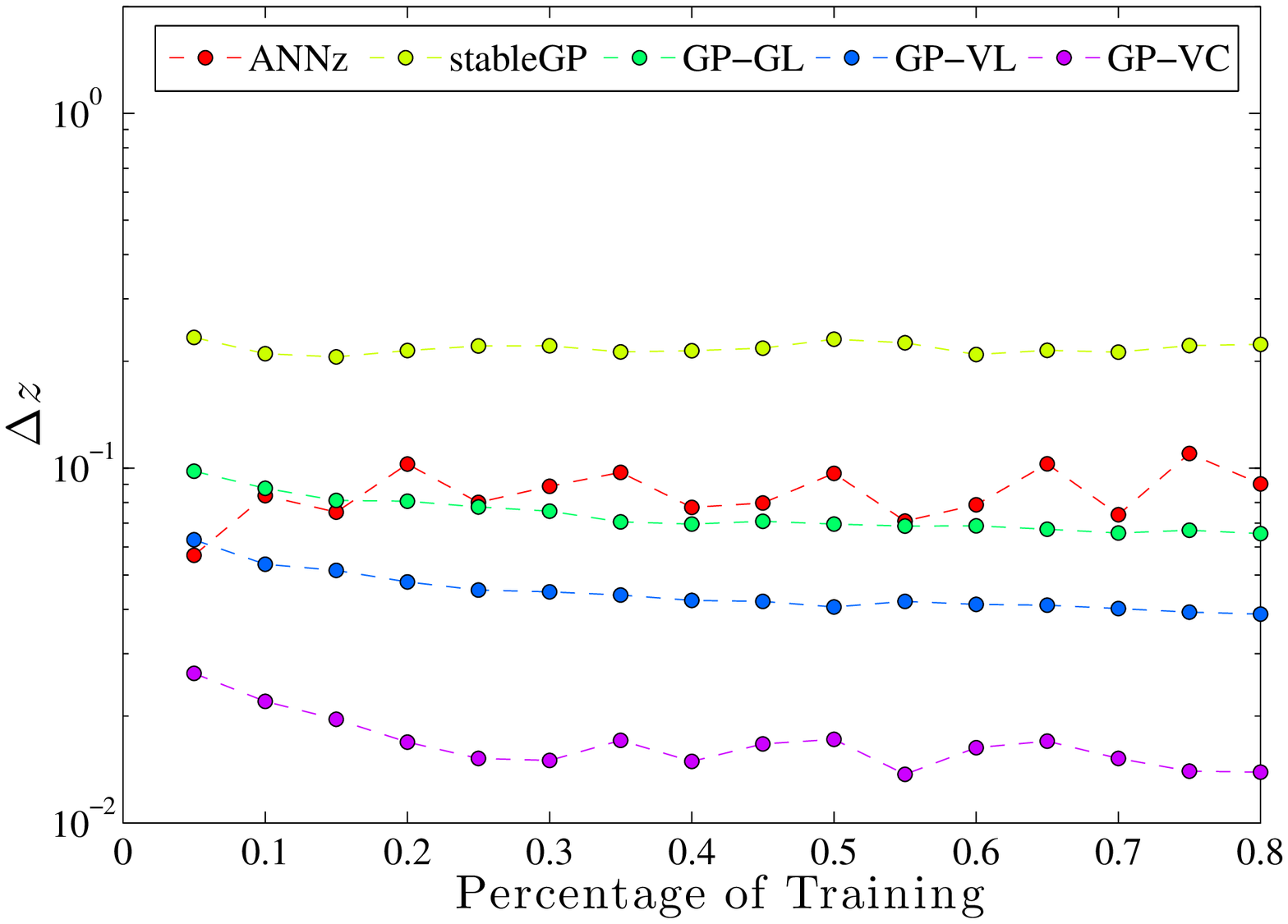}
                \caption{100 Basis Functions}
                \label{fig-training-percentage-100}
        \end{subfigure}
       \caption{$\Delta z$ as a function of training size for all the methods using a simple model ($m=10$) and a complex model ($m=100$) trained on the simulated data.}
	\label{fig-normal-balanced}
\end{figure}

Thus far, we have only considered the case of having a large (80 per cent of the total data) training sample. In practice, it is likely that the training-validation-test will be substantially smaller than the dataset for which photometric redshifts are required. Therefore, in this section the generalization performance of the models are tested by limiting the training sample size to different percentages of the dataset. The validation and test samples were fixed to the same samples used in previous experiments to ensure consistent reporting on the same test sample. The models were trained using various percentages from 5\% to 80\%, once using a small number of basis functions ($m=10$) and a second experiment using a larger number of basis functions ($m=100$) and the results are shown for both in Figures \ref{fig-training-percentage-10} and \ref{fig-training-percentage-100} respectively. GP-VC consistently outperforms the other models across the training range using both simple and complex models. {ANNz} on the other hand performs poorly and quickly overfits in the complex version of the test. It is worth noting that, unlike the other models, {ANNz} consistently shows an unsteady performance in all of the parameter tuning tests for the simulated data set. However, we note that on real noisy data this problem diminishes (Sec.~\ref{sec-experiments-sdss}).

\subsection{Size of the basis set}

Until now, we have limited the number of basis functions to 10, except for the last section to test the generalization performance of the models. In practice, the only limitation on the number of basis functions used for training the GP is computing resources. In this section we investigate how the accuracy of the photometric redshifts depends on the number of basis functions.

\begin{figure}
	\centering
	\includegraphics[width=\columnwidth]{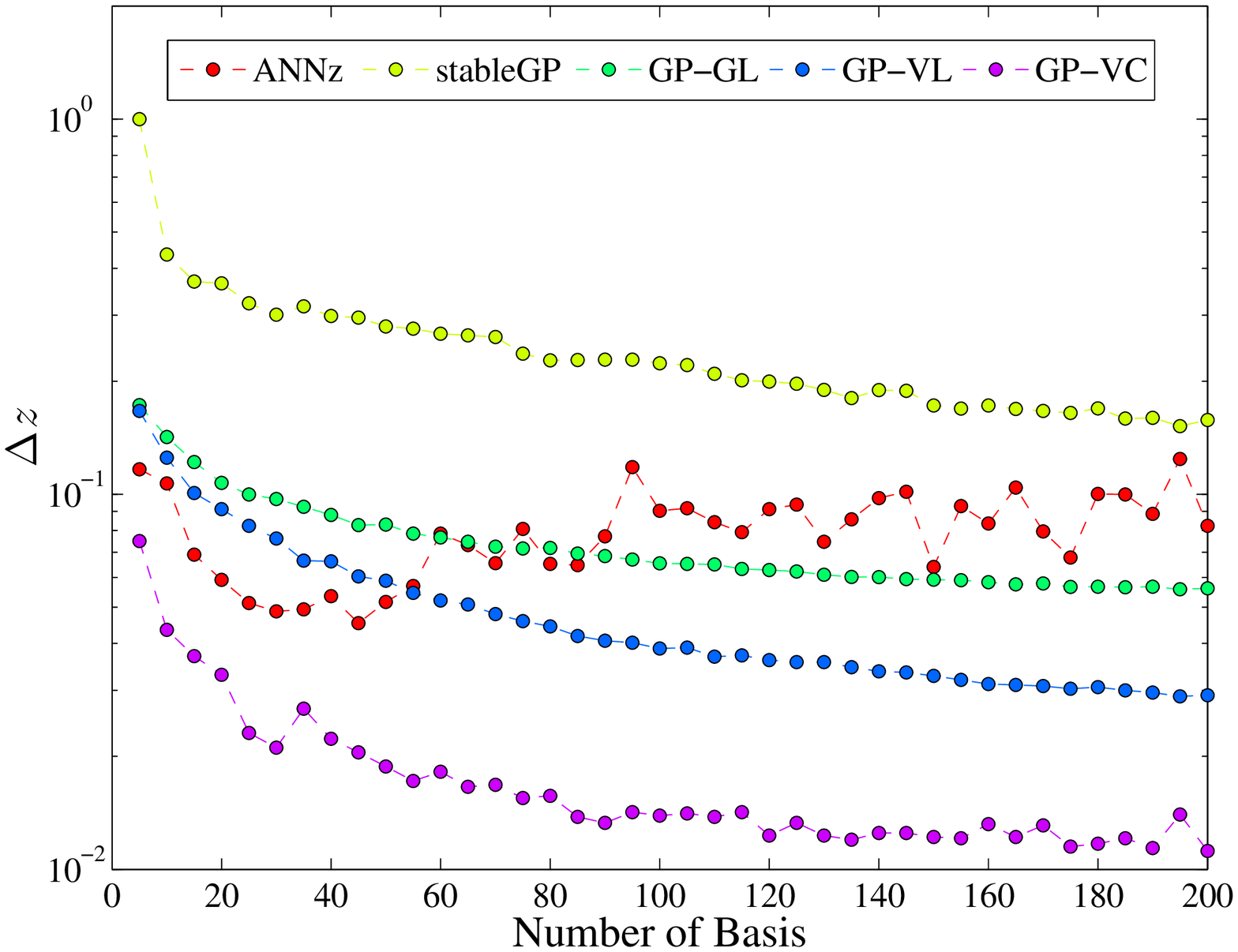}
	\caption{$\Delta z$ as a function of the number of basis functions for all the methods.}
	\label{fig-rmses}
\end{figure}

We cross-compare all of the models by varying the number of basis functions $m$ from 5 to 200 by an increment of 5 to study the relationship between accuracy and complexity. $\Delta z$ as a function of $m$ for all the models are shown in Figure \ref{fig-rmses}, the $y$-axis is shown on a log scale for the purpose of visualisation. The {\sc stableGP} method exhibits the worst performance across the board, especially when the number of basis functions is small. On the other hand, GP-VC consistently outperforms the rest, and most significantly when trained with few basis functions. {\sc ANNz} outperforms GP-GL and GP-VL, but it does not scale well with complexity as it starts to overfit after $m=30$. All the models were trained using a sum of squared errors objective with no cost-sensitive learning or a prior mean function optimization in this experiment. 

\begin{figure}
       \centering
       \includegraphics[width=\columnwidth]{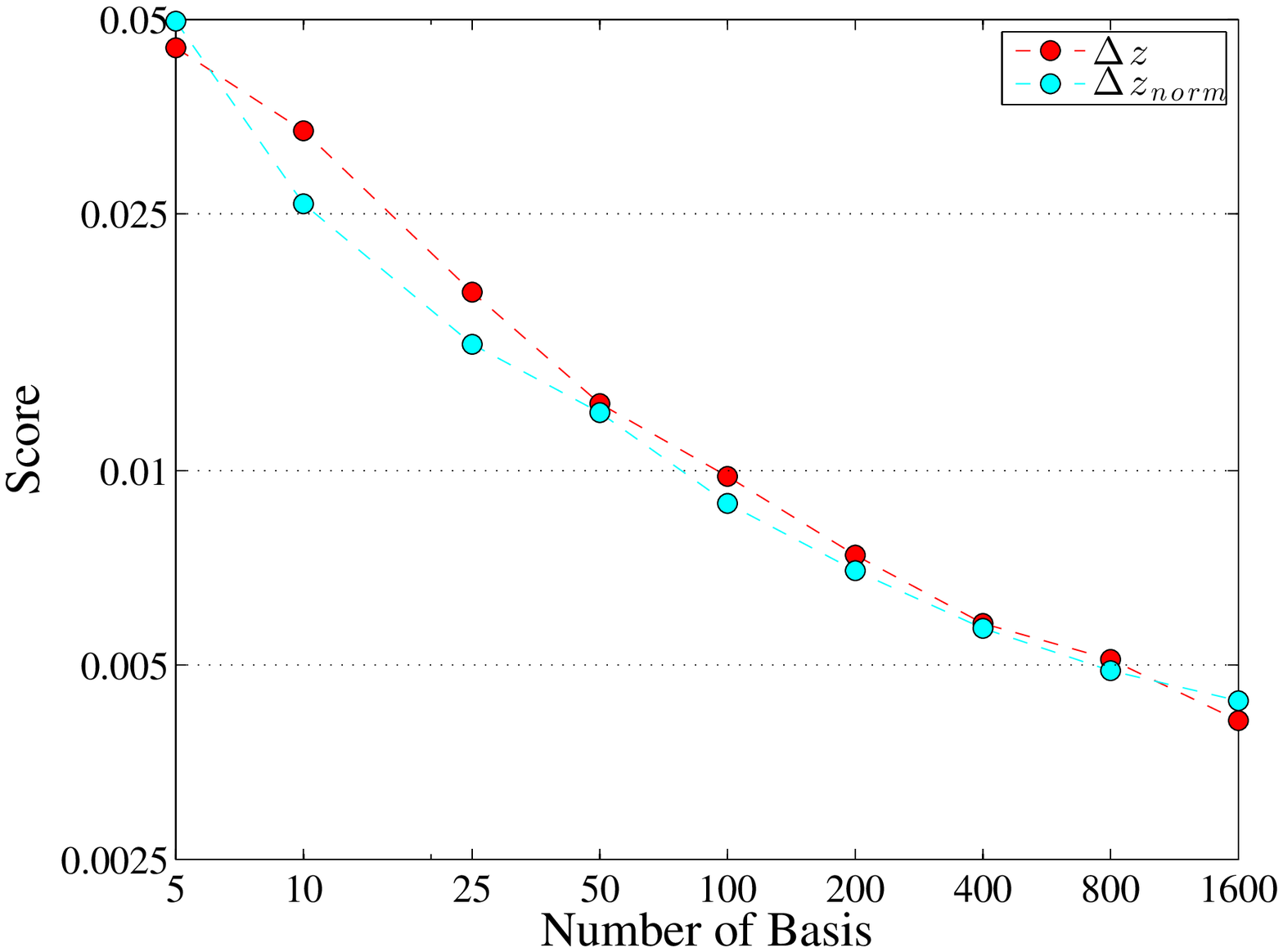}
        \caption{A log-log plot reporting the $\Delta z$ and $\Delta z_\textrm{norm}$ scores after training GP-VC models on the simulated data with 5, 10, 25, 50, 100, 200, 400, 800 and 1600 basis functions with joint linear optimization. The results for the $\Delta z$ were optimized using normal sum of squares whereas the results for the $\Delta z_\textrm{norm}$ were optimized using normalized weights}
       \label{fig-1600}
\end{figure}

In Figure \ref{fig-1600} we show the values of $\Delta z$ and $\Delta z_\textrm{norm}$ for the GP-VC approach using an extended range of basis functions of 5, 10, 25, 50, 100, 200, 400, 800 and 1600 with joint linear optimization, using both the normalized weights and normal sum of squares. With the GP-VC we obtain $\Delta z_\textrm{norm} = 0.0295$ with just $m=5$ basis functions, and when using $m=1600$ we obtain $\Delta z_\textrm{norm} = 0.0026$ and a maximum normalized error $\Delta z_\textrm{norm} = 0.0652$. We note that although the \emph{training} complexity costs require effort for large numbers of basis functions, once all parameters are inferred we enjoy effectively a linear basis model performance running over unseen (test) data. We therefore consider the performance for a realistic, yet large, number of functions. For an in depth analysis of feature selection, magnitude cuts and training sample size on photometric redshift the reader is referred to \cite{hoyle2015}.

\begin{figure}
        \centering
       
       \begin{subfigure}[b]{\columnwidth}
                \includegraphics[width=\columnwidth]{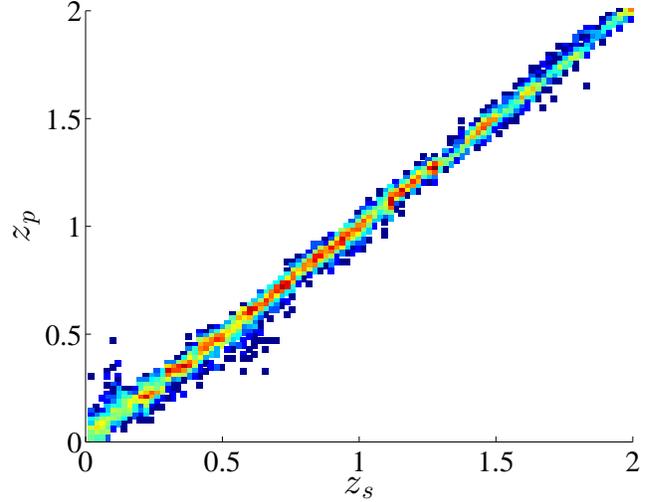}
        \caption{{\sc ANNz}}
        \end{subfigure}
        ~ 
        \begin{subfigure}[b]{\columnwidth}
                \includegraphics[width=\columnwidth]{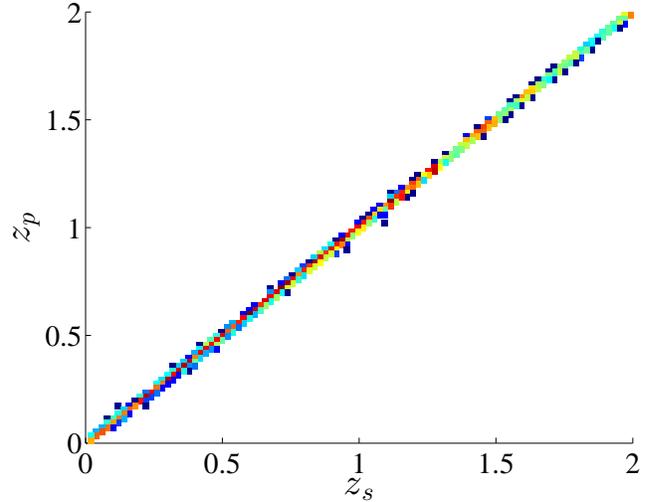}
        \caption{GP-VC}
        \end{subfigure}

       \caption{The density scatter plot for (a) the final {\sc ANNz} model using a committee of 5 networks with 8:16:16:1 architectures and (b) the final GP-VC model trained using $m=1600$ basis functions with a jointly optimized linear mean function on the simulated survey. }
       \label{fig-final-model}
\end{figure}

We also generated photometric redshifts from a committee of five neural networks using a two-layer architecture, each layer with twice the number of hidden units as the number of filters as recommended in \cite{Collister04} and has become a standard for most {\sc ANNz} users. The models were trained using the same training and validation samples, and the quoted results were calculated on the test sample. The results of the final GP-VC and {\sc ANNz} models are summarized in Table \ref{table-GP-ANN-simulated} and the density scatter plots for the final models are shown in Figure \ref{fig-final-model} for comparison. We find that both learn the underlying (simulated) relationship, GP-VC however provides a factor of $\sim 7$ improvement in the accuracy of $\Delta z$ and $\Delta z_\textrm{norm}$ over the commonly-used {\sc ANNz} architecture for the simulated data set.

 \begin{table*}
\caption{Performance measures for the final {\sc ANNz} model using a committee of 5 networks with 8:16:16:1 architectures and the final GP-VC model trained on the simulated survey using $m=1600$ basis functions with a jointly optimized linear function on the simulated survey.}
\begin{center}
\begin{tabular}{| l | c | c |  c | c |  c | c |  c | c |  c | c | }
     				&	$\Delta z$	&	$\Delta z_\textrm{norm}$	&	max$_{z}$ & max$_{norm}$		&	$\mu_{z}$&	$\mu_{norm}$	& $\sigma_{z}$ & $\sigma_{norm}$ & out$_{z}$&out$_{norm}$\\	\hline
	{\sc ANNz}		&	0.0262	&	0.0180		&	0.3696		&	0.3391&	-0.0004		&	-0.0007&	0.0262		&	0.0180&	\textbf{0.0433}		&	\textbf{0.0406}\\
	GP-VC		&	\textbf{0.0041} 	&	\textbf{0.0026}	&	\textbf{0.0764}		&	\textbf{0.0652}&	\textbf{0.0000}			&	\textbf{0.0000}&	\textbf{0.0041}		&	\textbf{0.0026}&	0.0480	&	0.0460\\\hline
  \end{tabular}
\end{center}
\label{table-GP-ANN-simulated}
\end{table*}

\section{Results on SDSS Data}
\label{sec-experiments-sdss}

In this section we compare {\sc ANNz} with GP-VC on data from the SDSS 12th Data Release. The data used for training was selected from all the galaxies in the database where photometric and spectroscopic data are available and any sources with missing data was excluded from training. The {\fontfamily{pcr}\selectfont modelMag} magnitudes were used with their associated error estimates. The following SQL statement was used to extract the data from the SDSS DR12 database using the CasJobs service provided by SDSS\footnote{\url{casjobs.sdss.org}}.

\begin{verbatim}
SELECT
p.objid,
p.modelMag_u, p.modelMag_g,
p.modelMag_r, p.modelMag_i,
p.modelMag_z, p.modelMagerr_u,
p.modelMagerr_g, p.modelMagerr_r,
p.modelMagerr_i, p.modelMagerr_z,
s.z as zspec, s.zErr as zspecErr,
s.survey as survey
INTO
mydb.modelmag_dataset
FROM
PhotoObjAll as p, SpecObj as s
WHERE
p.SpecObjID = s.SpecObjID AND
s.class = 'GALAXY' AND 
s.zWarning = 0 AND
p.mode = 1 AND
dbo.fPhotoFlags('PEAKCENTER') != 0 AND
dbo.fPhotoFlags('NOTCHECKED') != 0 AND
dbo.fPhotoFlags('DEBLEND_NOPEAK') != 0 AND
dbo.fPhotoFlags('PSF_FLUX_INTERP') != 0 AND
dbo.fPhotoFlags('BAD_COUNTS_ERROR') != 0 AND
dbo.fPhotoFlags('INTERP_CENTER') != 0
\end{verbatim}

We use similar image flags to the ones used in \citet{brescia2014catalogue}. To target the original spectroscopic limit of the SDSS, the models are also tested with a cutoff of $r<17.7$ applied. Four different data sets were created from the retrieved data:
\begin{enumerate}
  \item SDSS: This data set includes only sources from the SDSS but not the BOSS survey (817,604 sources) 
  \item SDSS with cut: This data set includes only the sources from the SDSS data set with $r<17.7$ (577,725 sources).
  \item SDSS+BOSS: This data set includes all sources from the BOSS and the SDSS surveys (2,120,465 sources). 
  \item SDSS+BOSS with cut: This data set includes only the sources from the SDSS+BOSS data set with $r<17.7$ (629,117 sources). 
\end{enumerate}

\begin{figure*}
        \centering
        \begin{subfigure}[b]{0.45\textwidth}
                \includegraphics[width=\textwidth]{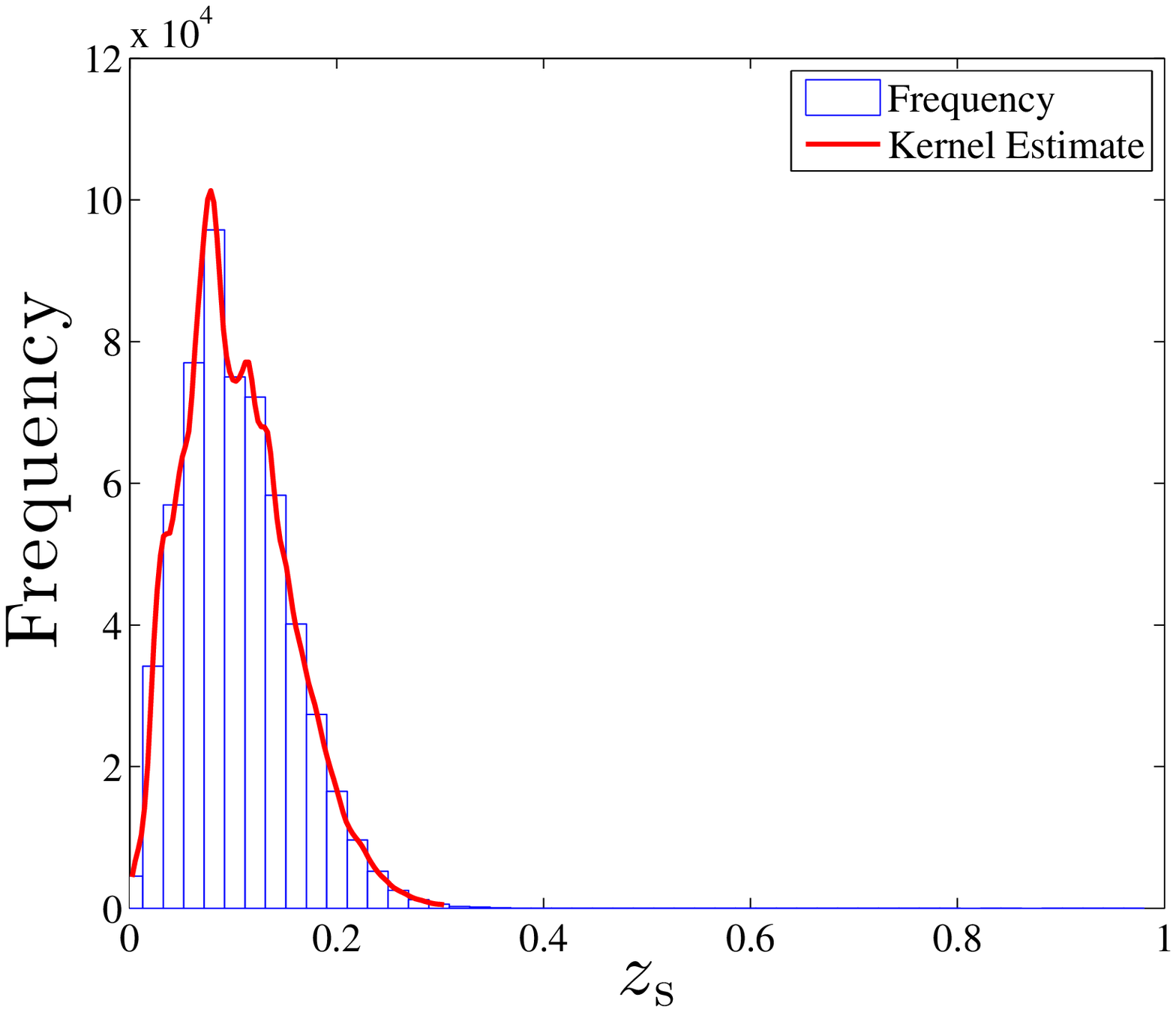}
        \end{subfigure}
        ~
        \begin{subfigure}[b]{0.45\textwidth}
                \includegraphics[width=\textwidth]{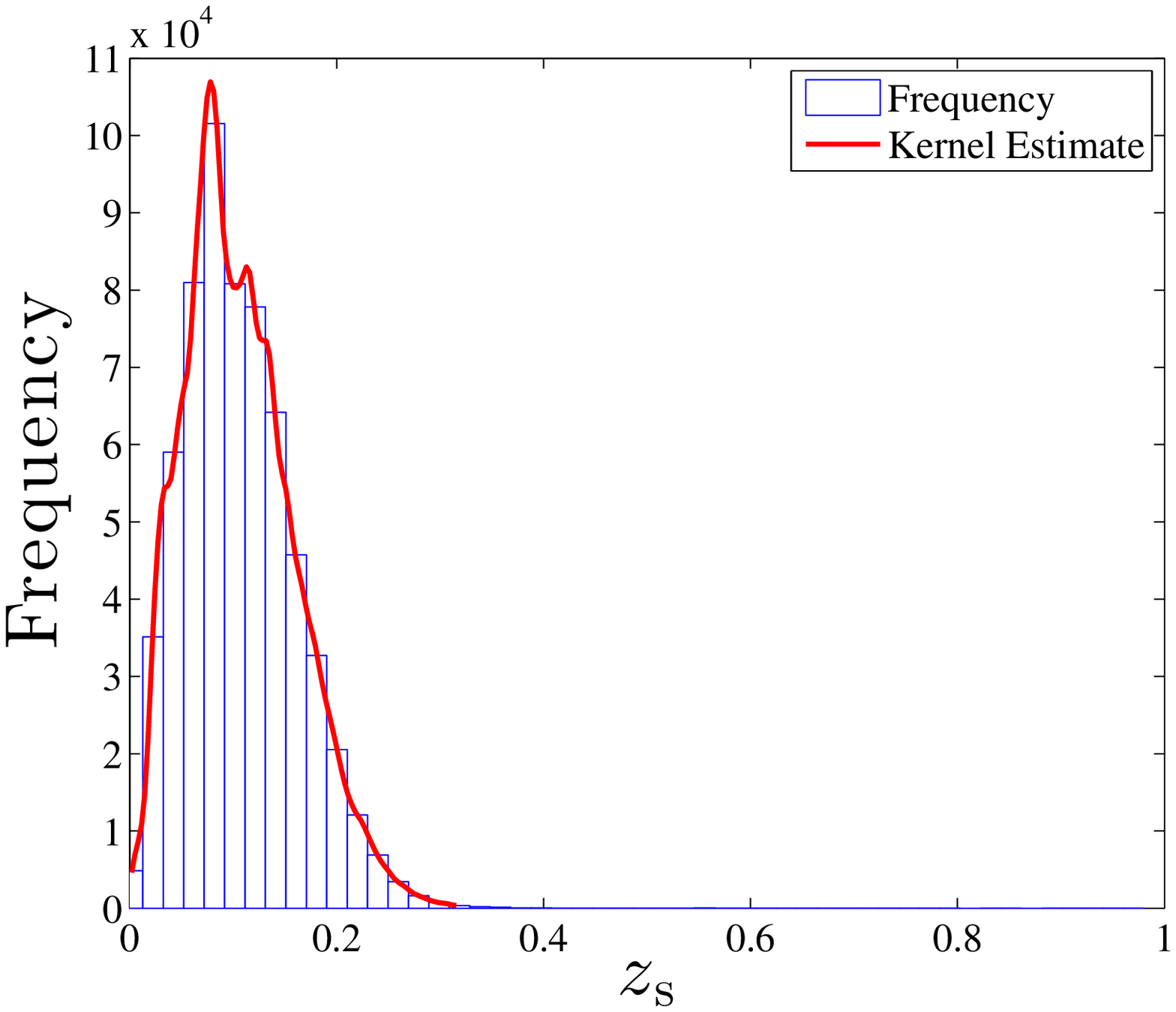}
        \end{subfigure}

       \begin{subfigure}[b]{0.45\textwidth}
                \includegraphics[width=\textwidth]{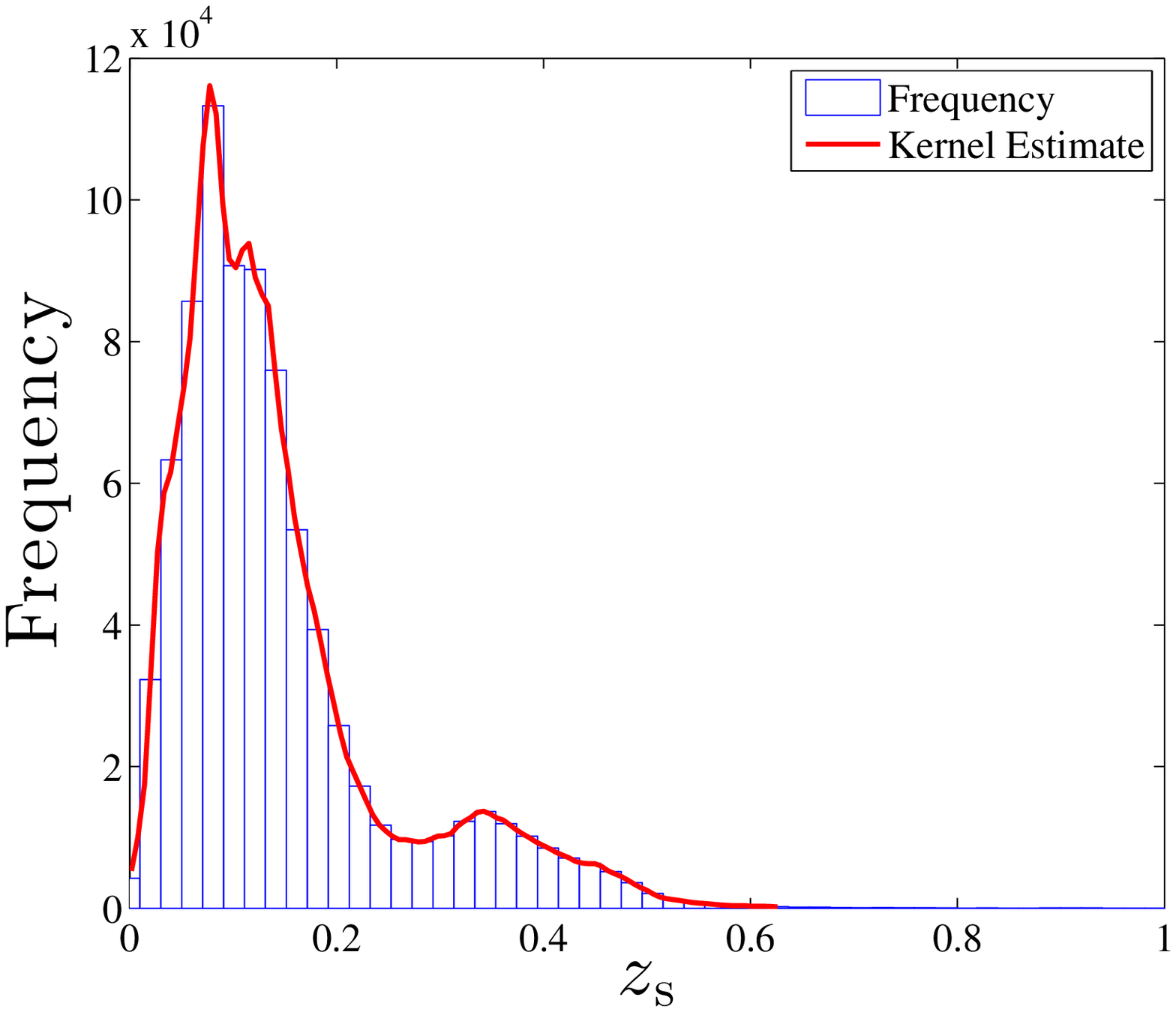}
                \caption{SDSS}
        \end{subfigure}
        ~
        \begin{subfigure}[b]{0.45\textwidth}
                \includegraphics[width=\textwidth]{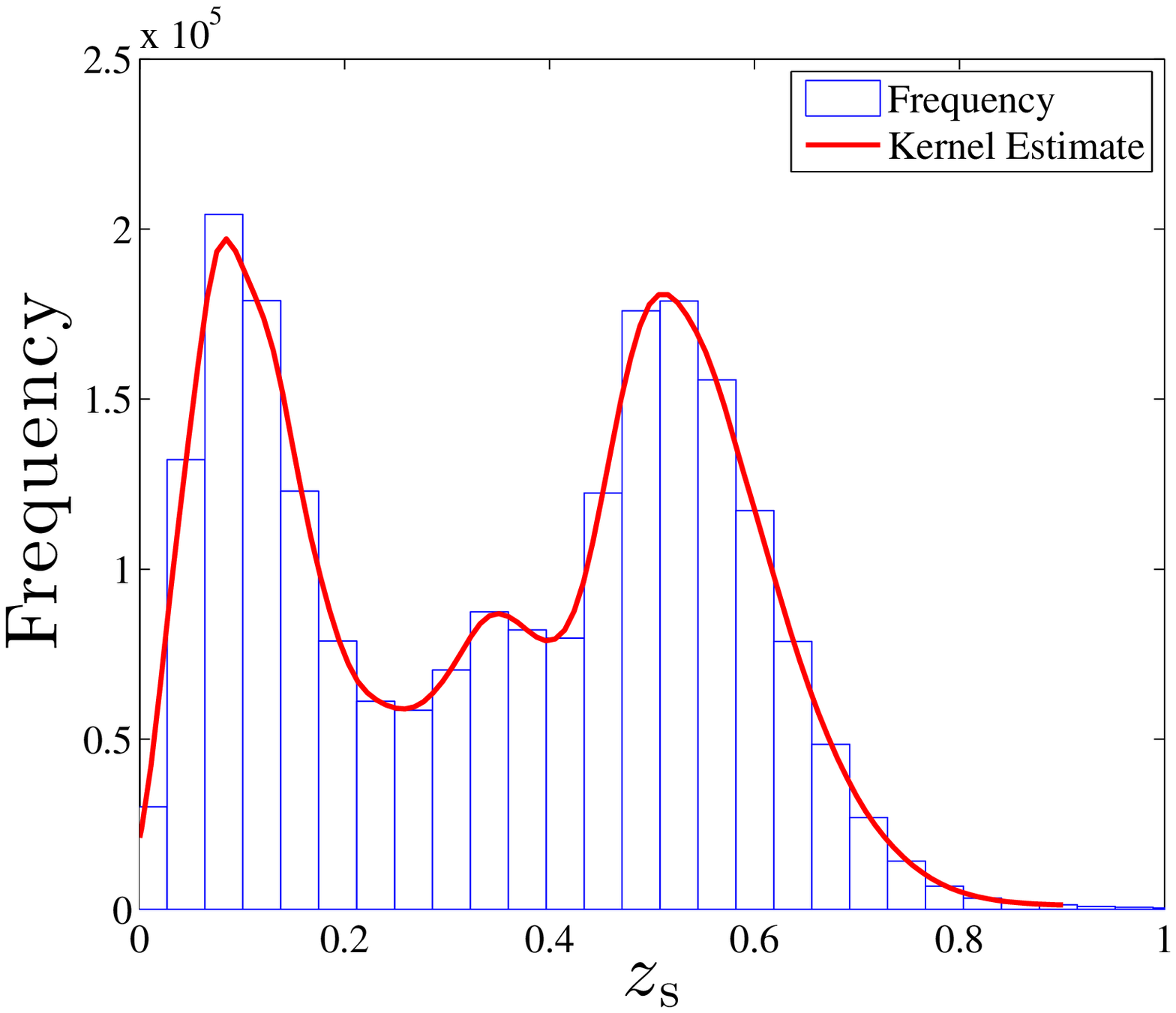}
                \caption{SDSS+BOSS}
        \end{subfigure}

        \caption{The distribution of the spectroscopic redshift in the (a) SDSS and (b) SDSS+BOSS datasets. The top figures show the distributions with the $r<17.7$ cut while the bottom figures show the distributions without the cut.}
        \label{fig-zpec-sdss}
\end{figure*}

The distribution of the spectroscopic redshifts of the data sets are shown in Figure \ref{fig-zpec-sdss}. Similar to the simulated data setup, we used 80\% for training, 10\% for validation and 10\% for testing in each data set.

\subsection{Varying the Degree of Freedom}

In this experiment, we compare {\sc ANNz} to GP-VC's performance based on the number of free parameters. This is achieved by setting the number of basis functions in GP-VC such that the number of free parameters are approximately equal to the number of free parameters in a two-layer neural network using the following assignment:

\begin{subequations}
\begin{align}
\label{eq-gpvc-degree}
\mathcal{D}_{g}		&=		m_{g}\left(d^{2}+d\right)+d+1,\\
\label{eq-ann-degree}
\mathcal{D}_{z}	&=		m_{z}^{2}+dm_{z}+3m_{z}+1,\\
\label{eq-gpvc-equals-ann}
m_{g}		&\approx 	\frac{m_{z}^{2}+dm_{z}+3m_{z}-d}{d^{2}+d},
\end{align}
\end{subequations}
where $\mathcal{D}_{g}$ is the degree of freedom in GP-VC, with joint prior mean function, $\mathcal{D}_{z}$ is the degree of freedom in {\sc ANNz}, $m_{g}$ is the number of basis functions in GP-VC and $m_{z}$ is the number of hidden units in {\sc ANNz}. The number of hidden units is set to be equal in both layers. Setting the number of basis functions in GP-VC according to \eqref{eq-gpvc-equals-ann}, ensures that $\mathcal{D}_{g} \approx \mathcal{D}_{z}$.

\begin{figure}
	\centering
	\includegraphics[width=\columnwidth]{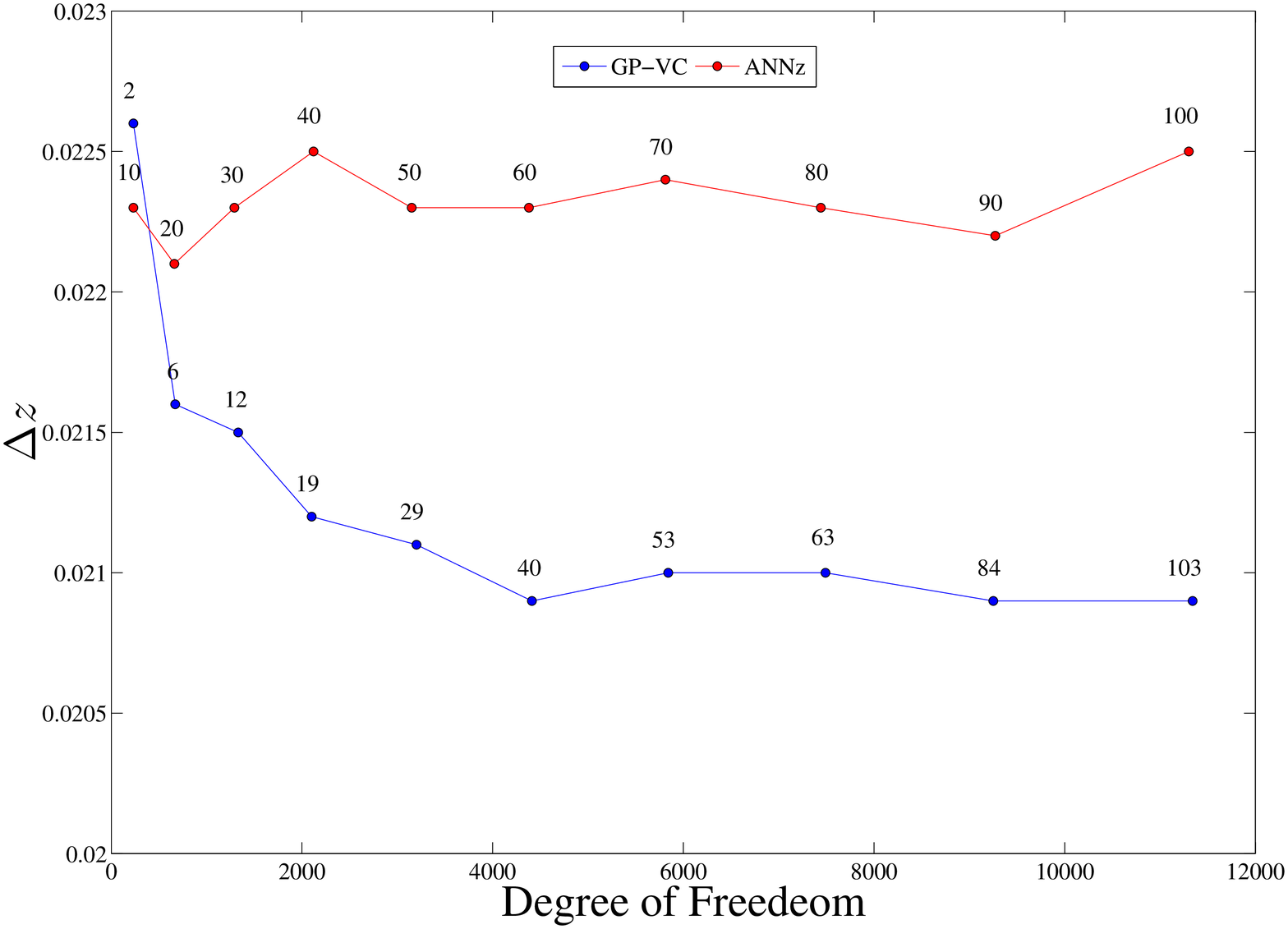}
	\caption{A comparison between GP-VC and {\sc ANNz} with two layers on the SDSS dataset using various degrees of freedom. The numbers on top of each point is the equivalent number of basis functions in GPVC and the number of hidden units for each layer in {\sc ANNz}.}
	\label{fig-ann-gpvc-sdss}
\end{figure}

In this test, we trained a two-layer {\sc ANNz} architecture using 10--100 hidden units and a matching GP-VC model based on Eq. \eqref{eq-gpvc-equals-ann} with joint mean optimization. The number of hidden units was set to be equal in both layers but only a single network was used to generate the predictions not a committee of five networks. Both models were trained on the SDSS data set and the results on the test set are shown in Figure \ref{fig-ann-gpvc-sdss}. The results are consistent with the results from the simulated data, the performance of {\sc ANNz} degrades as we increase the complexity of the network, whereas GP-VC is more robust and shows a steady improvement.

\subsection{Final Results}

\begin{figure*}
        \centering
       
       \begin{subfigure}[b]{0.35\textwidth}
                \includegraphics[width=\columnwidth]{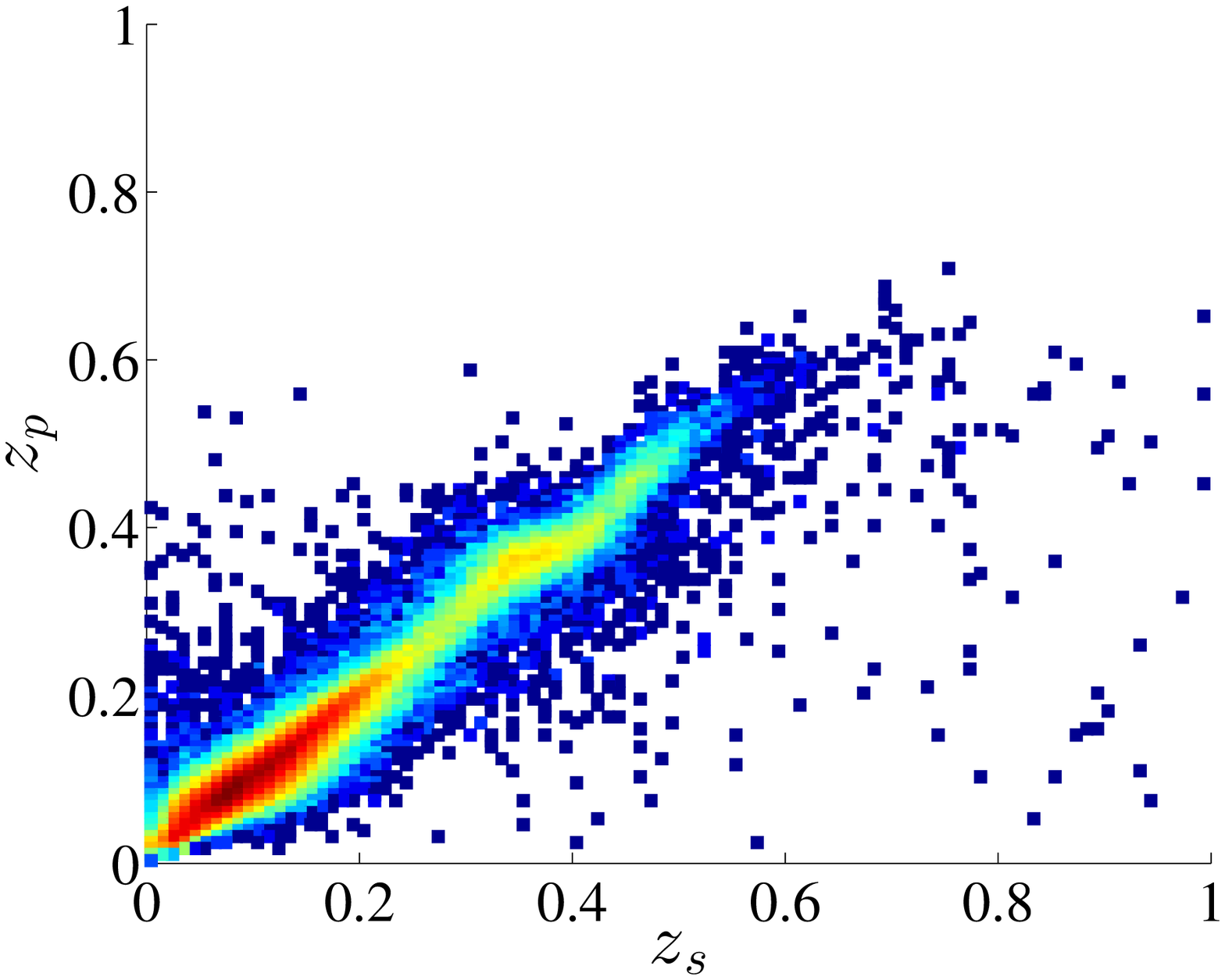}
        \end{subfigure}
        ~ 
        \begin{subfigure}[b]{0.35\textwidth}
                \includegraphics[width=\columnwidth]{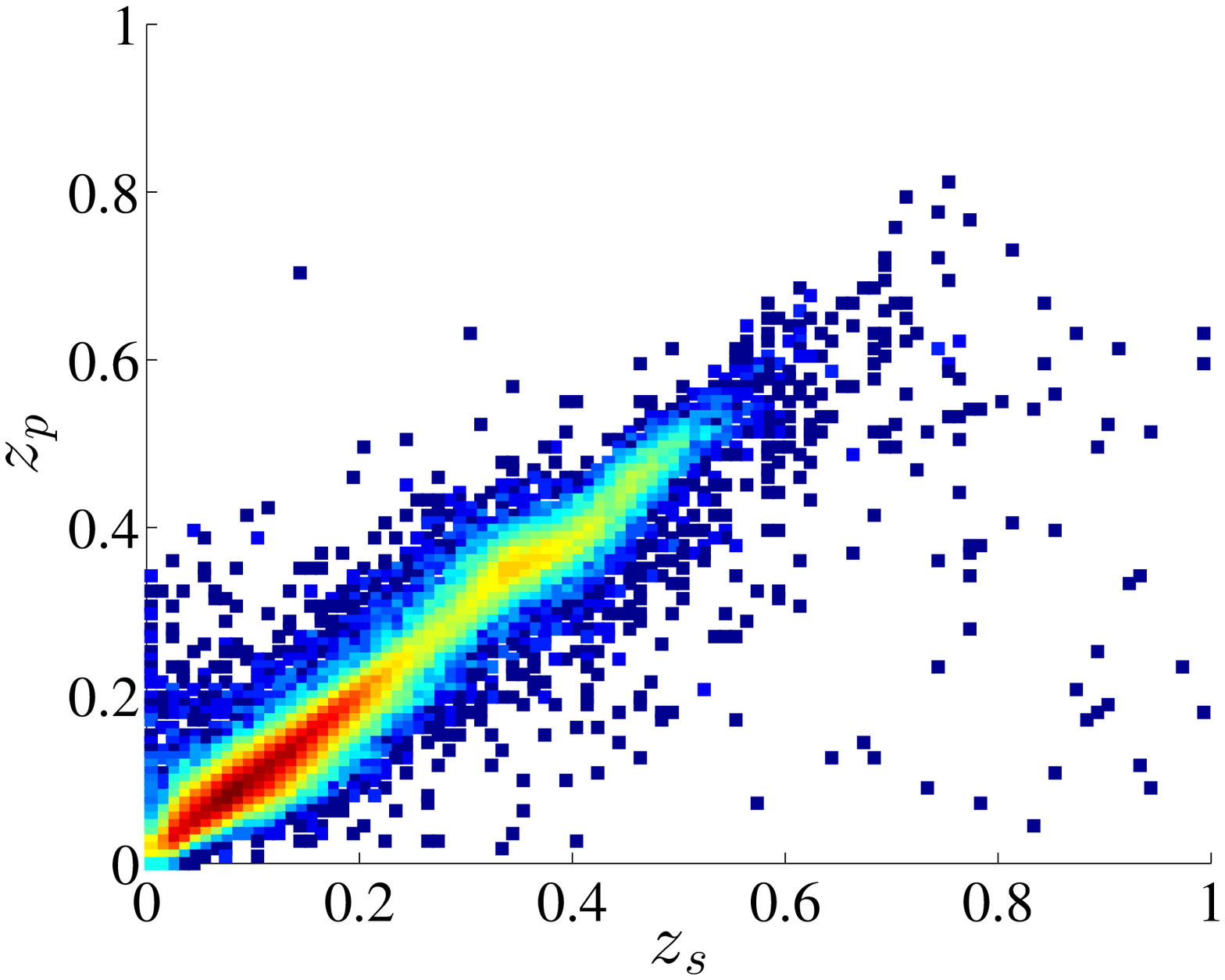}
        \end{subfigure}

        \begin{subfigure}[b]{0.35\textwidth}
                \includegraphics[width=\columnwidth]{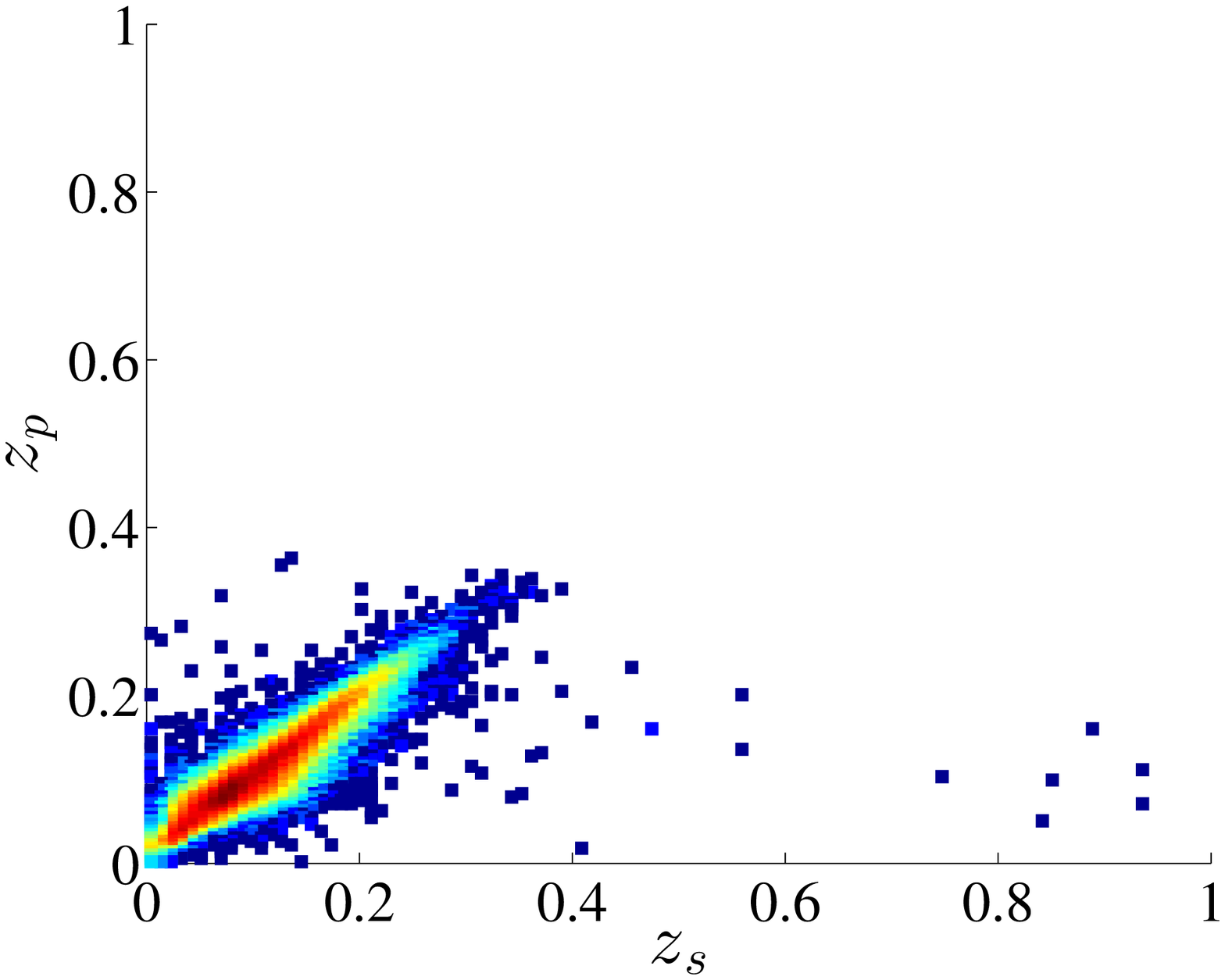}
        \caption{{\sc ANNZ}}
        \end{subfigure}
	~ 
        \begin{subfigure}[b]{0.35\textwidth}
                \includegraphics[width=\columnwidth]{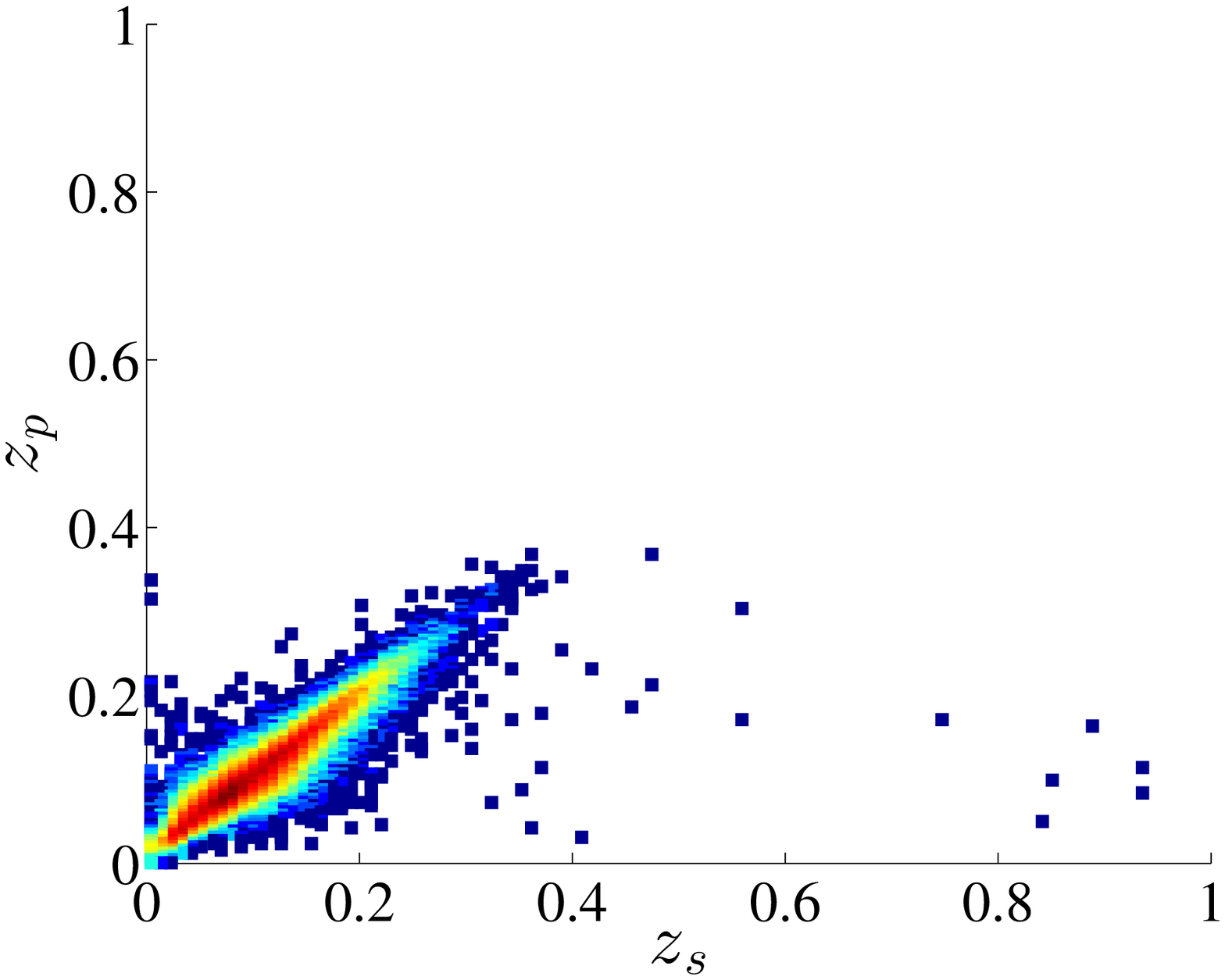}
        \caption{GP-VC}
        \end{subfigure}

       \caption{The density scatter plot for (a) the final {\sc ANNz} model using a committee of 5 networks with 5:10:10:1 architectures and (b) the final GP-VC model trained using $m=103$ basis functions with a jointly optimized linear mean function. The top figures show the plots for the SDSS data set with out the $r<17.7$ cut, while the bottom figures show the plots for the SDSS with cut.}
       \label{fig-final-model-sdss}
\end{figure*}

\begin{figure*}
        \centering

        \begin{subfigure}[b]{0.35\textwidth}
                \includegraphics[width=\columnwidth]{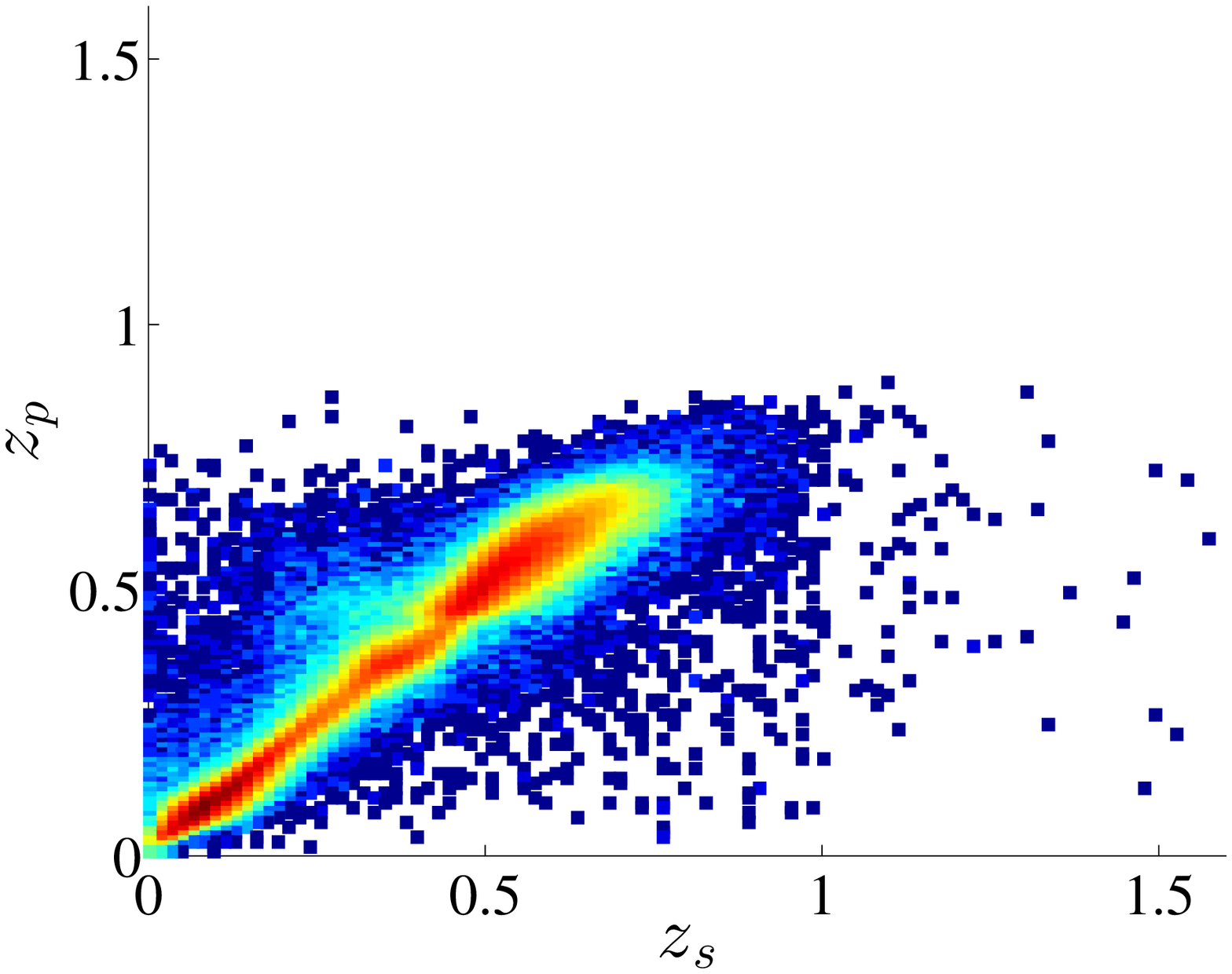}
        \end{subfigure}
        ~ 
        \begin{subfigure}[b]{0.35\textwidth}
                \includegraphics[width=\columnwidth]{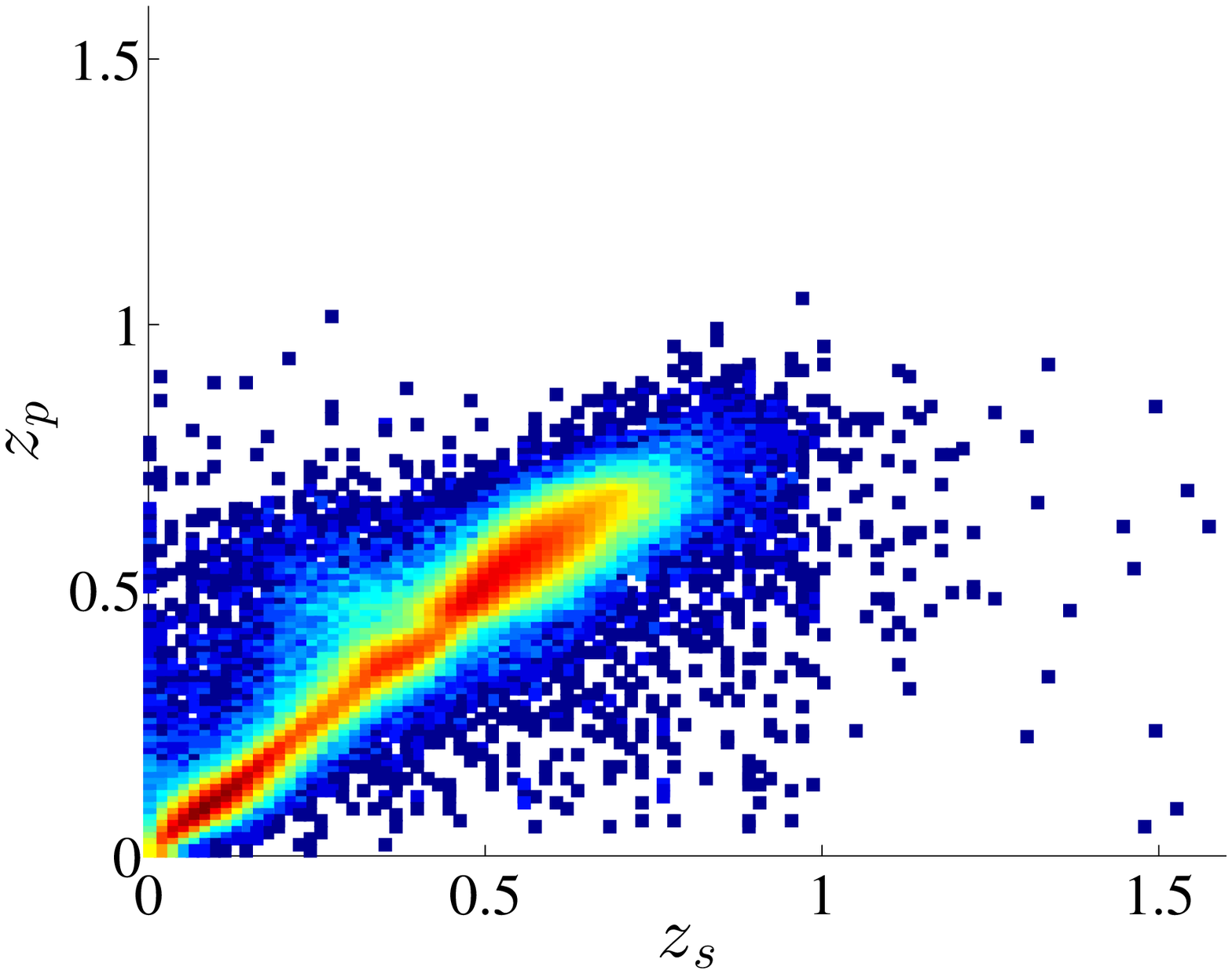}
        \end{subfigure}

        \begin{subfigure}[b]{0.35\textwidth}
                \includegraphics[width=\columnwidth]{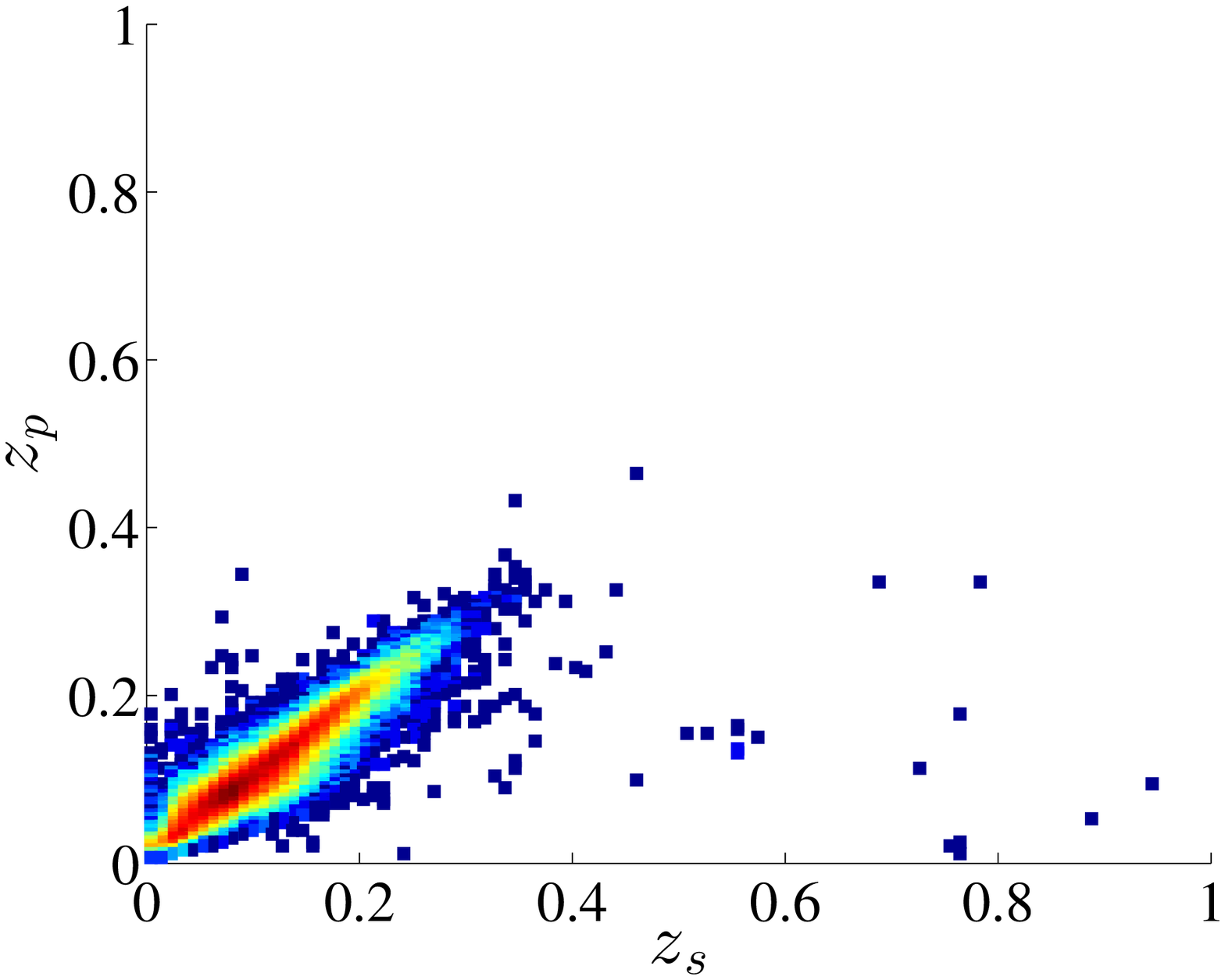}
        \caption{{\sc ANNZ}}
        \end{subfigure}
        ~ 
        \begin{subfigure}[b]{0.35\textwidth}
                \includegraphics[width=\columnwidth]{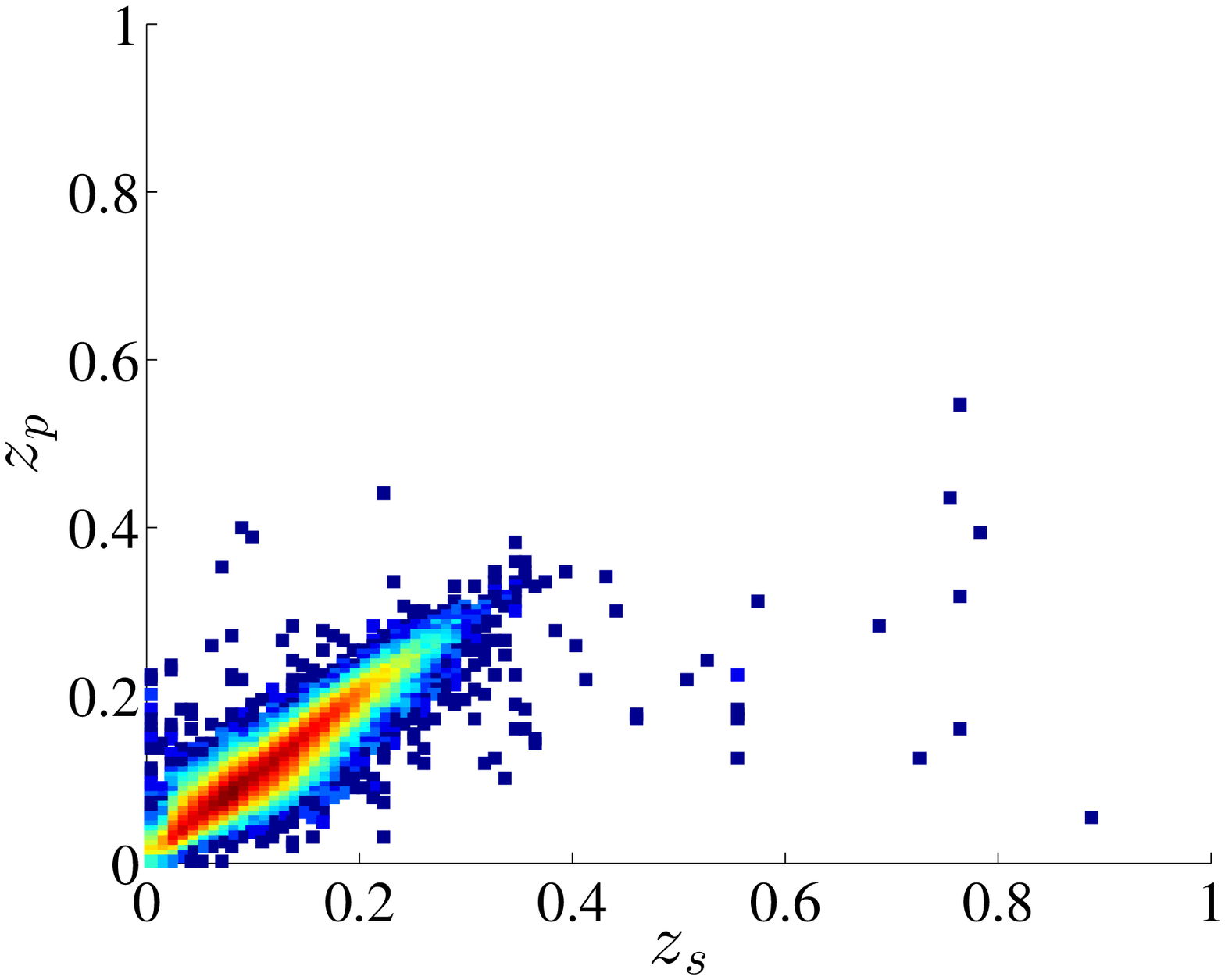}
        \caption{GP-VC}
        \end{subfigure}
        
       \caption{The density scatter plot for (a) the final {\sc ANNz} model using a committee of 5 networks with 5:10:10:1 architectures and (b) the final GP-VC model trained using $m=103$ basis functions with a jointly optimized linear mean function. The top figures show the plots for the SDSS+BOSS data set with out the $r<17.7$ cut, while the bottom figures show the plots for the SDSS+BOSS with cut. }
       \label{fig-final-model-boss}
\end{figure*}

In this experiment, photometric redshifts were generated using a committee of five two-layer architectures, each layer with twice the number of hidden units as the number of filters and compared to the best performing GP-VC model from the previous test. The experiment was carried out on the four data sets, the performance measures are reported in tables \ref{table-final-results-sdss}, \ref{table-final-results-sdss-cut}, \ref{table-final-results-boss} and \ref{table-final-results-boss-cut} for the SDSS, SDSS with cut, SDSS+BOSS and SDSS+BOSS with cut respectively. The density scatter plots for the SDSS and SDSS+BOSS data sets are shown in Figure \ref{fig-final-model-sdss} and \ref{fig-final-model-boss} respectively. 

Although not as stark as for the simulated data set, our GP-VC algorithm consistently outperforms {\sc ANNz} on the important metrics ($\Delta z$ and $\Delta z_{norm}$) and from examining the plots, the outliers are less extreme in the GP-VC case, with an increase in accuracy of around 5~per cent for the full SDSS+BOSS data set. However, we note that given the overall performance of GP-VC on the simulated data set, we expect the real power of our GP-VC to algorithm to be shown in the sparse data regime, and we reserve such an analysis to a future paper.

\begin{center}
\begin{table*}

\caption{Performance measures for the final {\sc ANNz} model using a committee of 5 networks with 5:10:10:1 architectures and the final GP-VC model using $m=103$ basis functions with a jointly optimized linear function on all the data sets.}
\begin{subtable}{\textwidth}
\centering

\begin{tabular}{| l | c | c |  c | c |  c | c |  c | c |  c | c | }
     				&	$\Delta z$	&	$\Delta z_\textrm{norm}$	&	max$_{z}$ & max$_{norm}$		&	$\mu_{z}$&	$\mu_{norm}$	& $\sigma_{z}$ & $\sigma_{norm}$ & out$_{z}$&out$_{norm}$\\	\hline
	{\sc ANNz}		&	0.0308	&	0.0249		&	0.8689		&	\textbf{0.4478}&	0.0000		&	-0.0007 &	0.0308		&	0.0249&	0.0327		&	0.0378\\
	{\sc GP-VC } 	&	\textbf{0.0302} 	&	\textbf{0.0244}		&	\textbf{0.8678}	&	0.5017 & 0.0000		&	\textbf{-0.0006}&	\textbf{0.0302}		&	\textbf{0.0244}&	\textbf{0.0316} 	&	\textbf{0.0365}\\\hline
  \end{tabular}

\caption{SDSS}
\label{table-final-results-sdss}
\end{subtable}

\begin{subtable}{\textwidth}\
\centering
\begin{tabular}{| l | c | c |  c | c |  c | c |  c | c |  c | c | }
     				&	$\Delta z$	&	$\Delta z_\textrm{norm}$	&	max$_{z}$ & max$_{norm}$		&	$\mu_{z}$&	$\mu_{norm}$	& $\sigma_{z}$ & $\sigma_{norm}$ & out$_{z}$&out$_{norm}$\\	\hline
	{\sc ANNz}		&	0.0221	&	0.0190		&	0.8710		&	0.4489&	0.0002		&	\textbf{-0.0002}&	0.0221		&	0.0190&	0.0397		&	0.0469\\
	{\sc GP-VC } 	&	\textbf{0.0209} 	&	\textbf{0.0179}		&	\textbf{0.8562}	&	\textbf{0.4413} & \textbf{0.0001}		&	-0.0003&	\textbf{0.0209}		&	\textbf{0.0179}&	\textbf{0.0386} 	&	\textbf{0.0456}\\\hline
  \end{tabular}

\caption{SDSS with cut}
\label{table-final-results-sdss-cut}
\end{subtable}

\begin{subtable}{\textwidth}
\centering

\begin{tabular}{| l | c | c |  c | c |  c | c |  c | c |  c | c | }
     				&	$\Delta z$	&	$\Delta z_\textrm{norm}$	&	max$_{z}$ & max$_{norm}$		&	$\mu_{z}$&	$\mu_{norm}$	& $\sigma_{z}$ & $\sigma_{norm}$ & out$_{z}$&out$_{norm}$\\	\hline
	{\sc ANNz}		&	0.0539	&	0.0386		&	\textbf{1.3113}		&	\textbf{0.7457}&	\textbf{-0.0000}		&	-0.0015 &	0.0539		&	0.0386&	0.0393		&	0.0346\\
	{\sc GP-VC } 	&	\textbf{0.0513} 	&	\textbf{0.0366}		&	1.4284	&	0.8601 & -0.0001		&	\textbf{-0.0013}&	\textbf{0.0513}		&	\textbf{0.0366}&	\textbf{0.0385} 	&	\textbf{0.0340}\\\hline
  \end{tabular}

\caption{SDSS+BOSS}
\label{table-final-results-boss}
\end{subtable}

\begin{subtable}{\textwidth}
\centering

\begin{tabular}{| l | c | c |  c | c |  c | c |  c | c |  c | c | }
     				&	$\Delta z$	&	$\Delta z_\textrm{norm}$	&	max$_{z}$ & max$_{norm}$		&	$\mu_{z}$&	$\mu_{norm}$	& $\sigma_{z}$ & $\sigma_{norm}$ & out$_{z}$&out$_{norm}$\\	\hline
	{\sc ANNz}		&	0.0222	&	0.0188		&	\textbf{0.8523}		&	\textbf{0.4421}&	-0.0000		&	-0.0004 &	0.0222		&	0.0188&	\textbf{0.0367}		&	0.0445\\
	{\sc GP-VC } 	&	\textbf{0.0207} 	&	\textbf{0.0178}		&	0.9831	&	0.5043 & -0.0000		&	\textbf{-0.0003}&	\textbf{0.0207}		&	\textbf{0.0178}&	0.0376 	&	0.0445\\\hline
  \end{tabular}

\caption{SDSS+BOSS with cut}
\label{table-final-results-boss-cut}
\end{subtable}

\label{table-final-results}
\end{table*}
\end{center}

\section{Summary and Future Work}
\label{sec-summary}

In this paper a sparse Gaussian process framework is presented and applied to photometric redshift estimation. The framework is able to out perform ANNz, sparse GP parametrized by a set of global hyper-parameters and low rank approximation GP. The performance increase is attributed to the handling of distribution bias via a weighting scheme integrated as part of the optimization objective, parametrizing each basis function with bespoke covariances, and integrating the learning of the prior mean function to enhance the extrapolation performance of the model. The methods were applied to a simulated dataset and SDSS DR12 where the proposed approach consistently outperforms the other models on the important metrics ($\Delta z$ and $\Delta z_{norm}$). We find that the model scales linearly in time with respect to the size of the data, and has a better generalization performance compared to the other methods even when presented with a limited training set. Results show that with only 30 per cent of the data, the model was able to reach accuracy close to that of using the full training sample. Even when data were selectively removed based on $RIZ$ magnitudes, the model was able to show the best recovery performance compared to the other models. The cost-sensitive learning component of the framework regularizes the predictions to limit the effect caused by the biased distribution of the output and allows for direct optimization of the survey objective (e.g. $z_{norm} = |z_\textrm{s} - z_\textrm{p}| / (1+z_{\rm s})$). Again, the algorithm consistently outperforms other approaches, including {\sc ANNz} and {\sc stableGP}, in all reported experiments. We also investigate how the size of the training sample and the basis set affects the accuracy of the photometric redshift prediction. We show that for the simulated set of galaxies, based on the work of \citet{jouvel09}, we are able to obtain a photometric redshift accuracy of $\Delta z_\textrm{norm} = 0.0026$ and $\mbox{max}_{norm}=0.0652$ using 1600 basis functions which is a factor of seven improvement over the standard {\sc ANNz} implementation. We find that GP-VC out-performed {\sc ANNz} on the real data from SDSS-DR12, with an improvement in accuracy of $\sim 5$\,per cent, even when restricted to have the same number of free parameters. In future work we will test the algorithm on a range of real data, and pursue investigations of how the algorithm performs over different redshift regimes and for different galaxy types.

\section*{Acknowledgments}
IAA acknowledges the support of King Abdulaziz City for Science and Technology.
MJJ and SNL acknowledge support from the UK Space Agency. The authors would like to thank the reviewers for their valuable comments.
\balance
\footnotesize{
\bibliographystyle{mn2e}
\bibliography{photoz}

\begin{thebibliography}{45}
\expandafter\ifx\csname natexlab\endcsname\relax\def\natexlab#1{#1}\fi

\bibitem[{{Abdalla} {et~al}\mbox{.}(2011){Abdalla}, {Banerji}, {Lahav}, \&
  {Rashkov}}]{abdalla11}
{Abdalla} F.~B., {Banerji} M., {Lahav} O., {Rashkov} V., 2011, \mnras, 417,
  1891

\bibitem[{{Alam} {et~al}\mbox{.}(2015){Alam}, {Albareti}, {Allende Prieto},
  {Anders}, {Anderson}, {Andrews}, {Armengaud}, {Aubourg}, {Bailey},
  {Bautista}, \& et~al.}]{SDSS3}
{Alam} S. {et~al.}, 2015, ArXiv e-prints

\bibitem[{{Ball} {et~al}\mbox{.}(2008){Ball}, {Brunner}, {Myers}, {Strand},
  {Alberts}, \& {Tcheng}}]{Ball2008}
{Ball} N.~M., {Brunner} R.~J., {Myers} A.~D., {Strand} N.~E., {Alberts} S.~L.,
  {Tcheng} D., 2008, \apj, 683, 12

\bibitem[{{Bolzonella}, {Miralles} \& {Pell{\'o}}(2000){Bolzonella},
  {Miralles}, \& {Pell{\'o}}}]{Hyperz}
{Bolzonella} M., {Miralles} J.-M., {Pell{\'o}} R., 2000, \aap, 363, 476

\bibitem[{{Bonfield} {et~al}\mbox{.}(2010){Bonfield}, {Sun}, {Davey}, {Jarvis},
  {Abdalla}, {Banerji}, \& {Adams}}]{bonfield10}
{Bonfield} D.~G., {Sun} Y., {Davey} N., {Jarvis} M.~J., {Abdalla} F.~B.,
  {Banerji} M., {Adams} R.~G., 2010, \mnras, 405, 987

\bibitem[{Bonnett {et~al}\mbox{.}(2015)Bonnett, Hartley, Amara, Leistedt,
  Becker, Bernstein, Bridle, Bruderer, Busha, Kind, {et~al.}}]{bonnett2015}
Bonnett C. {et~al.}, 2015, arXiv preprint arXiv:1507.05909

\bibitem[{{Brammer}, {van Dokkum} \& {Coppi}(2008){Brammer}, {van Dokkum}, \&
  {Coppi}}]{EAZY}
{Brammer} G.~B., {van Dokkum} P.~G., {Coppi} P., 2008, \apj, 686, 1503

\bibitem[{Brescia {et~al}\mbox{.}(2014)Brescia, Cavuoti, Longo, \&
  De~Stefano}]{brescia2014catalogue}
Brescia M., Cavuoti S., Longo G., De~Stefano V., 2014, Astronomy \&
  Astrophysics, 568, A126

\bibitem[{{Colless} {et~al}\mbox{.}(2003){Colless}, {Peterson}, {Jackson},
  {Peacock}, {Cole}, {Norberg}, {Baldry}, {Baugh}, {Bland-Hawthorn}, {Bridges},
  {Cannon}, {Collins}, {Couch}, {Cross}, {Dalton}, {De Propris}, {Driver},
  {Efstathiou}, {Ellis}, {Frenk}, {Glazebrook}, {Lahav}, {Lewis}, {Lumsden},
  {Maddox}, {Madgwick}, {Sutherland}, \& {Taylor}}]{2dfgrs}
{Colless} M. {et~al.}, 2003, ArXiv Astrophysics e-prints

\bibitem[{{Collister} \& {Lahav}(2004)}]{Collister04}
{Collister} A.~A., {Lahav} O., 2004, \pasp, 116, 345

\bibitem[{Cunha {et~al}\mbox{.}(2009)Cunha, Lima, Oyaizu, Frieman, \&
  Lin}]{Cunha2009}
Cunha C.~E., Lima M., Oyaizu H., Frieman J., Lin H., 2009, Monthly Notices of
  the Royal Astronomical Society, 396, 2379

\bibitem[{{Driver} {et~al}\mbox{.}(2011){Driver}, {Hill}, {Kelvin}, {Robotham},
  {Liske}, {Norberg}, {Baldry}, {Bamford}, {Hopkins}, {Loveday}, {Peacock},
  {Andrae}, {Bland-Hawthorn}, {Brough}, {Brown}, {Cameron}, {Ching}, {Colless},
  {Conselice}, {Croom}, {Cross}, {de Propris}, {Dye}, {Drinkwater}, {Ellis},
  {Graham}, {Grootes}, {Gunawardhana}, {Jones}, {van Kampen}, {Maraston},
  {Nichol}, {Parkinson}, {Phillipps}, {Pimbblet}, {Popescu}, {Prescott},
  {Roseboom}, {Sadler}, {Sansom}, {Sharp}, {Smith}, {Taylor}, {Thomas},
  {Tuffs}, {Wijesinghe}, {Dunne}, {Frenk}, {Jarvis}, {Madore}, {Meyer},
  {Seibert}, {Staveley-Smith}, {Sutherland}, \& {Warren}}]{GAMA}
{Driver} S.~P. {et~al.}, 2011, \mnras, 413, 971

\bibitem[{{Feldmann} {et~al}\mbox{.}(2006){Feldmann}, {Carollo}, {Porciani},
  {Lilly}, {Capak}, {Taniguchi}, {Le F{\`e}vre}, {Renzini}, {Scoville},
  {Ajiki}, {Aussel}, {Contini}, {McCracken}, {Mobasher}, {Murayama}, {Sanders},
  {Sasaki}, {Scarlata}, {Scodeggio}, {Shioya}, {Silverman}, {Takahashi},
  {Thompson}, \& {Zamorani}}]{ZEBRA}
{Feldmann} R. {et~al.}, 2006, \mnras, 372, 565

\bibitem[{{Firth}, {Lahav} \& {Somerville}(2003){Firth}, {Lahav}, \&
  {Somerville}}]{Firth2003}
{Firth} A.~E., {Lahav} O., {Somerville} R.~S., 2003, \mnras, 339, 1195

\bibitem[{Foster {et~al}\mbox{.}(2009)Foster, Waagen, Aijaz, Hurley, Luis,
  Rinsky, Satyavolu, Way, Gazis, \& Srivastava}]{foster2009}
Foster L. {et~al.}, 2009, Journal of Machine Learning Research, 10, 857

\bibitem[{{Geach}(2012)}]{Geach2012}
{Geach} J.~E., 2012, \mnras, 419, 2633

\bibitem[{Gibbs \& MacKay(1997)}]{gibbs97}
Gibbs M., MacKay D. J.~C., 1997, Efficient implementation of {G}aussian
  processes. Tech. rep., Technical report Cavendish Laboratory, Cambridge, UK

\bibitem[{{Hildebrandt} {et~al}\mbox{.}(2010){Hildebrandt}, {Arnouts}, {Capak},
  {Moustakas}, {Wolf}, {Abdalla}, {Assef}, {Banerji}, {Ben{\'{\i}}tez},
  {Brammer}, {Budav{\'a}ri}, {Carliles}, {Coe}, {Dahlen}, {Feldmann}, {Gerdes},
  {Gillis}, {Ilbert}, {Kotulla}, {Lahav}, {Li}, {Miralles}, {Purger},
  {Schmidt}, \& {Singal}}]{hildebrandt10}
{Hildebrandt} H. {et~al.}, 2010, \aap, 523, A31

\bibitem[{{Hogan}, {Fairbairn} \& {Seeburn}(2015){Hogan}, {Fairbairn}, \&
  {Seeburn}}]{Hogan2015}
{Hogan} R., {Fairbairn} M., {Seeburn} N., 2015, \mnras, 449, 2040

\bibitem[{{Hoyle} {et~al}\mbox{.}(2015){Hoyle}, {Rau}, {Zitlau}, {Seitz}, \&
  {Weller}}]{hoyle2015}
{Hoyle} B., {Rau} M.~M., {Zitlau} R., {Seitz} S., {Weller} J., 2015, \mnras,
  449, 1275

\bibitem[{{Ilbert} {et~al}\mbox{.}(2006){Ilbert}, {Arnouts}, {McCracken},
  {Bolzonella}, {Bertin}, {Le F{\`e}vre}, {Mellier}, {Zamorani}, {Pell{\`o}},
  {Iovino}, {Tresse}, {Le Brun}, {Bottini}, {Garilli}, {Maccagni}, {Picat},
  {Scaramella}, {Scodeggio}, {Vettolani}, {Zanichelli}, {Adami}, {Bardelli},
  {Cappi}, {Charlot}, {Ciliegi}, {Contini}, {Cucciati}, {Foucaud}, {Franzetti},
  {Gavignaud}, {Guzzo}, {Marano}, {Marinoni}, {Mazure}, {Meneux}, {Merighi},
  {Paltani}, {Pollo}, {Pozzetti}, {Radovich}, {Zucca}, {Bondi}, {Bongiorno},
  {Busarello}, {de La Torre}, {Gregorini}, {Lamareille}, {Mathez}, {Merluzzi},
  {Ripepi}, {Rizzo}, \& {Vergani}}]{Ilbert2006}
{Ilbert} O. {et~al.}, 2006, \aap, 457, 841

\bibitem[{Jolliffe(1986)}]{jolliffe1986}
Jolliffe I.~T., 1986, Principal Component Analysis. Springer-Verlag, New York,
  New York

\bibitem[{{Jouvel} {et~al}\mbox{.}(2009){Jouvel}, {Kneib}, {Ilbert},
  {Bernstein}, {Arnouts}, {Dahlen}, {Ealet}, {Milliard}, {Aussel}, {Capak},
  {Koekemoer}, {Le Brun}, {McCracken}, {Salvato}, \& {Scoville}}]{jouvel09}
{Jouvel} S. {et~al.}, 2009, \aap, 504, 359

\bibitem[{Kind \& Brunner(2013)}]{kind2013}
Kind M.~C., Brunner R.~J., 2013, Monthly Notices of the Royal Astronomical
  Society, 432, 1483

\bibitem[{{Laureijs} {et~al}\mbox{.}(2011){Laureijs}, {Amiaux}, {Arduini},
  {Augu{\`e}res}, {Brinchmann}, {Cole}, {Cropper}, {Dabin}, {Duvet}, {Ealet},
  \& et~al.}]{laureijs2011}
{Laureijs} R. {et~al.}, 2011, ArXiv e-prints

\bibitem[{{Le F{\`e}vre} {et~al}\mbox{.}(2013){Le F{\`e}vre}, {Cassata},
  {Cucciati}, {Garilli}, {Ilbert}, {Le Brun}, {Maccagni}, {Moreau},
  {Scodeggio}, {Tresse}, {Zamorani}, {Adami}, {Arnouts}, {Bardelli},
  {Bolzonella}, {Bondi}, {Bongiorno}, {Bottini}, {Cappi}, {Charlot}, {Ciliegi},
  {Contini}, {de la Torre}, {Foucaud}, {Franzetti}, {Gavignaud}, {Guzzo},
  {Iovino}, {Lemaux}, {L{\'o}pez-Sanjuan}, {McCracken}, {Marano}, {Marinoni},
  {Mazure}, {Mellier}, {Merighi}, {Merluzzi}, {Paltani}, {Pell{\`o}}, {Pollo},
  {Pozzetti}, {Scaramella}, {Tasca}, {Vergani}, {Vettolani}, {Zanichelli}, \&
  {Zucca}}]{LeFevre2013}
{Le F{\`e}vre} O. {et~al.}, 2013, \aap, 559, A14

\bibitem[{{Le F{\`e}vre} {et~al}\mbox{.}(2015){Le F{\`e}vre}, {Tasca},
  {Cassata}, {Garilli}, {Le Brun}, {Maccagni}, {Pentericci}, {Thomas},
  {Vanzella}, {Zamorani}, {Zucca}, {Amorin}, {Bardelli}, {Capak},
  {Cassar{\`a}}, {Castellano}, {Cimatti}, {Cuby}, {Cucciati}, {de la Torre},
  {Durkalec}, {Fontana}, {Giavalisco}, {Grazian}, {Hathi}, {Ilbert}, {Lemaux},
  {Moreau}, {Paltani}, {Ribeiro}, {Salvato}, {Schaerer}, {Scodeggio},
  {Sommariva}, {Talia}, {Taniguchi}, {Tresse}, {Vergani}, {Wang}, {Charlot},
  {Contini}, {Fotopoulou}, {L{\'o}pez-Sanjuan}, {Mellier}, \&
  {Scoville}}]{LeFevre2015}
{Le F{\`e}vre} O. {et~al.}, 2015, \aap, 576, A79

\bibitem[{LeCun {et~al}\mbox{.}(1998)LeCun, Bottou, Orr, \&
  Mueller}]{lecun1998}
LeCun Y., Bottou L., Orr G.~B., Mueller K.-R., 1998, Lecture Notes in Computer
  Science, 1524, 9

\bibitem[{{Lilly} {et~al}\mbox{.}(2009){Lilly}, {Le Brun}, {Maier}, {Mainieri},
  {Mignoli}, {Scodeggio}, {Zamorani}, {Carollo}, {Contini}, {Kneib}, {Le
  F{\`e}vre}, {Renzini}, {Bardelli}, {Bolzonella}, {Bongiorno}, {Caputi},
  {Coppa}, {Cucciati}, {de la Torre}, {de Ravel}, {Franzetti}, {Garilli},
  {Iovino}, {Kampczyk}, {Kovac}, {Knobel}, {Lamareille}, {Le Borgne}, {Pello},
  {Peng}, {P{\'e}rez-Montero}, {Ricciardelli}, {Silverman}, {Tanaka}, {Tasca},
  {Tresse}, {Vergani}, {Zucca}, {Ilbert}, {Salvato}, {Oesch}, {Abbas},
  {Bottini}, {Capak}, {Cappi}, {Cassata}, {Cimatti}, {Elvis}, {Fumana},
  {Guzzo}, {Hasinger}, {Koekemoer}, {Leauthaud}, {Maccagni}, {Marinoni},
  {McCracken}, {Memeo}, {Meneux}, {Porciani}, {Pozzetti}, {Sanders},
  {Scaramella}, {Scarlata}, {Scoville}, {Shopbell}, \& {Taniguchi}}]{Lilly2009}
{Lilly} S.~J. {et~al.}, 2009, \apjs, 184, 218

\bibitem[{Lima {et~al}\mbox{.}(2008)Lima, Cunha, Oyaizu, Frieman, Lin, \&
  Sheldon}]{Lima2008}
Lima M., Cunha C.~E., Oyaizu H., Frieman J., Lin H., Sheldon E.~S., 2008,
  Monthly Notices of the Royal Astronomical Society, 390, 118

\bibitem[{Mercer(1909)}]{mercer1909}
Mercer J., 1909, Philosophical transactions of the royal society of London.
  Series A, containing papers of a mathematical or physical character, 415

\bibitem[{Nocedal(1980)}]{jorge1980}
Nocedal J., 1980, Mathematics of Computation, 35, 773

\bibitem[{Rasmussen \& Williams(2006)}]{rasmussen2006gaussian}
Rasmussen C., Williams C., 2006, Gaussian Processes for Machine Learning,
  Adaptative computation and machine learning series. University Press Group
  Limited

\bibitem[{Rau {et~al}\mbox{.}(2015)Rau, Seitz, Brimioulle, Frank, Friedrich,
  Gruen, \& Hoyle}]{rau2015}
Rau M.~M., Seitz S., Brimioulle F., Frank E., Friedrich O., Gruen D., Hoyle B.,
  2015, arXiv preprint arXiv:1503.08215

\bibitem[{Roberts {et~al}\mbox{.}(2013)Roberts, Osborne, Ebden, Reece, Gibson,
  \& Aigrain}]{roberts2012rs}
Roberts S., Osborne M., Ebden M., Reece S., Gibson N., Aigrain S., 2013,
  Philosophical Transactions of the Royal Society A: Mathematical, Physical and
  Engineering Sciences, 371, 20110550

\bibitem[{{S{\'a}nchez} {et~al}\mbox{.}(2014){S{\'a}nchez}, {Carrasco Kind},
  {Lin}, {Miquel}, {Abdalla}, {Amara}, {Banerji}, {Bonnett}, {Brunner},
  {Capozzi}, {Carnero}, {Castander}, {da Costa}, {Cunha}, {Fausti}, {Gerdes},
  {Greisel}, {Gschwend}, {Hartley}, {Jouvel}, {Lahav}, {Lima}, {Maia},
  {Mart{\'{\i}}}, {Ogando}, {Ostrovski}, {Pellegrini}, {Rau}, {Sadeh}, {Seitz},
  {Sevilla-Noarbe}, {Sypniewski}, {de Vicente}, {Abbot}, {Allam}, {Atlee},
  {Bernstein}, {Bernstein}, {Buckley-Geer}, {Burke}, {Childress}, {Davis},
  {DePoy}, {Dey}, {Desai}, {Diehl}, {Doel}, {Estrada}, {Evrard},
  {Fern{\'a}ndez}, {Finley}, {Flaugher}, {Frieman}, {Gaztanaga}, {Glazebrook},
  {Honscheid}, {Kim}, {Kuehn}, {Kuropatkin}, {Lidman}, {Makler}, {Marshall},
  {Nichol}, {Roodman}, {S{\'a}nchez}, {Santiago}, {Sako}, {Scalzo}, {Smith},
  {Swanson}, {Tarle}, {Thomas}, {Tucker}, {Uddin}, {Vald{\'e}s}, {Walker},
  {Yuan}, \& {Zuntz}}]{sanchez14}
{S{\'a}nchez} C. {et~al.}, 2014, \mnras, 445, 1482

\bibitem[{Schmidt(2005)}]{schmidt2005}
Schmidt M., 2005, minfunc: unconstrained differentiable multivariate
  optimization in {M}atlab

\bibitem[{Smola \& Vapnik(1997)}]{smola1997}
Smola A., Vapnik V., 1997, Advances in neural information processing systems,
  9, 155

\bibitem[{Snelson \& Ghahramani(2006)}]{snelson2005}
Snelson E., Ghahramani Z., 2006, in Advances in Neural Information Processing
  Systems 18, Weiss Y., Sch\"{o}lkopf B., Platt J., eds., MIT Press, pp.
  1257--1264

\bibitem[{Tipping(2001)}]{tipping2001}
Tipping M.~E., 2001, Journal of Machine Learning Research, 1, 211

\bibitem[{Tsiligkaridis \& Hero(2013)}]{tsiligkaridis2013}
Tsiligkaridis T., Hero A., 2013, Signal Processing, IEEE Transactions on, 61,
  5347

\bibitem[{Vanzella {et~al}\mbox{.}(2004)Vanzella, Cristiani, Fontana, Nonino,
  Arnouts, Giallongo, Grazian, Fasano, Popesso, Saracco,
  {et~al.}}]{vanzella2004photometric}
Vanzella E. {et~al.}, 2004, Astronomy \& Astrophysics, 423, 761

\bibitem[{{Way} {et~al}\mbox{.}(2009){Way}, {Foster}, {Gazis}, \&
  {Srivastava}}]{Way2009}
{Way} M.~J., {Foster} L.~V., {Gazis} P.~R., {Srivastava} A.~N., 2009, \apj,
  706, 623

\bibitem[{Weiss, McCarthy \& Zabar(2007)Weiss, McCarthy, \& Zabar}]{weiss2007}
Weiss G., McCarthy K., Zabar B., 2007, in DMIN, Stahlbock R., Crone S.~F.,
  Lessmann S., eds., CSREA Press, pp. 35--41

\bibitem[{Zhang, Leithead \& Leith(2005)Zhang, Leithead, \&
  Leith}]{zhang2005time}
Zhang Y., Leithead W.~E., Leith D.~J., 2005, in Decision and Control, 2005 and
  2005 European Control Conference. CDC-ECC'05. 44th IEEE Conference on, IEEE,
  pp. 3711--3716

\end{thebibliography}
}

\label{lastpage}
\end{document}